\documentclass[12pt]{article}

\usepackage[T1]{fontenc}
\usepackage{lscape,multirow}
\usepackage{graphics,amsmath, amssymb,multirow, mathrsfs,graphicx,bm,colordvi,verbatim}
\usepackage{setspace, natbib}
\usepackage{caption, tabularx}
\usepackage{enumitem,colortbl}
\usepackage{url}
\usepackage{tablefootnote, ragged2e}
\usepackage{mathptmx}
\usepackage{pdflscape}
\usepackage{amsthm}
\usepackage{chngcntr}
\usepackage{changepage}
\usepackage{tikz}
\usepackage{tgpagella}
\usepackage{longtable}
\usepackage{pdflscape}
\usepackage{afterpage}

\usepackage{hyperref}
\hypersetup{
    colorlinks=true,
    linkcolor=blue,
    citecolor=blue,
    filecolor=magenta,
    urlcolor=blue
}

\usepackage{booktabs}

\usepackage{dcolumn}
\newcolumntype{d}[1]{D..{#1}}

\newcolumntype{Y}{>{\centering\arraybackslash}X}

\newcommand{\sym}[1]{\rlap{#1}}%
\usepackage{siunitx}
\usepackage[left=1in,right=1in,top=1in, bottom=1in]{geometry}

\newcommand{\st}{\begin{eqnarray}}
    \newcommand{\nd}{\end{eqnarray}}
\newcommand{\stt}{\begin{eqnarray*}}
    \newcommand{\ndd}{\end{eqnarray*}}

\newtheorem{theorem}{Theorem}
\newtheorem{assumption}{Assumption}
\newtheorem{corollary}{Corollary}
\newtheorem{definition}[theorem]{Definition}

\newtheorem{proposition}{Proposition}

\usepackage{setspace}
\renewcommand{\thefootnote}{\fnsymbol{footnote}}

\graphicspath{{./PEfig/}}

\begin{document}
\begin{titlepage}
\title{\bf Fundamentals of Perpetual Futures\thanks{We are grateful to Lin William Cong, Urban Jermann, Shimon Kogan, Stefan Nagel, Tim Roughgarden, Adrien Verdelhan, as well as conference participants at the 2024 Utah Winter Finance Conference and seminar participants at a16z Crypto, Bank of Israel, Hebrew University, Reichman University, the University of Chicago, and the Virtual Derivatives Workshop for their insightful feedback and helpful comments. 

Songrun He is at Washington University in St. Louis (\url{h.songrun@wustl.edu}).  Asaf Manela is at Washington University in St. Louis and Reichman University (\url{amanela@wustl.edu}). Omri Ross is at University of Copenhagen (\url{omri@di.ku.dk}), and Chief Blockchain Scientist at the eToro Group. Victor von Wachter is at University of Copenhagen (\url{victor.vonwachter@di.ku.dk}).}}
\author{Songrun He\quad\quad Asaf Manela \quad\quad Omri Ross \quad\quad Victor von Wachter}
        \date{First draft December 2022. This draft September 2026.}

        \maketitle
\begin{abstract}

Perpetual futures are the most popular cryptocurrency derivatives. Perpetuals offer leveraged exposure to their underlying without rollover or direct ownership. Unlike fixed-maturity futures, perpetuals are not guaranteed to converge to the spot price. To minimize the gap between perpetual and spot prices, long investors periodically pay shorts a funding rate proportional to this gap. We derive no-arbitrage prices for perpetual futures in frictionless markets and bounds in markets with trading costs. Empirically, deviations from these prices in crypto are larger than in traditional currency markets, comove across currencies, and diminish over time. An implied arbitrage strategy yields high Sharpe ratios.
\end{abstract}
~\\  
        {\bf JEL Classification}: G11, G12, G13 \\
        {\bf Keywords}: Perpetual futures, crypto, random-maturity arbitrage, funding rate, clamp

        \thispagestyle{empty}
\end{titlepage}

\newpage
\setcounter{page}{1}
\renewcommand{\thefootnote}{\arabic{footnote}}

\setstretch{1.5}
\doublespacing
\section{Introduction}
Perpetual futures are, by far, the most popular derivative traded in cryptocurrency markets, generating a daily volume of more than \$100 billion.
Prior to the recent collapse of FTX, perpetual futures were among the most actively traded products on the exchange, with the now-bankrupt hedge fund Alameda Research taking the other side of many such leveraged trades. Despite their central role in crypto markets, and their potential adoption by traditional derivative markets, there is relatively little work studying these derivatives. In this paper we ask: what are the theoretical fundamental values of perpetual futures, and how large are deviations from these fundamentals empirically?

Perpetuals are derivatives that allow investors to speculate on or hedge against cryptocurrency price fluctuations using high leverage, without needing to acquire the cryptocurrencies or to roll them over.
Like traditional fixed-maturity futures, at initiation, a long and short counterparty agree on an initial futures price without exchanging any money, other than posting margin with the exchange. Both parties can enter or exit the contract at any time, with profits or losses continuously calculated and allocated to each side's margin account based on prevailing market prices.

Unlike fixed-maturity futures, perpetuals do not expire. This feature enhances the liquidity of the contract, as there is no staggering of contracts with different maturities traded on the exchange, and only a single perpetual futures contract per underlying asset is listed. Moreover, this instrument does not require market participants to roll over their futures positions and is traded 24/7.

Because they have no set expiration date, perpetuals are not guaranteed to converge to the spot price of their underlying asset at any given time, and the usual no-arbitrage prices for perpetuals are not applicable without additional restrictions.
To minimize the gap between perpetual futures and spot prices, long position holders periodically pay short position holders a 'funding rate' proportional to this gap to incentivize trades that narrow it.
For example, when the futures price exceeds the spot price, arbitrageurs who borrow cash to long the spot and simultaneously short the futures would receive the funding rate. Their trades would tend to increase the spot price and decrease the futures price.
A narrow gap means that perpetual futures offer effective exposure to variation in the spot price of the underlying asset to hedging and speculating investors. Typically, the funding rate is paid every eight hours and approximately equals the average futures-spot spread over the preceding eight hours.
Note that the strategy just sketched, commonly referred to as 'funding rate arbitrage,' is not risk-free even disregarding margin requirements and trading costs, simply because there is no predetermined expiration date when the trade would be unwound at a profit.

We derive no-arbitrage prices for perpetual futures in frictionless markets and derive no-arbitrage bounds in markets with trading costs.
The theoretical perpetual futures price is proportional to the spot price of the underlying, with a constant of proportionality determined by the interaction of interest rates and the exchange's funding rate mechanism. A key feature of this mechanism is the `clamp' - when futures and spot prices are close, the funding rate is fixed at a small positive value, but when they diverge significantly, the funding rate increases linearly with the price gap to incentivize convergence. The interest rate captures the cost of borrowing cash to finance holding the underlying, while the funding rate intensity captures the benefit of shorting the futures. Thus, intuitively, the futures-spot spread is larger when the cash borrowing interest rate is large relative to the funding rate.

Our derivation relies on a generalized notion of arbitrage that we call \emph{random-maturity arbitrage}. Unlike traditional riskless arbitrage, which fixes the liquidation date in advance, we allow the strategy to be unwound at a bounded stopping time determined by the evolution of prices.
At first glance, one might object that such prices are not truly based on riskless arbitrage. Note, however, that riskless no-arbitrage pricing is usually just a useful fiction \citep{pedersen2019efficiently}. In real-world futures markets, arbitrageurs must maintain a margin account during the entire period in which the arbitrage trade is open. Temporary worsening of apparent arbitrage opportunities can lead to liquidations and losses. As the saying goes, an arbitrageur must remain liquid longer than the market stays irrational. Thus, even arbitrage opportunities that appear to be riskless in theory may be risky in practice.

Our no-arbitrage prices provide a useful benchmark for perpetual futures and simultaneously prescribe a strategy to exploit divergence from these fundamental values. Our analysis allows for jumps in both futures and spot prices. 
Motivated by the theoretical understanding, we study the empirical deviations of the perpetual futures price from the spot. We find substantial and time-varying deviations, with perpetual futures trading above the no-arbitrage benchmark on average and converging to our suggested benchmark over time.

Around this benchmark, however, perpetual futures often differ significantly from spot prices.
The mean \emph{absolute} deviation is about 52\% to 90\% per year across different cryptocurrencies, which is considerably larger than the deviations documented in traditional currency markets by \cite*{du_deviations_2018}.
We find strong comovement of the futures-spot gap across different cryptocurrencies. This comovement can be due to commonality in funding and market liquidity faced by arbitrageurs who operate in multiple cryptocurrencies. Common sentiment could also drive the difference in futures demand relative to the spot. Overall, the deviation's magnitude is comparable to our theoretical no-arbitrage bound calibrated to actual trading fees.

Deviations decline on average about 22 percentage points a year, consistent with an increase in arbitrage capital and competition among arbitrageurs \citep{kondor_risk_2009}. The narrowing gap provides an additional perspective on the downfall of Alameda Research and Three Arrows Capital. According to news reports and interviews, both hedge funds seem to have pivoted from such arbitrage activity around late 2021 to early 2022 and started taking more directional bets on cryptocurrencies, with both direct unhedged crypto holdings and investments in crypto startups. The large declines in crypto prices in 2022 subsequently exhausted their capital and led to their bankruptcies.\footnote{See, for example, Forbes, November 19, 2022, on \href{https://www.forbes.com/sites/jeffkauflin/2022/11/19/how-did-sam-bankman-frieds-alameda-research-lose-so-much-money}{How Did Sam Bankman-Fried's Alameda Research Lose So Much Money?}, Odd Lots, November 17, 2022, on \href{https://www.bloomberg.com/news/articles/2022-11-17/odd-lots-podcast-understanding-sam-bankman-fried-s-ftx-crypto-collapse}{Understanding the Collapse of Sam Bankman-Fried's Crypto Empire}, and Hugh Hendry's interview on December 3, 2022 of Kyle Davies on the \href{https://www.youtube.com/watch?v=TzGdkB0xbCE}{Collapse of Three Arrows Capital}.}

To understand the economics of the futures-spot spread, we consider a trading strategy motivated by the random-maturity arbitrage theory. Whenever the futures-spot spread exceeds the theoretical bound for a given trading cost tier, we open the trading position and close it when the futures-spot spread returns to its theoretical relationship under no trading costs. We find that empirically, the random maturity arbitrage strategy generates a sizable Sharpe ratio even under high trading costs. For example, for Bitcoin perpetual futures, the strategy generates a Sharpe ratio of 3.35 under high trading costs typical of retail investors, and up to 11.65 for highly active market makers who pay no such fees. After additionally accounting for effective bid-ask spreads, the corresponding Sharpe ratios remain 3.27 and 10.46. The strategy's performance is even better for Ether and other cryptocurrencies and delivers significant alphas relative to the 3-factor model of \citet*{liu2022common} and the 5-factor model of \citet*{cong2022value}.

The strong performance of our simple trading strategy, which exploits deviations from the random-maturity no-arbitrage bounds we derive, provides compelling evidence for the usefulness of these bounds as a benchmark for perpetual futures prices. The strategy generates high Sharpe ratios, even for investors subject to the highest trading costs on Binance. This indicates that the no-arbitrage prices serve as an effective anchor for perpetual futures, with profitable random maturity arbitrage opportunities arising when market prices diverge significantly from these fundamental values. Thus, the success of the trading strategy underscores the practical value and validity of our theoretical no-arbitrage results.

What explains these large no-arbitrage deviations? We organize the evidence with a simple model in which end users demand perpetual futures for leveraged exposure, while arbitrageurs take the other side and hedge in the spot market subject to trading costs and limited balance-sheet capacity. The model delivers two predictions. First, relaxing arbitrageurs’ capital constraints should compress futures-spot deviations. Second, end-user demand should have a larger price impact when arbitrage capacity is constrained.

Both predictions receive support in the data. We exploit Binance’s April 2022 introduction of portfolio margining, which allowed arbitrageurs to net collateral across their spot and futures positions and thereby increased the capital efficiency of the arbitrage trade. Deviations compress significantly following its introduction. We also find that perpetual-futures order flow has substantially greater price impact when initial deviations are large, when arbitrage capacity is more likely to bind. Together, these results point to limited arbitrage capacity as an important force behind the large deviations we document. Finally, we find that past returns positively predict futures-spot deviations. Following high spot returns, perpetual futures tend to trade at higher prices relative to the no-arbitrage benchmark, consistent with extrapolative demand for leveraged perpetual exposure.

The existing literature on perpetual futures mainly focuses on empirical evidence. See e.g. \citet*{alexander2020bitmex}, \citet*{de2022arbitrage}, \citet*{ferko2022trades}, and \citet*{streltsov2022perpetual}. \citet*{christin2022crypto} present a novel study on the strategy of longing the spot and shorting the perpetual futures to understand the demand pressure in the market. They document a significant return to this trading strategy and attribute the profits to differences of opinion and leverage constraints in the market. We find that our dynamic real-time trading strategy outperforms their directional bet even under a high trading cost scenario. Our paper provides a general explanation for their findings as perpetual futures are more expensive relative to the theoretical no-arbitrage price for most of the recent sample period across major cryptocurrencies. Therefore, longing the spot and shorting the futures would be consistent with arbitrage trading directions.  

On the theory side, \citet*{angeris2022primer} provide a theoretical no-arbitrage analysis of perpetuals, which relies on a strong assumption that the payoff from the perpetual is a fixed function of the underlying spot price and focuses on the security design problem. Their approach would not be able to analyze arbitrage trading in case of a deviation from the no-arbitrage price. \cite{ackerer_perpetual_2023} derive no-arbitrage prices for various perpetual contracts in both discrete and continuous-time by assuming the existence of an equivalent martingale measure. Our work advances this literature in three key ways: First, we derive no-arbitrage prices that incorporate realistic funding specifications, including the clamp feature used by major exchanges like Binance. Second, we provide explicit arbitrage bounds and trading strategies for markets with transaction costs. Third, we demonstrate empirically that our benchmark improves arbitrage strategy performance compared to existing approaches. We further relate our work to both papers below.

Also related is a recent literature on fixed maturity futures in the crypto market. \citet*{schmeling2022crypto} provide a comprehensive analysis of the carry of fixed-maturity crypto futures, with the carry defined following the general definition of \citet*{koijen2018carry}. They document volatile convenience yields for fixed-maturity crypto futures and attribute this to trend-chasing small investors and the relative scarcity of arbitrage capital. \citet*{cong2022staking} provide a novel link between the volatile convenience yield and staking, service flow, and transaction convenience of the underlying tokens. They show that the large deviation from uncovered interest rate parity can be reconciled with transaction convenience. Our paper focuses on perpetual futures rather than fixed maturity futures and extends fixed-maturity to random-maturity no-arbitrage pricing.

More broadly, our paper contributes to the understanding of frictions and arbitrage in cryptocurrency markets. \citet*{makarov2020trading} study price deviations across exchanges. They find large gaps across countries, highlighting the important role played by capital controls and slow-moving arbitrage capital as in \citet{duffie2010presidential}. Our analysis focuses on the price wedge between the spot and the futures market. We find that even within an exchange, futures prices deviate from their theoretical arbitrage-free values. These results indicate there are significant limits to arbitrage as in \citet{gromb2018dynamics} for cryptocurrencies in the early years. Finally, our finding that deviations from the no-arbitrage benchmark in crypto markets are highly correlated across currencies suggests that a common risk factor drives this variation as documented by \citet*{lustig2011common} in traditional currency markets.

The rest of the paper is organized as follows: Section \ref{sec:background} provides the institutional details and history of perpetual futures. Section \ref{sec:theory} presents the no-arbitrage analysis of the perpetual futures market and derives the theoretical price of perpetual futures. Section \ref{sec:empirical} demonstrates the empirical futures-spot deviation, implements the theory-motivated arbitrage strategy, and examines execution costs and liquidation risk.
Section \ref{sec:explaining} studies the economic mechanism behind the deviations and tests predictions concerning arbitrage capacity and the price impact of end-user demand.
Section \ref{sec:conclusion} concludes.

\section{An Introduction to Perpetual Futures}\label{sec:background}
The idea of perpetual futures was first introduced by \citet{shiller1993measuring}. The goal was to set up a perpetual claim on the cash flows of an illiquid asset. For example, the underlying illiquid asset could be a house that generates rents as cash flows. The purpose of perpetual futures is to enable price discovery for the underlying with an illiquid or hard-to-measure price. Perpetual futures have no expiration date but cash is exchanged between the long and the short side: after buying the perpetual futures, the long side is entitled to receive the cash flow from the short side and they settle the price difference when exiting the position.

\begin{figure}[!htb]
    \begin{center}
    \includegraphics[width = \textwidth]{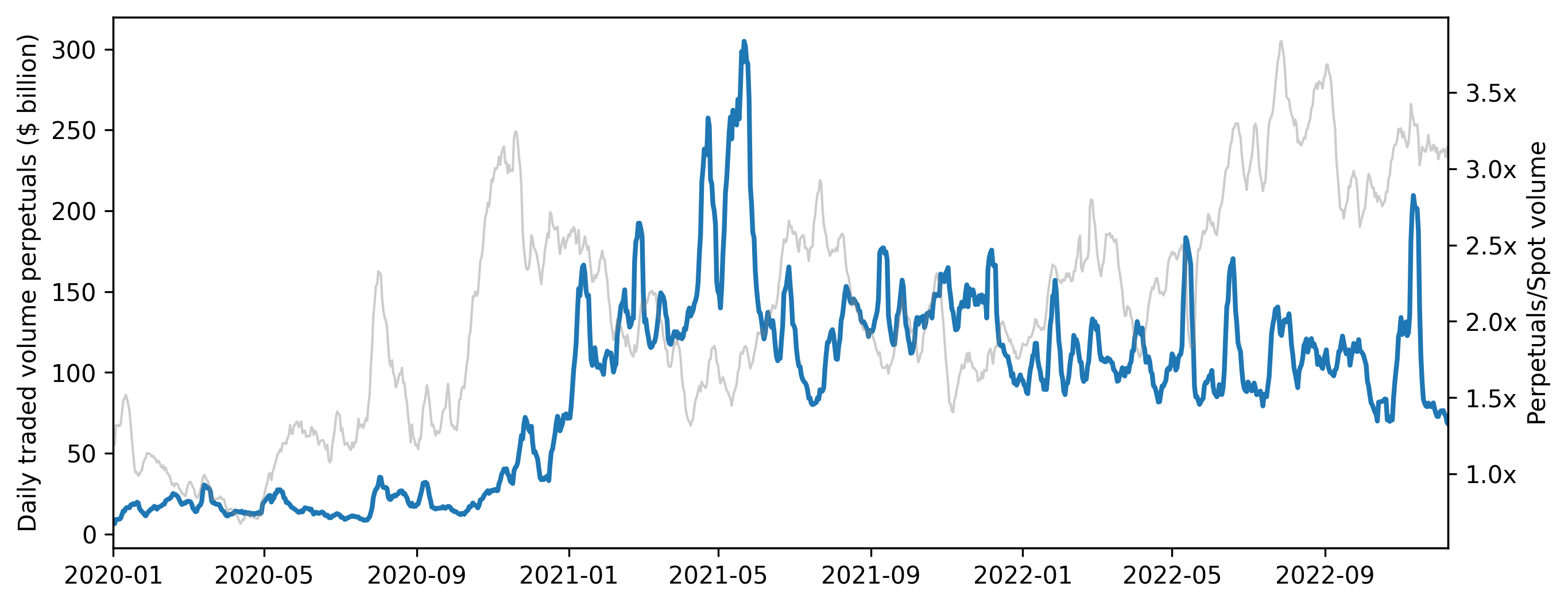}
    \end{center}
    \caption{Total trading volumes of perpetual futures across exchanges}
    \label{fig:vol_exch}
    \bigskip
    \small
    The figure displays the 7-day moving average daily traded volume for perpetual futures across all exchanges in blue. The median daily volume is \$17.8 bn. (2020), \$132.0 bn. (2021) and \$101.9 bn. (2022). This translates to a yearly volume of \$8,551 bn. (2020), \$51,989 bn. (2021) and \$39,306 bn. (2022) respectively. Additionally, the grey line depicts the ratio between the traded volume in perpetuals and spot markets. In 2022, perpetual futures markets consistently traded between 2x and 3.5x spot volume. The data is obtained from CoinGecko, a crypto data specialist. We exclude exchanges that are known for misrepresenting data (e.g., forms of wash trading).
\end{figure}

Perpetual futures in crypto markets similarly have no expiration date and cash is exchanged between the long and the short side, but their purpose is different from Shiller's original idea. First, unlike, e.g., the real estate market, crypto has no inherent dividend or cash flow. Second, the price discovery argument of Shiller is most applicable to settings where spot prices are difficult to measure. Crypto spot prices, however, can be measured from active trading on different exchanges, and decentralized exchanges such as Bancor \citep{hertzog2017bancor} or Uniswap \citep{adams2021uniswap} offer price discovery for assets with minimal liquidity. The major role perpetual futures play in the crypto space is to offer an effective leveraged trading vehicle to hedge against or speculate on the underlying spot price movement, which makes the market more complete. Perpetual futures can also facilitate tax management. For example, an investor holding an appreciated token can reduce or hedge the associated price exposure by taking an offsetting perpetual futures position rather than selling the underlying token. In jurisdictions where capital gains on the token are recognized upon disposition, this may allow investors to defer realization of the embedded capital gain while retaining the underlying position. 

Crypto perpetual futures were first introduced by BitMEX in 2016, which have gained great popularity in the crypto space since their inception. It initially served as an effective hedging tool for crypto miners. It was later adopted by crypto speculators interested in leveraged exposure.
Nowadays, based on data from CoinGecko, the median total daily trading volume of perpetual futures across all exchanges is 101.9 billion in the year 2022 which is about $2\times$ to $3\times$ the total spot trading volume across these exchanges. Figure \ref{fig:vol_exch} presents the 7-day moving average of the total trading volume of perpetual futures across all exchanges. We see a significant rise in trading of perpetual futures around January 2021 and the total volume stabilizes at a level above \$100 billion per day following the rise.

More recently, decentralized perpetual futures venues have emerged as significant competitors to centralized exchanges. A prominent example is Hyperliquid\footnote{\url{https://hyperliquid.gitbook.io/hyperliquid-docs}.}, a high-performance decentralized exchange built on its own Layer~1 blockchain with a fully on-chain order book. Launched in 2023 in the wake of the FTX collapse, Hyperliquid aims to combine the speed and usability of centralized venues with the transparency of decentralized finance. Its rapid growth in trading volume, including its ability to process large-scale liquidations during the October 2025 crypto flash crash, illustrates a broader trend toward on-chain derivatives infrastructure that could reshape the competitive landscape of perpetual futures markets. The contract specification on Hyperliquid mirrors that of Binance, including the clamp feature. Our theoretical results, therefore, apply directly to this exchange as well.

\citet*{cong2022crypto} document significant wash trading behavior among crypto exchanges because of competition, the ranking mechanism, and lack of regulation. They estimate that over 70\% of crypto trading volume is not real. \citet*{amiram2020competition} confirm this conclusion using new data and an extended methodology. Therefore, in calculating trading volume, we exclude exchanges that are known to misrepresent data.

\begin{figure}[!htb]
\includegraphics[width=\textwidth]{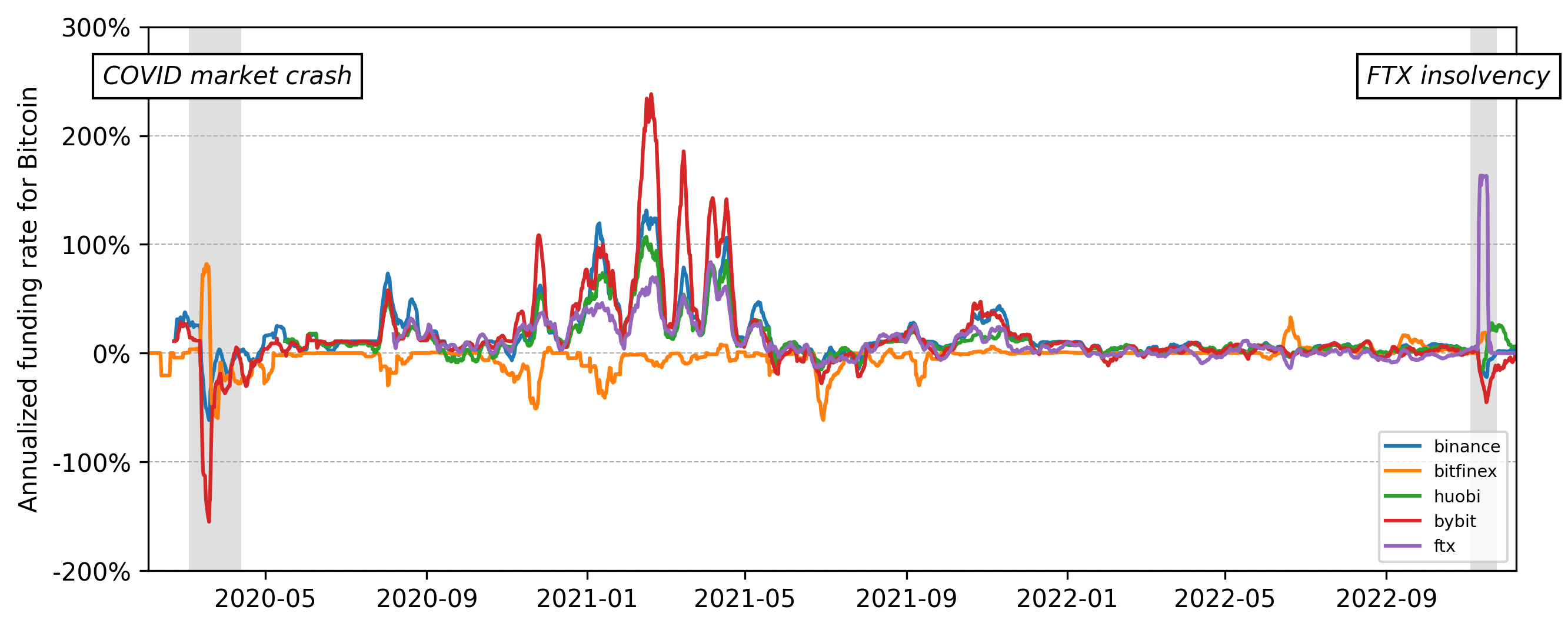}
\caption{Bitcoin annualized funding rate across exchanges}\label{fund_exchange}
\bigskip
\small
The figure presents the 7-day moving average annualized funding rate for Bitcoin (BTC) across exchanges. The data is obtained from Glassnode, an analytics platform. The data covers two major market turbulences: the COVID stock market crash from February to April 2020 and the FTX insolvency in November 2022. The FTX collapse led to significant negative funding rates on all solvent exchanges. Thus, the perpetual futures prices were lower than the spot prices on the solvent exchanges. Vice versa on the insolvent FTX. The funding rates become more volatile after the event, indicating increased uncertainty for the market participants.
\end{figure}

A key feature of crypto perpetual futures is the funding rate, which is the cash exchanged between the long and short counterparties. Its goal is to keep the futures price close to the underlying spot so that the futures can be an effective hedging tool for spot price movement. The funding rate is typically paid every 8 hours. Its value is approximately a weighted average of the prior 8 hours' price gap of the futures and the spot. If the futures price is above the spot, the funding rate will be positive, meaning the long side of the futures needs to pay the short side. This incentivizes traders to short the futures, and in doing so, to move its price back in line with the spot. On the other hand, when the futures price is below the spot, the funding rate turns negative, which means the short side needs to pay the long side. Therefore, the funding rate is the key mechanism keeping the futures price close to the spot price.\footnote{\url{https://www.binance.com/en/support/faq/360033525031} provides a detailed explanation of how the funding rate is calculated. We summarize the mechanism in Appendix \ref{sec:fund_binance}.} Note that perpetual futures cannot be replicated by rolling over 8-hour maturity futures. For the latter, every 8 hours, the price is guaranteed to converge to the underlying spot, while for the perpetual futures this is not the case.

Figure \ref{fund_exchange} presents the annualized 7-day moving average of Bitcoin perpetual funding rates across leading exchanges from January 2020 to December 2022.
The funding rate is positive when the futures-to-spot spread is positive, and negative otherwise. Funding rates tend to be similar across exchanges due to cross-exchange arbitrage activity, but can diverge during extreme liquidity episodes. During March 2020, as COVID-19 started to spread and liquidity evaporated, funding rates turned substantially negative in most exchanges. Funding rates turned highly positive during the crypto bull run of early 2021. The last episode highlighted is the collapse of FTX, which was the 4th largest crypto exchange at the time. The figure shows that Bitcoin futures prices were substantially higher than spot prices at FTX, but the opposite was true on other exchanges. This pattern is consistent with FTX investors liquidating their short futures positions quickly, either voluntarily to reduce their exposure to the failing exchange, or involuntarily as the exchange liquidated their underfunded positions.

Two forces account for the remaining dispersion in funding rates in Figure \ref{fund_exchange}. First, there are segmentations in arbitrage across exchanges because exchanges do not offer cross-margining across venues. An arbitrageur exploiting a funding-rate differential must therefore fund and margin the two legs separately, tying up capital on each exchange and limiting the amount of arbitrage capital that can respond to a given discrepancy. Second, exchanges differ substantially in liquidity. From 2020 to 2022, Binance accounts for approximately 49\% of Bitcoin perpetual trading volume across the five exchanges in the figure, compared with only 0.2\% for Bitfinex. Funding rates on a relatively illiquid venue such as Bitfinex can therefore be more sensitive to exchange-specific order flow and trading conditions.

The comparison between Binance and Bybit can further illustrate this mechanism. Their funding rates are highly correlated, at 0.74, consistent with active cross-exchange arbitrage, yet meaningful divergences remain during periods of unusually large funding payments. We apply the effective spread estimator of \citet{roll1984simple} to trade-level data from the two venues and find that the effective spread of the Bitcoin perpetual on Bybit exceeds that on Binance on every day in our sampled subset, by a factor ranging from two to ten.\footnote{We sample seven days spanning the episodes in Figure \ref{fund_exchange}, using the trade-by-trade archives that Binance and Bybit publish for the BTC perpetual.} Because of the differences in liquidity and arbitrage capital, the same order flow can move the perpetual futures price further on Bybit compared to Binance. Thus, although arbitrage keeps prices and funding rates closely linked across exchanges in normal times, it does not eliminate venue-specific deviations. When liquidity is thinner, and arbitrage capital is more constrained, a given imbalance in demand can generate a larger price response and, consequently, a larger funding-rate differential.

Therefore, the residual cross-exchange dispersion in Figure \ref{fund_exchange} reflects the limits of arbitrage across segmented trading venues, which can be a promising avenue for future research.

\section{Arbitrage in Perpetual Futures}\label{sec:theory}
We next derive no-arbitrage prices for perpetual futures relative to spot prices. Unlike traditional futures, perpetual futures have no expiration date. To analyze arbitrage in this market, we generalize the traditional notion of risk-free arbitrage, where arbitrageurs have a guaranteed positive payoff at a certain time in the future. We first describe the payoff structure of perpetual futures. We then introduce a generalized notion of an arbitrage opportunity with a certain positive payoff but at an uncertain future time.

\begin{definition}[Perpetual futures]\label{def:perpetual}
A perpetual futures contract with price $\{F_t\}_{t=0}^\infty$ written on an underlying asset with spot price $\{S_{t}\}_{t = 0}^\infty$ is a swap agreement between the long and the short side. Define the premium index at time $s$ as $p_s \equiv \frac{F_s-S_s}{F_s}.$ There is no exchange of principal to open the position. Once open, both the long side and the short side can terminate their position at any time $t$. To stay in the contract, both sides have to meet two requirements. (1) For each unit of the perpetual, the long and short sides must exchange an $\mathcal{F}_s-$adapted cash flow $\Phi(p_s) F_s ds,$ referred to as the funding payment, at each time $s\in (0,t)$. The funding rate function $\Phi(\cdot): \mathbb{R}\rightarrow \mathbb{R}$ is a univariate function of the premium index. If $\Phi(p_s)$ is positive, the long pays the short, and vice versa. (2) Both sides need to post mark-to-market margins to cover any losses in their margin accounts.

\end{definition}

Let $(\Omega,\mathcal{F},\{\mathcal{F}_t\}_{t\geq 0},\mathbb{P})$ be a
filtered probability space, where $\mathcal{F}_t$ represents the information available
to the arbitrageur at time $t$. The traditional notions of arbitrage typically consider a guaranteed positive payoff at a certain future date. For example, \cite{cochrane2009asset} defines arbitrage as follows. Let $X_t$ denote the net payoff from a trading strategy if all positions are liquidated at time $t$.

\begin{definition}[Riskless Arbitrage]\label{def:riskless}
A riskless arbitrage is a trading strategy with nonpositive initial cost and a
deterministic future time $T<\infty$ such that
(1) $X_T\geq 0$ almost surely and
(2) $\Pr(X_T>0)>0$.
\end{definition}

There is no certain future time, however, when a perpetual futures contract expires, and its price is guaranteed to converge to some fixed function of the spot price. Therefore, a generalized notion of arbitrage is required. We define \emph{random-maturity arbitrage} opportunities as zero-cost strategies with a guaranteed positive payoff at a bounded uncertain future time. Stated formally:

\begin{definition}[Random-maturity arbitrage]\label{def:random_maturity_arb}
A random-maturity arbitrage is a trading strategy with nonpositive initial cost together
with an $\{\mathcal{F}_t\}$-stopping time $\tau$ such that, for some deterministic
$\bar{T}<\infty$,
(1) $\tau\leq\bar{T}$ almost surely,
(2) $X_\tau\geq 0$ almost surely, and
(3) $\Pr(X_\tau>0)>0$.
\end{definition}

Definition \ref{def:random_maturity_arb} generalizes traditional arbitrage in the sense that there is a guaranteed positive payoff but at an uncertain future time. Because $\tau$ is a stopping time, the decision to unwind at time $t$ can depend only on information available up to time $t$. Riskless arbitrage is the special case in which $\tau\equiv T$ is deterministic. The following corollary specializes this definition for perpetual futures:

\begin{corollary}\label{cor1}
In the perpetual futures market, if a strategy (1) has $0$ cost at time $0$, and (2) for any price path of the futures and the spot, $\{F_t\}_{t=0}^\infty$ and $\{S_t\}_{t = 0}^\infty$, there exists a bounded unwinding time $\tilde{\tau}$ such that the strategy's discounted payoff is positive, then this strategy is a random-maturity arbitrage.
\end{corollary}

Throughout our no-arbitrage analysis, we abstract from exchange default risk and assume that the exchange remains solvent. If the exchange can default before the unwinding stopping time, funding payments, gains on the futures position, or posted collateral may not be fully recovered. The convergence strategy would then no longer have a guaranteed nonnegative payoff and therefore need not constitute a random-maturity arbitrage under our definition. Thus, our no-arbitrage results should be interpreted as conditional on exchange solvency.

Our analysis is quite general and allows for jumps in both futures and spot prices. We next show that when there is no random-maturity arbitrage, the futures price is a fixed constant times the spot price. First, we state three key assumptions needed for our results:\footnote{For a function $f$, we define its generalized (set-valued) inverse as: $f^{-1}(y)=\{x: f(x)=y\}.$ }

\begin{assumption}
The risk-free short rate $r$ in the cash market is constant. \label{assp1}
\end{assumption}

\begin{assumption}
The ratio between the price of perpetual futures and the spot is bounded: $0 < \underline{R} \leq \frac{F_t(\omega)}{S_t(\omega)}\leq \overline{R} < \infty$, $\forall t, \omega$.\label{assp2}

\end{assumption}

\begin{assumption}
The funding rate function $\Phi(\cdot)$ is weakly increasing in its argument, and its level set at the risk-free rate, $\Phi^{-1}(r)$, is a nonempty compact interval $[x_-, x_+]$ with $x_+ < 1$.\label{assp3}
\end{assumption}

Assumption \ref{assp1} guarantees that there is no roll-over risk for the arbitrageurs. In Appendix \ref{sec:stochastic}, we relax this assumption and show that stochastic short rates leave the structure of the arbitrage payoff intact up to an empirically small rebalancing term. Assumption \ref{assp2} says that the ratio between the perpetual futures price and the spot price is bounded away from zero and infinity. Assumption \ref{assp3} states that the design of the contract needs to guarantee that there is a reward for convergence trade. The reward should be weakly increasing when the deviation becomes larger. The condition that $\Phi^{-1}(r)$ is a nonempty compact interval ensures that the funding rate reaches the risk-free rate over a well-defined range of deviations, while $x_+<1$ ensures that the corresponding equilibrium futures price is positive. With these three assumptions, we can state our result as follows.

\begin{proposition}[No-arbitrage price: General Case]\label{prop1}
Suppose $\Phi^{-1}(r)$ is a single point. Under Assumptions \ref{assp1}, \ref{assp2}, and \ref{assp3}, there is an absence of random-maturity arbitrage opportunities if and only if
\begin{equation}
    F_t = (1 - \Phi^{-1}(r))^{-1} S_t.\label{benchmark}
\end{equation}
\end{proposition}

\begin{proof}
The proof is provided in Appendix \ref{sec:proof}.
\end{proof}

We next provide the economic intuition behind Proposition \ref{prop1} and derive the payoff expressions that we use below. Start off with the conjecture that in equilibrium, the futures price is a linear function of the spot, $F_t = \lambda S_t$, for some constant $\lambda$. If this conjecture is true, it implies both arbitrage strategies laid out in Table \ref{tabcase1} should always have $0$ payoff. 

\begin{table}[!htb]
\begin{center}
\footnotesize
\begin{tabularx}{\textwidth}{llYY}
\toprule
& & $F_0 > \lambda S_0$ & $F_0 < \lambda S_0$ \\
\midrule 
Actions & & Long $\lambda$ spot, short 1 futures & Long 1 futures, short $\lambda$ spot \\
\midrule
& Futures & 0 & 0\\
Time $0$& Spot & $-\lambda S_0$ & $+\lambda S_0$ \\
& Cash & $+\lambda S_0$ & $-\lambda S_0$ \\
\midrule
\multirow{4}{*}{Time $t$}& Futures & $-\int_{0}^t e^{-r s} dF_s$ & $\int_{0}^t e^{-r s} dF_s$ \\
& Funding & $\int_0^t \Phi(p_s) F_s e^{-r s}ds$ & $-\int_0^t \Phi(p_s) F_s e^{-r s}ds$ \\
& Spot & $\lambda\int_0^t e^{-r s}dS_s$ & $-\lambda \int_0^t e^{-r s}dS_s$ \\
& Cash & $-\lambda\int_0^t e^{-r s}r S_s ds$ & $\lambda \int_0^t e^{-r s}r S_s ds$ \\
\cmidrule{2-4}
& \multirow{2}{*}{Payoff} & $-\int_0^t e^{-r s} d(F_s - \lambda S_s)+$ & $\int_0^t e^{-r s} d(F_s - \lambda S_s)-$ \\
& & $\int_{0}^t e^{-r s} \left(\Phi(p_s) F_s - r\lambda S_s\right)ds $ & $\int_{0}^t e^{-r s} \left( \Phi(p_s) F_s - r\lambda S_s\right)ds $ \\
\bottomrule
\end{tabularx}
\end{center}
\caption{Discounted payoffs to arbitrage strategies in perpetual futures and spot markets}\label{tabcase1}
\bigskip
\small
This table presents the costs and benefits of two arbitrage trading strategies: (1) long $\lambda$ spot and short 1 futures; (2) long 1 futures and short $\lambda$ spot. In the last row, the payoff from exiting the position is the discounted payoff from futures and spot price changes, proceeds from the cash market, and the funding rate.
\end{table}

We can write down the payoff of the first strategy as:
\begin{equation}
    -\int_0^t e^{-r s} d(F_s - \lambda S_s)+\int_{0}^t e^{-r s} \left(\Phi(p_s) F_s - r\lambda S_s\right)ds = \int_0^t e^{-rs} (\Phi(p_s) - r) F_s ds = 0,
\end{equation}
where the first equality follows from the condition $F_t = \lambda S_t,\, \forall t$. From the second equality, we obtain the key condition for pinning down the value of $\lambda$:
\begin{equation}
    \Phi \left(\frac{F_t-S_t}{F_t}\right) = r, 
\end{equation}
which states that in equilibrium, the funding rate must equal the risk-free rate. When $\Phi^{-1}(r)$ can be mapped to a single point, it pins down the equilibrium $\lambda$:
\begin{equation}
\lambda = (1 - \Phi^{-1}(r))^{-1}.
\end{equation}
Thus, $\lambda$ depends on the risk-free rate $r$. Under Assumption
\ref{assp1}, $r$ is constant, so $\lambda(r)$ is constant over time; for
notational simplicity, we henceforth write $\lambda\equiv\lambda(r)$. Therefore, the relationship between F and S is exactly the one given in Equation \eqref{benchmark}.

If Equation \eqref{benchmark} always holds, then both strategies in Table \ref{tabcase1} have zero payoff forever. Indeed, in this case the perpetual futures is redundant given the spot and cash markets. Since the spot and cash markets admit no arbitrage, adding the perpetual futures cannot create a random-maturity arbitrage opportunity.

Next, consider a deviation from this benchmark, where $F_0 > \lambda S_0$ (the opposite case has similar intuition). In this case, we consider the first strategy in Table \ref{tabcase1}. 
For a fixed position in the convergence trade not to be an arbitrage opportunity, it must be that for all times $t$ it does not generate a positive payoff by time $t$, which requires that:
\begin{equation}
     \underbrace{F_0 - \lambda S_0}_{\text {initial spread }} + \underbrace{\int_0^t e^{-rs} (\Phi(p_s) - r) F_s ds}_{\text {funding payments }} \leq
    \underbrace{e^{-r t} 
    \left(F_t - \lambda S_t \right).}_{\text {discounted spread at unwinding }} \label{breakdown}
\end{equation}
Our proof shows that this cannot occur under the maintained assumptions.

To gain an intuition for these results, consider the three terms of Equation \eqref{breakdown}. The first term is the initial spread. The second term is the present value of cumulated funding payments from initiation to unwinding. These first two terms account for gains to the arbitrageur in this scenario. The last term is the spread at unwinding discounted to the present, which is the source of tension here. For fixed-maturity futures, this spread is guaranteed to be zero at expiration. However, an arbitrageur in perpetual futures faces the risk that when they wish to unwind the trade, the spread may be large.

Our key insight is that a positive and even modestly large spread can still leave the arbitrageur with a positive payoff because funding payments also increase with the spread. As long as the spread does not diverge to infinity, there will be some future bounded unwinding time $t$ when the accumulated funding rate payments overcome any finite potential losses at unwinding.

Note that if we relax the assumption that the short rate is constant for arbitrageurs, this may introduce an additional risk to four parts in Table \ref{tabcase1}: (1) mark-to-market changes in futures price, (2) funding-rate payment, (3) reinvestment in the spot market, and (4) rollover of the short-term debt. In Appendix \ref{sec:stochastic}, we show that a stochastic short rate changes the payoff to the arbitrage strategies by a single additional term, a small rebalancing cash flow that accrues as the hedge ratio tracks the interest rate, and widens the no-arbitrage bound of Proposition \ref{prop3} accordingly. This extension relates to the literature examining the differences between futures and forwards in the context of stochastic interest rates, such as \citet{cox1981relation} and \citet{grinblatt1996relative}. They show that marking-to-market introduces an additional term in the futures price related to the covariance between the underlying asset and interest rate. 

Empirically, however, the interest rate changes required to rationalize the deviations we document are one to two orders of magnitude larger than observed movements in financing rates. Moreover, Appendix \ref{sec:stochastic} further shows that the additional payoff exposure generated by stochastic rates is a small rebalancing cash flow arising as the hedge ratio adjusts with the short rate.
Thus, stochastic interest rate risk is unlikely to account for the deviations we observe.

\begin{figure}[!htb]
    \begin{center}
    \includegraphics[width = 0.6\textwidth]{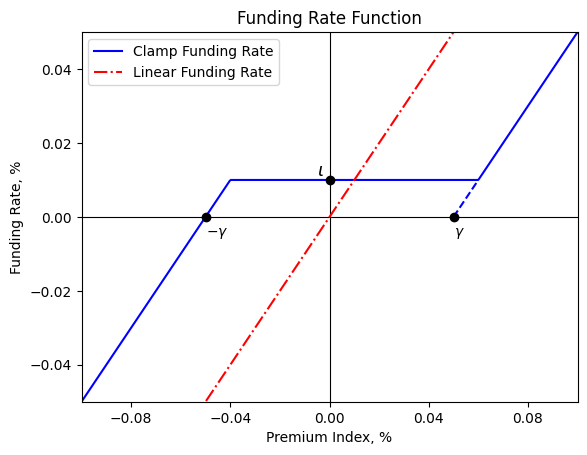}
    \end{center}
\caption{The Funding Rate Function $\Phi$ on Binance}\label{figfr}
\bigskip
\small
This figure plots the funding rate as a function of the average premium index, $\frac{F-S}{F}$.
\end{figure}

For completeness, we next consider the edge case where for a given interest rate, the inverse of the funding rate function maps to an interval rather than a point. From Figure \ref{figfr}, we can see that this can happen on Binance if the interest rate exactly equals $\iota$.
\begin{proposition}[No-arbitrage price: Edge Case]\label{prop2}
Suppose $\Phi^{-1}(r)$ is an interval. Under Assumptions \ref{assp1}, \ref{assp2}, and \ref{assp3}, there is an absence of random-maturity arbitrage opportunities if and only if
\begin{equation}
F_t \in \left[\left(1 - \inf\left\{\Phi^{-1}(r)\right\}\right)^{-1}S_t, \left(1 - \sup\left\{\Phi^{-1}(r)\right\}\right)^{-1}S_t \right].\label{benchmarkint}
\end{equation}
\end{proposition}
\begin{proof}
The proof is provided in Appendix \ref{sec:proof}.
\end{proof}

Having derived no-arbitrage prices for perpetual futures for a general specification of the funding rate function $\Phi$, we next specialize these results for the case of Binance, which is currently the largest perpetual futures exchange by volume and the source of our data below.

The funding rate on Binance can be written as:
\begin{equation}
    \Phi\left(\frac{F-S}{F}\right)=\kappa \frac{F-S}{F}+\operatorname{clamp}\left(\iota-\kappa \frac{F-S}{F},-\gamma, \gamma\right),
\end{equation}
where $\frac{F - S}{F}$ is the premium index of futures to spot, $\kappa$ is a scaling constant, $clamp(x, l, h) \equiv \min(\max(x, l), h)$ bounds $x$ between $l$ and $h$, $\iota$ denotes the interest component in the funding rate, and $\gamma$ is a parameter determining the width of the clamp. Currently on Binance, $\kappa = 1$, $\gamma = 0.05\%$, and $\iota = 0.01\%$. We plot $\Phi$ as a function of the premium index in Figure \ref{figfr}.

\begin{corollary}[No-arbitrage price for Binance]\label{corollary2}
The funding rate function on Binance satisfies Assumption $\ref{assp3}$. Furthermore, under Assumptions $\ref{assp1}$ and $\ref{assp2}$, there is an absence of random-maturity arbitrage opportunities if and only if
\begin{align}
    & F_t=\frac{\kappa}{\kappa+\operatorname{sgn}(\iota -r) \cdot \gamma-r} S_t, \quad \text{when $r \neq \iota$,}\label{benchmark_binance} \\
    & F_t \in \left[\frac{\kappa}{\kappa + \gamma - r}S_t, \frac{\kappa}{\kappa - \gamma - r}S_t \right], \quad \text{when $r = \iota$}.
\end{align}
where $sgn(\cdot)$ is the sign function which equals $+1(-1)$ if its argument is positive (negative).
\end{corollary}

\begin{proof}
The proof follows from applying Propositions \ref{prop1} and \ref{prop2} to the Binance funding rate function. We provide detailed scenario analysis in Appendix \ref{sec:proof}, which also gives the divergence speed of the deviation in this specific case.
\end{proof}

Empirically, the arbitrageurs face trading costs when opening or terminating their positions. Suppose they face a round-trip trading cost of $C$. We can derive no-arbitrage bounds for perpetual futures.

\begin{proposition}[No-arbitrage bounds with trading costs]\label{prop3}
Under Assumptions \ref{assp1}, \ref{assp2}, and \ref{assp3}, and write
\[
    \Phi^{-1}(r)=[x^-,x^+], \qquad
    \lambda^L \equiv (1-x^-)^{-1}, \qquad
    \lambda^U \equiv (1-x^+)^{-1}.
\]
When there is a round-trip trading cost $C$ of executing the trading strategy, the absence of random-maturity arbitrage opportunities implies that the perpetual futures price will lie within the following bounds relative to the spot:
\begin{equation}
    F_t \in
    \left[
        \lambda^L S_t-C,\,
        \lambda^U S_t+C
    \right].\label{eq_no_arb_bound}
\end{equation}
\end{proposition}
\begin{proof}
The proof is provided in Appendix \ref{sec:proof}.
\end{proof}

For the Binance funding-rate specification, Corollary \ref{corollary2} implies that Proposition \ref{prop3} specializes to
\begin{align}
    & F_t \in \left[\frac{\kappa}{\kappa + \text{sgn}\left(\iota - r\right)\cdot\gamma - r} S_t - C, \frac{\kappa}{\kappa + \text{sgn}\left(\iota - r\right)\cdot\gamma - r} S_t + C  \right], \quad \text{when $r \neq \iota$}, \label{eq_no_arb_bound_binance_point} \\
    & F_t \in \left[\frac{\kappa}{\kappa + \gamma - r}S_t - C, \frac{\kappa}{\kappa - \gamma - r}S_t + C\right], \quad \text{when $r = \iota$}. \label{eq_no_arb_bound_binance_interval}
\end{align}

In the presence of trading costs, when the deviation of the perpetual futures price from $\lambda S_t$ is larger than the round-trip trading costs $C$, arbitrageurs would have a strong incentive to trade perpetual futures toward the price $\lambda S_t$. Proposition \ref{prop3} also prescribes a trading strategy to exploit the futures-spot divergence in markets with different levels of trading costs. In the next section, we provide an empirical analysis of the futures-spot deviation and present arbitrage trading strategies motivated by our theory.

Before turning to the data, we pause to relate our results to two recent theoretical contributions. 
\citet{angeris2022primer} consider an inverse problem to our setting. It assumes there is no arbitrage and that the perpetual futures price is a function of the spot, and derives the funding rate process that is consistent with the pricing formula. This takes a perspective of security design. However, empirically, funding rate mechanisms are predetermined and fixed for traders. Moreover, because of the limits of arbitrage, futures prices can deviate from the fixed function set by the designer. Our approach prescribes a strategy to exploit divergence from the fundamental price and offers a closer alignment with real-world trading conditions.

Compared to \citet{ackerer_perpetual_2023}, our initial draft, disseminated December 2022 (see \url{https://arxiv.org/abs/2212.06888v1}), considers the case where the funding rate is a linear function of the premium index, as illustrated by the dashed line in Figure \ref{figfr}. It shows that the absence of random-maturity arbitrage opportunities implies the price of perpetual futures lies within the following bound relative to the spot:
\begin{equation}
    |F_t - \left(1 + \frac{r}{\kappa}\right) S_t | \leq C,
\end{equation}
and that in the case without trading costs, the price would be 
\begin{equation}
    F_{t}=\left(1+\frac{r}{\kappa}\right)S_{t}. \label{eq:our_old_formula}
\end{equation}

\citet[\url{https://arxiv.org/abs/2310.11771v1}]{ackerer_perpetual_2023}, dated October 2023, use a different set of assumptions and solution method to derive no-arbitrage prices for a class of perpetual contracts. In their paper, the price relevant for the case we analyzed is:
\begin{equation}
    F_{t}=\frac{\kappa}{\kappa-r}S_{t}. \label{eq:their_formula}
\end{equation}
Their paper further pointed out an inconsistency in the cash flow specification in our original derivation. After correcting this issue, Equation \eqref{eq:their_formula} above is consistent with their derivation in this case, and they deserve credit for deriving the correct formula first.\footnote{We thank Damien Ackerer, Julien Hugonnier, and Urban Jermann for the comments that helped clarify the issue.} However, their framework does not allow for the clamp feature, an empirically important component of the funding rate specifications employed by major perpetual futures exchanges.\footnote{According to \citet{han2024primer}, the exchanges which have perpetual futures with the clamp feature account for more than 75\% of open interest for Bitcoin perpetual futures.}

Another advantage of our derivation is that it makes explicit the arbitrage trading strategy in case of deviations and provides bounds in the case of trading costs which we use in our empirical analysis. The approach also enables the analysis of no-arbitrage trading for the clamped funding rate, a specification less understood by practitioners so far according to \citet{han2024primer}.

Finally, we differ in our assumptions about price deviations. We assume that the \emph{ratio} between the futures and spot is bounded, whereas \cite{ackerer_perpetual_2023} assume that a transversality condition holds such that the futures price (and therefore the spot) grows at a slow enough rate relative to the funding payments intensity $\kappa$. But even under their assumptions, it can be shown that strategies aimed at arbitrage of deviations from their benchmark prices are necessarily random-maturity arbitrage strategies, as we emphasize here.

These pricing formulas imply different predictions for the point toward which the futures-spot basis should converge. In Section \ref{sec:strategy}, we directly compare these convergence targets in the data.

\section{Data and Empirical Analysis}\label{sec:empirical}

We conduct an empirical analysis of perpetual futures arbitrage strategies. We first describe the data and measure the deviations of the crypto futures-spot spread from the no-arbitrage benchmark. We then implement a trading strategy that exploits deviations from random-maturity no-arbitrage bounds. Finally, we examine implementation frictions by incorporating exchange fees and effective bid-ask spreads and assessing the interim losses relevant for liquidation risk.

\subsection{Data}\label{sec:data}
We focus on the five largest cryptocurrencies excluding stablecoins: Bitcoin (BTC), Ether (ETH), BNB (BNB), Dogecoin (DOGE), and Cardano (ADA) with a total market cap of \$1.83 trillion, which account for 73.2\% of the total market share of the Crypto market by March 2024.

For each token, we obtain perpetual futures and spot prices at a 1-hour frequency from Binance. Binance is by far the leading exchange in the crypto realm: it accounts for about half of the perpetual futures volume of the exchanges in Figure \ref{fund_exchange} and for the largest share of spot trading volume, and, as discussed in Section \ref{sec:background}, its perpetuals are also more liquid than those of the smaller venues in the figure. Another major benefit of using the Binance data is that every part of our trading strategies can be completed within the same platform without delay in transferring funds. As such, it approximates the real-world investment opportunities faced by traders.

We retrieve the perpetual funding rates from Binance, where the funding rate is paid every 8 hours. So we have futures and spot prices every hour and realized funding rate payments at 8:00, 16:00, and 0:00 GMT each day. The perpetual and spot tradings occur 24 hours per day and 7 days a week so there are no after-market hours.

Table \ref{tab2} lists the starting and ending dates of our data for each cryptocurrency. Our data ends on 2024-03-11, covering the FTX fallout.

\begin{table}[!htb]
\begin{center}
\begin{tabularx}{1\columnwidth}{@{\hskip\tabcolsep\extracolsep\fill}l*{3}{l}}
\toprule
Asset & Start date & End date & N \\
\midrule
BTC & 2020-01-08 & 2024-03-11 & 36,578 \\
ETH & 2020-01-08 & 2024-03-11 & 36,578 \\
BNB & 2020-02-10 & 2024-03-11 & 35,777 \\
DOGE & 2020-07-10 & 2024-03-11 & 32,152 \\
ADA & 2020-01-31 & 2024-03-11 & 36,017 \\
\bottomrule
\end{tabularx}
\end{center}
\caption{Sample Descriptions}
\bigskip
\small
This table presents the sample start and end dates for the five cryptocurrencies and their total number of observations.
\label{tab2}
\end{table}

We obtain trading costs from Binance's website\footnote{\url{https://www.binance.com/en/fee/trading} provides data on trading fees in perpetual futures and the spot market}. In general, trading costs for the spot market are significantly higher than those for perpetual futures for similar trading volume because futures are typically traded with leverage. Fee tiers are attributed to the 30-day trading volume. We attribute high trading costs with a 30-day spot trading volume above \$1 million and futures trading volume above \$15 million (small individual trader). Medium trading costs are attributed to a 30-day spot trading volume above \$150 million and futures trading volume above \$1 billion (small funds). Low trading costs are attributed to a 30-day spot trading volume above \$2 billion and futures trading volume above 12.5 billion (large funds). Zero fees can be negotiated with customized contracts, for example, for market makers. We consider trading costs for makers instead of takers because institutions typically trade maker orders.\footnote{Takers trade market orders while makers trade limit orders. Takers take liquidity from the market while makers make the market or provide liquidity to the market.} 

Throughout our empirical analysis, $\rho$ denotes the annualized interest-rate wedge that would rationalize the observed futures-spot spread. We denote its lower and upper no-arbitrage bounds implied by different tiers of trading costs by $\rho_l$ and $\rho_u$, respectively. Section \ref{sec:deviation} formally defines $\rho$ and derives these bounds.

Table \ref{tab3} presents the specification of different trading costs. It also shows the random-maturity arbitrage bound for the deviation between the perpetual futures and the no-arbitrage price ($\rho_l$ and $\rho_u$). The bounds become wider as the trading costs increase. Detailed explanations of the trading costs specifications are also provided in Appendix Figure \ref{trading_cost_spot} and Figure \ref{trading_cost_perp}.

\begin{table}[!htb]
\begin{center}
\begin{tabularx}{1\columnwidth}{@{\hskip\tabcolsep\extracolsep\fill}l*{4}{r}}
\toprule
Fee tier & Spot & Futures & $\rho_l$ & $\rho_u$ \\
\midrule
No & 0.0000\% & 0.0000\% & $-0.0\%$ & $0.0\%$ \\
Low & 0.0225\% & 0.0018\% & $-53.2\%$ & $53.2\%$ \\
Medium & 0.0450\% & 0.0072\% & $-114.3\%$ & $114.3\%$ \\
High & 0.0675\% & 0.0144\% & $-179.4\%$ & $179.4\%$ \\
\bottomrule
\end{tabularx}
\end{center}
\caption{Exchange Fee Tiers}\label{tab3}
\bigskip
\small
Fee tiers are assigned based on past 30-day trading volume. The Spot and Futures columns report one-way maker trading fees, denoted
$c_S$ and $c_F$, respectively, for a single transaction, expressed as a percentage of traded notional. The corresponding proportional round-trip cost of the two-legged arbitrage strategy is $c=2(c_S+c_F)$. High fees correspond to a 30-day trading volume above \$1mn in spot and above \$15mn in perpetuals, typically an individual trader. Medium fees apply to traders with 30-day trading volumes above \$150mn in spot and above \$1bn in perpetuals (small funds). Low fees are attributed to a 30-day trading volume above \$2bn in spot and above \$12.5bn in perpetuals (large funds). The no-fee tier can be negotiated with customized contracts, for example, for market makers. The $\rho_l$ and $\rho_u$ columns report the theory-implied lower and upper no-arbitrage bounds for $\rho$, respectively, expressed in annualized percentage points and calculated using Equation \eqref{tc_empirical}.
\end{table}

In addition to the direct trading fees, traders also face bid-ask spread and price impact costs in our arbitrage strategies. To assess the importance of such indirect trading costs, we obtain from Kaiko a high-frequency sample with one month of order book snapshots taken every 30 seconds. With this data, we estimate that the average bid-ask spread for perpetual futures is less than 0.05 bps, and the price impact of a trade with a size equal to 0.5 million is less than 0.3 bps. Combined, these indirect costs are less than 0.35 bps, which is substantially smaller than our direct trading fees. To study how trading costs evolve over the full sample, Section \ref{sec:tradingcosts} uses Binance's public trade records to construct daily effective bid-ask spreads and incorporates them into the random-maturity arbitrage strategy.

To measure the deviation from the no-arbitrage price, we also obtain the interest rate data from Aave, a leading open-source DeFi liquidity protocol. Customers on Aave can either be suppliers or borrowers of cryptocurrencies. Because of the anonymity and decentralization of the DeFi system, all borrowings are over-collateralized and the collateral can serve as an additional supply to the system for borrowing. The customers can also supply spare currencies to the system to earn an interest rate. The supply interest rate is typically different from the borrowing interest rate. Both interest rates and their wedge are determined algorithmically based on the market conditions of supply and demand. We use interest rates from this platform because we believe it is a good proxy for the funding condition in the crypto market. Appendix \ref{sec:robustness} shows that our results are robust to using the interest rate from the traditional financial market or using the margin borrowing rate from Binance.

Our theory indicates using a risk-free rate available to the arbitrageurs. Therefore, we consider the interest rates on the stablecoins, which are not subject to volatile spot price movements and are the currency of denomination for perpetual futures margins. There are three major stablecoins traded on Aave: USDT, USDC, and DAI. To get a robust measure of the risk-free rate, we take an average of the three interest rates to arrive at our final risk-free supply and borrowing rate for the arbitrageurs. 

We plot the time-series evolution of the interest rate in Appendix Figure \ref{fig:interest_rate}. During the early years, we see higher interest rate volatility due to the funding liquidity of the DeFi platform. In later samples, interest rates are more stable and approach interest rates in traditional financial markets.

Our interest rate data starts from 2020-01-08. Therefore, for coins with perpetual data available before this date (BTC and ETH), we begin the analysis from 2020-01-08. For other coins (BNB, DOGE, ADA), we begin the analysis from the time when they have data available.

\subsection{Deviations of perpetual futures from no-arbitrage benchmarks}\label{sec:deviation}

The focus of our empirical analysis is the annualized deviation $\rho$, defined as the interest rate spread that would rationalize an observed futures-spot spread. Under the clamp specification of the funding rate from Binance, we have:
\begin{equation}
    F = \frac{\kappa}{\kappa + \text{sgn}\left(\iota - r\right)\cdot\gamma - (r + \rho)} S. \label{eqrho}
\end{equation}
For example, an estimated $\rho$ of one percentage point would mean that borrowing costs in the cash market faced by arbitrageurs would have to be one percentage point higher than the prevailing interest rate $r$ in the cash market for the futures-spot gap to be arbitrage-free.

Rearranging Equation \eqref{eqrho}, we obtain the following equation for $\rho$:
\begin{equation}
    \rho = \kappa\left(\frac{F-S}{F} \right) + \text{sgn}\left(\iota - r\right)\gamma - r. \label{eqdef}
\end{equation}
This equation shows that deviations are closely related to the premium index $\frac{F-S}{F}$, adjusted for the clamp parameter $\gamma$ and the interest rate $r$.
This definition is the same in spirit to \cite{du_deviations_2018}, who measure covered interest parity (CIP) deviations with the wedge that would equate the dollar borrowing rate and the synthetic dollar borrowing rate.\footnote{Alternatively, we may define $\rho$ as the wedge that satisfies 
    $$
    F = \frac{\kappa}{\kappa + \text{sgn}\left(\iota - (r + \rho)\right)\cdot\gamma - (r + \rho)} S.
    $$ 
    Using this definition, after rearranging terms, $\rho$ simplifies to:
    $$
    \rho = \kappa(\frac{F - S}{F}) + clamp(\iota - \kappa \frac{F - S}{F}, - \gamma, \gamma) - r = \Phi - r.
    $$
    This result is closely related to the key equation used to determine the fundamental price, $\Phi = r$. We adopt the first definition for its greater tractability when incorporating trading costs, but both have their merits.}

Furthermore, for each tier of trading cost in Table \ref{tab3}, let $c_S$ and $c_F$ denote the one-way maker fees in the spot and perpetual futures markets, respectively. Since the arbitrage strategy trades both legs at initiation and again at unwinding, its proportional round-trip exchange-fee cost is $c=2(c_S+c_F)$. We map this cost into a threshold value of $\rho$ using the bound from Proposition \ref{prop3}, specialized to a proportional trading cost $c$,
 \begin{align}
     & F - \frac{\kappa}{\kappa + \text{sgn}(\iota - r)\gamma - r} S = c F. \label{eqfsc}
 \end{align}
Equations \eqref{eqrho} and \eqref{eqfsc} together imply that the corresponding threshold for $\rho$ is:
\begin{equation}
    \rho =  (\kappa + \text{sgn}(\iota - r)\gamma - r) c \approx \kappa c, \label{tc_empirical}
\end{equation}
where the last approximation follows from the empirical fact that $\kappa$ is much larger than $\gamma$ and $r$.\footnote{Incorporating $\gamma$ and $r$ makes a negligible difference for the threshold $\rho$.} Therefore, we calculate the upper and lower trading thresholds in Table \ref{tab3} as $\kappa c$ and $-\kappa c$.

Each hour, we calculate $\rho$ defined in Equation \eqref{eqdef} using data on the perpetual futures price, spot price, and the crypto risk-free interest rate. As shown in Table \ref{tabcase1}, when the perpetual futures price is above the spot, an arbitrageur would short the futures and long the spot, and would finance her position by borrowing in the cash market. On the other hand, when the perpetual futures price is below the spot, an arbitrageur would long the futures, short the spot, and invest the proceeds from shorting in the cash market. 

\begin{figure}[!htb]
\begin{center}
\noindent\includegraphics[width=\textwidth]{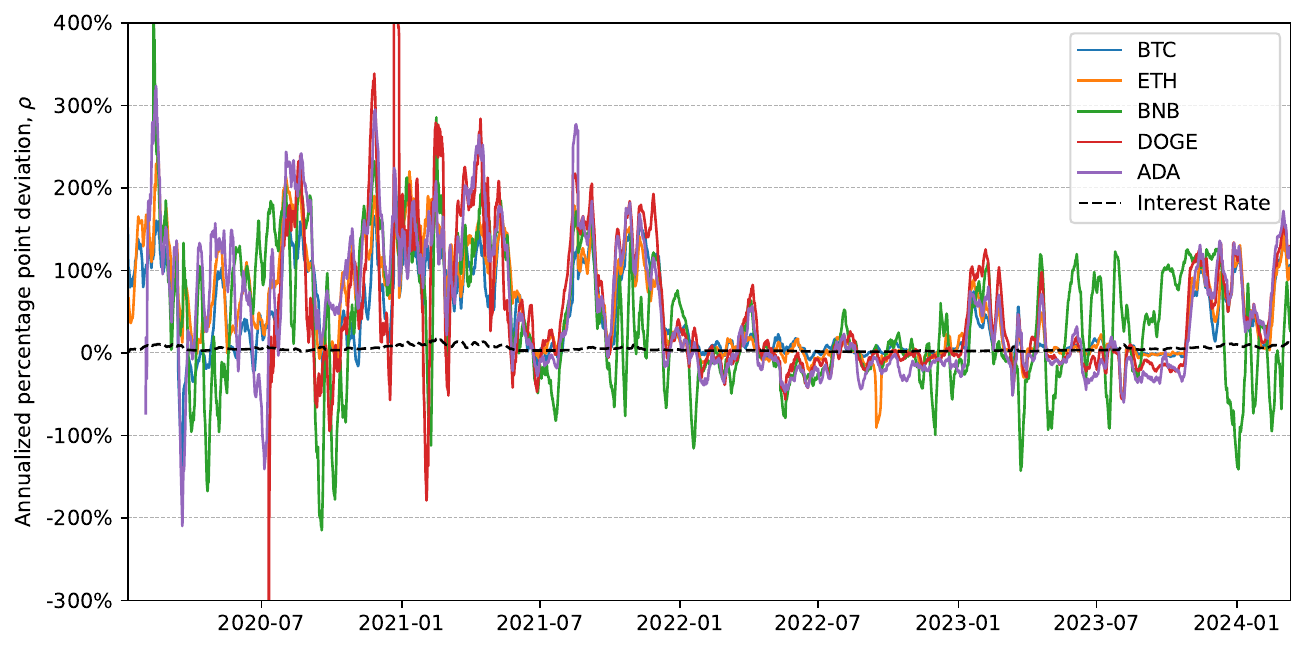}
\end{center}
\caption{Deviations of Perpetual Futures from No-arbitrage Benchmarks}\label{fig1}
\bigskip
\small
This figure presents the 7-day moving averages of the annualized deviation of the perpetual futures-spot spread from the no-arbitrage benchmark for the five cryptocurrencies. We also plot the 7-day moving average of the annualized interest rate for comparison.
\end{figure}

Table \ref{tab_sum_rho} reports summary statistics for $\rho$ in our data. As can be seen, the deviation from the no-arbitrage price is predominantly positive across a substantial portion of the sample, suggesting perpetual futures are more expensive than the no-arbitrage price on average. However, there is considerable variation in these deviations, with the standard deviation exceeding twice the mean deviation for most tokens. The mean absolute deviation is about 60\% to 90\% per year across different cryptocurrencies. 

\begin{table}[!htb]
\begin{center}
\begin{tabularx}{1\columnwidth}{@{\hskip\tabcolsep\extracolsep\fill}l*{8}{c}}
\toprule
\multicolumn{9}{c}{Panel A: Full Sample} \\
\midrule
& \multicolumn{4}{c}{$\rho$} & \multicolumn{4}{c}{$|\rho|$} \\
\cmidrule{2-5} \cmidrule{6-9}
Asset &  Mean &  Median &   Std &  p-value &  Mean &  Median &   Std &  p-value \\
\midrule
BTC & 0.42 & 0.19 & 0.84 & 0.00 & 0.52 & 0.25 & 0.78 & 0.00 \\
ETH & 0.52 & 0.26 & 0.88 & 0.00 & 0.63 & 0.33 & 0.81 & 0.00 \\
BNB & 0.38 & 0.31 & 1.23 & 0.00 & 0.90 & 0.68 & 0.92 & 0.00 \\
DOGE & 0.50 & 0.23 & 1.90 & 0.00 & 0.86 & 0.49 & 1.77 & 0.00 \\
ADA & 0.51 & 0.27 & 1.27 & 0.00 & 0.84 & 0.53 & 1.08 & 0.00 \\
\midrule
\multicolumn{9}{c}{Panel B: Subsample Comparison} \\
\midrule
& \multicolumn{4}{c}{$\rho$ Before 2022} & \multicolumn{4}{c}{$\rho$ After 2022} \\
\cmidrule{2-5} \cmidrule{6-9}
Asset &  Mean &  Median &   Std &  p-value &  Mean &  Median &   Std &  p-value \\
\midrule
BTC & 0.69 & 0.62 & 1.08 & 0.00 & 0.17 & 0.08 & 0.42 & 0.00 \\
ETH & 0.92 & 0.85 & 1.01 & 0.00 & 0.16 & 0.06 & 0.53 & 0.00 \\
BNB & 0.69 & 0.81 & 1.53 & 0.00 & 0.11 & 0.07 & 0.81 & 0.09 \\
DOGE & 0.99 & 0.95 & 2.83 & 0.00 & 0.18 & 0.06 & 0.62 & 0.01 \\
ADA & 0.99 & 0.94 & 1.60 & 0.00 & 0.09 & 0.03 & 0.66 & 0.11 \\
\bottomrule
\end{tabularx}
\end{center}
\caption{Summary Statistics for Deviations from No-arbitrage Prices}
\bigskip
\small
The table summarizes statistics for $\rho$ across various crypto tokens. Panel A presents the summary statistics of $\rho$ and $|\rho|$ across the whole sample. Panel B presents the summary statistics for $\rho$ before and after 2022-01-01. p-value corresponds to the Newey-West adjusted p-value, testing the null hypothesis that the corresponding mean of $\rho$ is zero. The sample periods are shown in Table \ref{tab2}.
\label{tab_sum_rho}
\end{table}

Subsample analysis reveals that the mean deviation and variation in deviation shrink over time. After 2022, the average deviation reduces by almost 80\% with volatility shrinking by more than 50\% for most tokens. Figure \ref{fig1} depicts the 7-day moving average of $\rho$ for each of the five examined cryptocurrencies. It reveals considerable comovement in futures-spot spreads across these cryptocurrencies, suggesting a common factor at play that drives the observed divergence. This observation is corroborated by Table \ref{tab_corr}, which demonstrates high correlations among futures-spot deviations across the different cryptocurrencies. This trend underscores the interconnectedness of cryptocurrency markets and is consistent with the comovement of traditional currencies documented by \citet*{lustig2011common}.

\afterpage{
\begin{landscape}
\renewcommand{\arraystretch}{1.2}
\begin{table}
\begin{center}
\footnotesize
\begin{tabularx}{1\columnwidth}{@{\hskip\tabcolsep\extracolsep\fill}ll*{16}{r}}
\toprule
 & & \multicolumn{4}{c}{Arbitrage Deviation} & \multicolumn{5}{c}{Funding Rate} & \multicolumn{5}{c}{Spot Return}\\
 \cmidrule{3-6} \cmidrule{7-11} \cmidrule{12-16}
{} & & $\rho_{ETH}$ & $\rho_{BNB}$ & $\rho_{DOGE}$ & $\rho_{ADA}$ & $fr_{BTC}$ & $fr_{ETH}$ & $fr_{BNB}$ & $fr_{DOGE}$ & $fr_{ADA}$ & $R_{BTC}$ & $R_{ETH}$ & $R_{BNB}$ & $R_{DOGE}$ & $R_{ADA}$ \\
\midrule
\multirow[t]{5}{*}{Arbitrage Deviation} & $\rho_{BTC}$ & 0.88 & 0.53 & 0.73 & 0.79 & 0.67 & 0.58 & 0.45 & 0.40 & 0.53 & -0.02 & 0.01 & 0.03 & 0.09 & 0.02 \\
 & $\rho_{ETH}$ &  & 0.58 & 0.69 & 0.82 & 0.64 & 0.72 & 0.48 & 0.41 & 0.58 & 0.06 & 0.08 & 0.07 & 0.11 & 0.08 \\
 & $\rho_{BNB}$ &  &  & 0.50 & 0.53 & 0.45 & 0.45 & 0.82 & 0.37 & 0.43 & 0.01 & 0.05 & 0.03 & 0.05 & 0.06 \\
 & $\rho_{DOGE}$ &  &  &  & 0.70 & 0.46 & 0.46 & 0.41 & 0.61 & 0.50 & 0.07 & 0.09 & 0.11 & -0.00 & 0.07 \\
 & $\rho_{ADA}$ &  &  &  &  & 0.57 & 0.56 & 0.44 & 0.45 & 0.73 & 0.07 & 0.10 & 0.10 & 0.10 & 0.08 \\
\multirow[t]{5}{*}{Funding Rate} & $fr_{BTC}$ &  &  &  &  &  & 0.86 & 0.65 & 0.63 & 0.74 & 0.08 & 0.09 & 0.15 & 0.13 & 0.10 \\
 & $fr_{ETH}$ &  &  &  &  &  &  & 0.63 & 0.58 & 0.74 & 0.12 & 0.16 & 0.18 & 0.17 & 0.14 \\
 & $fr_{BNB}$ &  &  &  &  &  &  &  & 0.55 & 0.60 & 0.04 & 0.09 & 0.11 & 0.10 & 0.10 \\
 & $fr_{DOGE}$ &  &  &  &  &  &  &  &  & 0.68 & 0.12 & 0.14 & 0.23 & -0.01 & 0.13 \\
 & $fr_{ADA}$ &  &  &  &  &  &  &  &  &  & 0.10 & 0.15 & 0.21 & 0.16 & 0.12 \\
\multirow[t]{4}{*}{Spot Return} & $R_{BTC}$ &  &  &  &  &  &  &  &  &  &  & 0.84 & 0.71 & 0.46 & 0.70 \\
 & $R_{ETH}$ &  &  &  &  &  &  &  &  &  &  &  & 0.73 & 0.43 & 0.74 \\
 & $R_{BNB}$ &  &  &  &  &  &  &  &  &  &  &  &  & 0.34 & 0.65 \\
 & $R_{DOGE}$ &  &  &  &  &  &  &  &  &  &  &  &  &  & 0.42 \\
\bottomrule
\end{tabularx}
\end{center}
\caption{Correlation of arbitrage deviation ($\rho$), funding rate ($fr$), and spot returns ($R$)}\label{tab_corr}
\bigskip
\small
This table presents the correlation among different cryptocurrencies' futures-spot deviation $\rho$ (rows 3 to 7), funding rate (rows 8 to 12), and spot returns (rows 13 to 16).
\end{table}
\end{landscape}
}

One hypothesis for this significant comovement in $\rho$ across various cryptocurrencies pertains to time-varying funding constraints encountered by arbitrageurs in the market \citep{brunnermeier2009market, garleanu2011margin}. Because arbitrageurs often serve as marginal traders in all markets, their funding constraints are potentially reflected in the $\rho$ across all major cryptocurrency markets. Our theoretical and empirical analyses underscore the importance of measuring $\rho$. Doing so not only sheds light on potential market stress but also provides a gauge for the shadow costs associated with the trading and funding constraints arbitrageurs face.

Conversely, from the demand side, shared sentiment across various cryptocurrencies could also contribute to explaining the concurrent movement of $\rho$ across different markets. Our theoretical analysis suggests that arbitrageurs will only meet the market demand if the price deviation surpasses the combined trading and funding costs. As a result, the overall market sentiment in the futures market relative to the spot market becomes evident in the price difference between futures and spot prices. If the sentiment across different markets is influenced by a common factor, we would anticipate a significant comovement in the futures-spot spread. 

Moreover, arbitrageurs are willing to absorb the idiosyncratic demand first because combining the arbitrage trades across different idiosyncratic deviations would generate a high Sharpe ratio strategy as in \citet{kozak2018interpreting}. Therefore, in equilibrium, after arbitrageurs have absorbed the idiosyncratic demand, we anticipate the emergence of a common factor in deviations across different cryptocurrencies, driven by limits to arbitrage.

In Section \ref{sec:explaining}, we delve deeper into explaining the futures-spot spread. Our results demonstrate that the past returns of each cryptocurrency serve as significant explanatory variables for the time-series variation in the spread.

Interestingly, the correlation between deviations and spot market returns is quite modest. Although spot market returns are highly correlated among themselves, in line with the strong market factor results from \citet{liu2022common}, and no-arbitrage price deviations are highly correlated among themselves, it appears that different forces drive these distinct phenomena.

Furthermore, we find that after the year 2022, futures-spot spreads become much closer and anchored to $0$ compared to earlier samples. This suggests the market is becoming increasingly efficient.

Last but not least, by design, $\rho$ is also highly correlated with the funding rate. The funding rate does not correlate perfectly with $\rho$ because in real-world implementations: (1) the funding rate is calculated as a weighted average of futures-spot price deviations, with larger weight given to more recent observations; (2) in calculating the funding rate, Binance does not just consider the quote price; they also use an impact margin notional to consider the price impact of the trade.\footnote{See \url{https://www.binance.com/en/support/faq/360033525031} for details of the funding rate calculation.}

\subsection{Random-maturity Arbitrage Strategy}\label{sec:strategy}

We next provide a trading strategy motivated by our random-maturity arbitrage theory. We first consider the zero-cost benchmark. Whenever $\rho$ deviates from zero, we open the convergence position: we short the perpetual and long the spot when $\rho>0$, and take the opposite position when $\rho<0$. We close the position when $\rho$ first returns to zero. Section \ref{sec:tradingcosts} extends this strategy to account for exchange fees and effective bid-ask spreads.

\begin{figure}[!htb]
\begin{center}
\includegraphics[width=\textwidth]{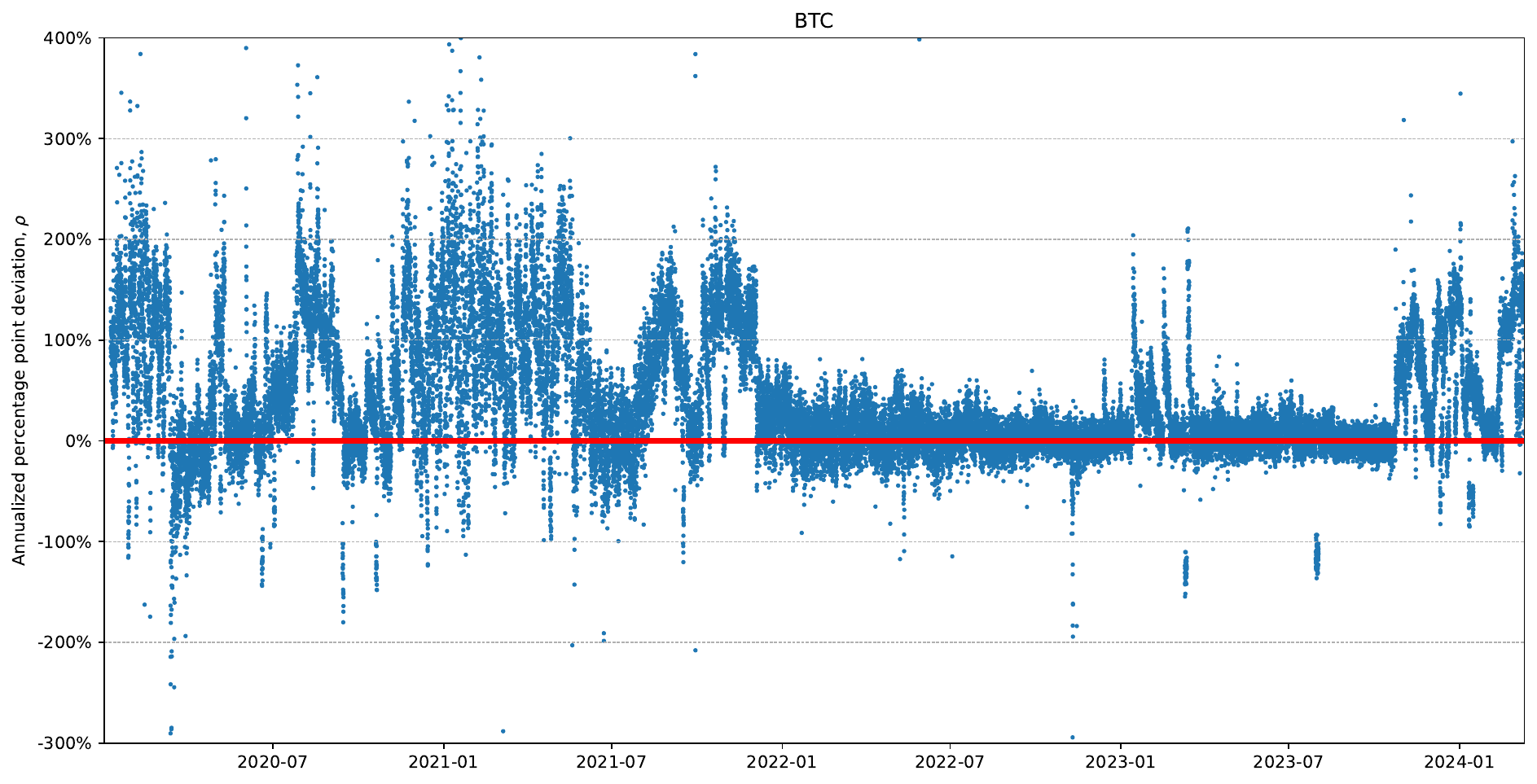}
\end{center}
\caption{Random-maturity Arbitrage Strategy for Bitcoin}\label{fig_btc_gas}
\bigskip
\small
Deviation from the no-arbitrage price ($\rho$) and trading rules of the random-maturity arbitrage strategy we implement for BTC. Each blue dot in the figure represents the deviation of futures from the no-arbitrage price calculated using Equation \ref{eqdef} at hourly frequency. The number is annualized for ease of interpretation. The red line represents the zero benchmark.
\end{figure}

Figure \ref{fig_btc_gas} provides an illustration of the trading strategy. When the deviation is above the red line, the strategy opens a long position in the spot and a short position in perpetual. Conversely, when the deviation is below the red line, the strategy is short the spot and long the perpetual. 

Table \ref{port_mh_theory_unrestricted} presents the performance of the trading strategy over the years. We find that the trading strategy generates a high Sharpe ratio for all five tokens. The large returns with low volatility and drawdowns suggest that our theory provides useful benchmarks. A deviation from the no-arbitrage price indeed implies a near-arbitrage opportunity.

\begin{table}[!htb]
\begin{center}
\footnotesize
\begin{tabularx}{\textwidth}{@{\hskip\tabcolsep\extracolsep\fill}ll*{6}{r}}
\toprule
 &  & 2020 & 2021 & 2022 & 2023 & 2024 & All \\
\midrule
BTC & SR & 7.18 & 12.36 & 37.85 & 15.37 & 11.84 & 11.65 \\
 & Return & 49.62 & 64.81 & 60.02 & 41.68 & 21.44 & 52.55 \\
 & Volatility & 6.91 & 5.24 & 1.59 & 2.71 & 1.81 & 4.51 \\
 & MaxDD & -3.82 & -2.00 & -0.36 & -0.26 & -0.17 & -3.82 \\
 & OtC time & 8.35 & 9.17 & 2.97 & 3.71 & 11.44 & 4.92 \\
 & N & 8,616 & 8,760 & 8,760 & 8,760 & 1,682 & 36,578 \\
ETH & SR & 10.43 & 12.04 & 25.92 & 12.82 & 13.90 & 12.77 \\
 & Return & 67.53 & 75.56 & 70.73 & 44.81 & 25.11 & 62.83 \\
 & Volatility & 6.48 & 6.27 & 2.73 & 3.49 & 1.81 & 4.92 \\
 & MaxDD & -2.62 & -2.28 & -0.67 & -0.29 & -0.16 & -2.62 \\
 & OtC time & 11.19 & 8.80 & 3.02 & 3.68 & 8.54 & 5.05 \\
 & N & 8,616 & 8,760 & 8,760 & 8,760 & 1,682 & 36,578 \\
BNB & SR & 15.30 & 16.89 & 34.91 & 17.66 & 22.64 & 17.66 \\
 & Return & 159.04 & 166.86 & 134.73 & 87.08 & 102.83 & 134.74 \\
 & Volatility & 10.40 & 9.88 & 3.86 & 4.93 & 4.54 & 7.63 \\
 & MaxDD & -1.96 & -1.66 & -0.83 & -0.97 & -0.25 & -1.96 \\
 & OtC time & 8.50 & 6.44 & 3.32 & 6.33 & 6.60 & 5.46 \\
 & N & 7,814 & 8,760 & 8,760 & 8,760 & 1,682 & 35,776 \\
DOGE & SR & 18.14 & 12.78 & 45.37 & 19.58 & 12.29 & 14.90 \\
 & Return & 409.29 & 207.91 & 191.43 & 62.89 & 35.36 & 181.12 \\
 & Volatility & 22.56 & 16.27 & 4.22 & 3.21 & 2.88 & 12.16 \\
 & MaxDD & -13.90 & -5.61 & -0.29 & -1.65 & -0.29 & -13.90 \\
 & OtC time & 3.90 & 6.91 & 2.57 & 4.60 & 12.19 & 4.13 \\
 & N & 4,189 & 8,760 & 8,760 & 8,760 & 1,682 & 32,151 \\
ADA & SR & 14.11 & 13.07 & 45.97 & 34.87 & 13.03 & 19.76 \\
 & Return & 147.63 & 138.28 & 210.28 & 145.53 & 42.29 & 155.16 \\
 & Volatility & 10.46 & 10.58 & 4.57 & 4.17 & 3.25 & 7.85 \\
 & MaxDD & -3.47 & -5.69 & -0.24 & -0.50 & -0.18 & -5.69 \\
 & OtC time & 7.74 & 6.28 & 2.70 & 3.46 & 13.90 & 4.32 \\
 & N & 8,054 & 8,760 & 8,760 & 8,760 & 1,682 & 36,016 \\
\bottomrule
\end{tabularx}
\end{center}
\caption{Performance of Random Maturity Arbitrage Trading Strategy Over Time}\label{port_mh_theory_unrestricted}
\bigskip
\small
This table presents the Sharpe ratios, annualized returns (\%), standard deviations (\%), maximum drawdowns (\%), and the average open-to-close time (hours) of the unrestricted random-maturity arbitrage trading strategies under zero trading costs for five different cryptocurrencies. The returns reported are excess returns from the random maturity arbitrage strategies.
\end{table}

An important question is whether zero is in fact the convergence target favored by the data. This is particularly relevant for distinguishing our benchmark from \citet{ackerer_perpetual_2023}. Under Corollary \ref{corollary2}, the clamp-adjusted no-arbitrage price corresponds to $\rho = 0$. In contrast, the no-clamp benchmark of \citet{ackerer_perpetual_2023}, given by Equation \eqref{eq:their_formula}, implies
$$
    \rho = \operatorname{sgn}(\iota-r)\gamma.
$$
Thus, in annualized units, their benchmark corresponds to $\rho=+54.75$ percentage points when $r<\iota$, which occurs in 93\% of our sample hours, and $\rho=-54.75$ percentage points when $r>\iota$.

\begin{figure}[!htb]
    \begin{center}
    \includegraphics[width=0.95\textwidth]{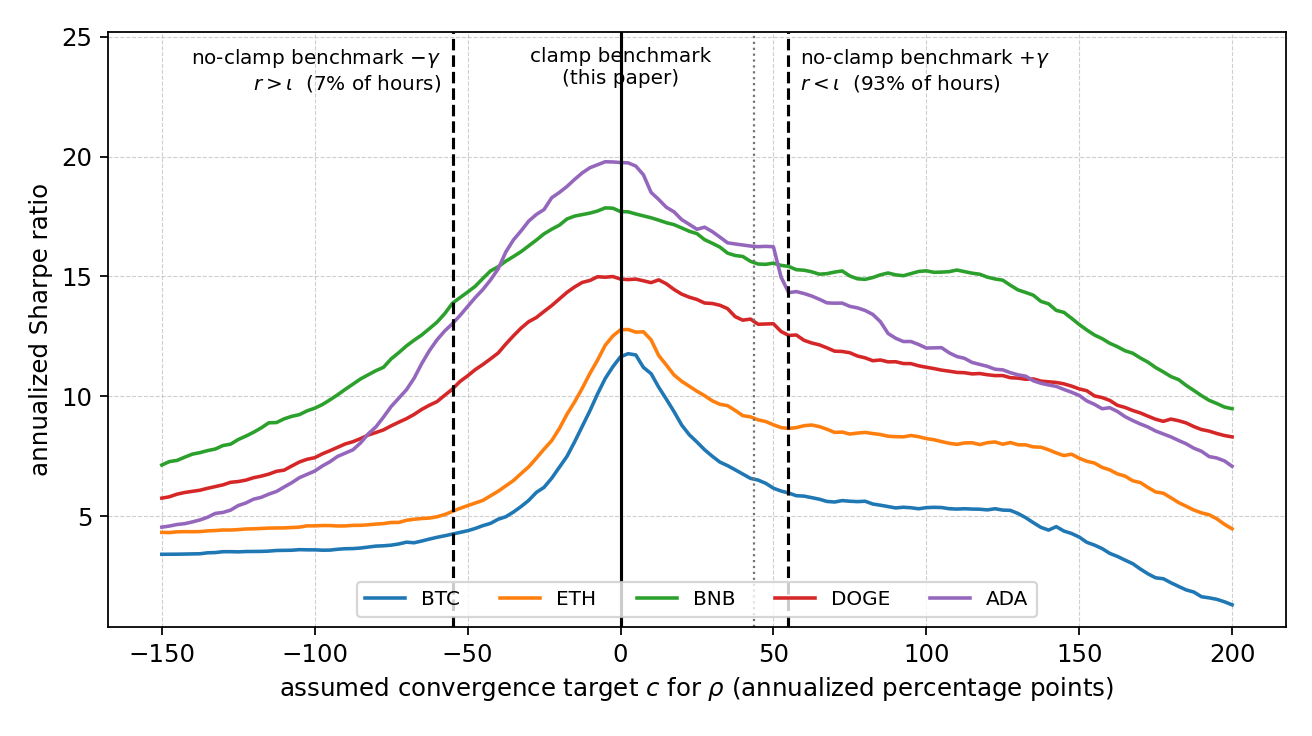}
    \end{center}
    \caption{The convergence target implied by the data}\label{fig:target}
    \bigskip
    \small
    This figure plots the annualized Sharpe ratio of the zero-cost convergence strategy that is short the perpetual when $\rho_t > c$ and long it when $\rho_t < c$, as a function of the assumed convergence target $c$, in annualized percentage points of $\rho$. The solid vertical line is the clamp-consistent benchmark of Corollary \ref{corollary2}, at which $\rho$ is defined to be zero. The dashed line is the no-clamp benchmark $F = \frac{\kappa}{\kappa - r}S$ of Equation \eqref{eq:their_formula}, which sits at $+54.75$. The dotted line at $+43.8$ is the intermediate variant that replaces $\gamma$ with $\iota$. Sample periods are those of Table \ref{tab2}.
\end{figure}

Rather than imposing either convergence target, we estimate the target
preferred by the data. For each candidate target $c$, we implement the
zero-cost convergence strategy that trades against deviations from $c$
and calculate its annualized Sharpe ratio. Figure \ref{fig:target}
plots the resulting Sharpe-ratio profile. For all five currencies, the
Sharpe ratio is maximized close to zero. Thus, the data-implied convergence point lies close to the clamp-adjusted benchmark derived in Corollary \ref{corollary2}.

Table \ref{port_strategy_comparison} provides a complementary strategy-level comparison. The first alternative ignores the clamp feature and instead uses the no-clamp fundamental price \eqref{eq:our_old_formula} provided in our earlier draft, or equivalently, in \citet{ackerer_perpetual_2023}. Consistent with Figure \ref{fig:target}, accounting for the clamp substantially improves arbitrage performance: our baseline strategy almost doubles the Sharpe ratio for Bitcoin and
also improves the Sharpe ratio for each of the other four currencies.

In the early years of our sample, the infrastructure for shorting cryptocurrencies was underdeveloped. To address this concern, we consider a long-spot-only strategy, in which we set the portfolio weights at $0$ if the strategy shorts the spot and longs the futures. We find that even with this restriction, the strategy still generates a high Sharpe ratio in all cryptocurrencies. 

The last alternative we consider is based on \citet{christin2022crypto} who propose a carry strategy in the perpetual futures market where the trader longs the spot, shorts the perpetual, and holds on to the strategy throughout the sample. We present the performance of this strategy in the last column of Table \ref{port_strategy_comparison}. We find that the long-only strategy significantly outperforms the carry strategy. These results highlight the usefulness of our fundamental price. Given this benchmark, we can identify deviations in real-time and dynamically rebalance the position to take advantage of the deviation rather than making a directional bet as in the Carry strategy.

\begin{table}
\begin{center}
\footnotesize
\begin{tabularx}{\textwidth}{@{\hskip\tabcolsep\extracolsep\fill}ll*{4}{r}}
\toprule
  &  & \multicolumn{4}{c}{Trading Strategies} \\
\cmidrule{3-6}
 &  & Correct & No Clamp & Long-only & Carry \\
\midrule
BTC & SR & 11.65 & 6.06 & 8.20 & 2.24 \\
 & Return & 52.55 & 27.52 & 31.37 & 10.18 \\
 & Volatility & 4.51 & 4.54 & 3.83 & 4.55 \\
 & MaxDD & -3.82 & -3.82 & -3.82 & -5.33 \\
 & $\alpha$ & 58.82 & 31.22 & 33.30 & 7.89 \\
 & $t_{\alpha}$ & 19.69 & 11.09 & 18.39 & 4.96 \\
 & Active \% & 100.00 & 100.00 & 74.05 & 100.00 \\
 & OtC time & 4.92 & 16.23 & 7.28 & - \\
ETH & SR & 12.77 & 8.29 & 8.64 & 2.66 \\
 & Return & 62.83 & 40.98 & 38.02 & 13.22 \\
 & Volatility & 4.92 & 4.94 & 4.40 & 4.96 \\
 & MaxDD & -2.62 & -2.62 & -2.62 & -2.80 \\
 & $\alpha$ & 72.39 & 49.86 & 41.45 & 10.77 \\
 & $t_{\alpha}$ & 17.45 & 11.02 & 16.21 & 5.31 \\
 & Active \% & 100.00 & 100.00 & 74.61 & 100.00 \\
 & OtC time & 5.05 & 13.60 & 7.53 & - \\
BNB & SR & 17.66 & 14.71 & 11.28 & -0.66 \\
 & Return & 134.74 & 112.80 & 64.80 & -5.14 \\
 & Volatility & 7.63 & 7.67 & 5.74 & 7.76 \\
 & MaxDD & -1.96 & -1.96 & -1.66 & -33.61 \\
 & $\alpha$ & 145.62 & 116.29 & 68.54 & -7.13 \\
 & $t_{\alpha}$ & 13.94 & 11.91 & 12.54 & -2.21 \\
 & Active \% & 100.00 & 100.00 & 63.30 & 100.00 \\
 & OtC time & 5.46 & 7.46 & 6.91 & - \\
DOGE & SR & 14.90 & 12.68 & 9.29 & 1.01 \\
 & Return & 181.12 & 154.69 & 96.76 & 12.39 \\
 & Volatility & 12.16 & 12.20 & 10.41 & 12.31 \\
 & MaxDD & -13.90 & -13.90 & -13.90 & -14.37 \\
 & $\alpha$ & 221.37 & 195.57 & 112.52 & 6.94 \\
 & $t_{\alpha}$ & 11.27 & 9.13 & 11.91 & 2.80 \\
 & Active \% & 100.00 & 100.00 & 64.91 & 100.00 \\
 & OtC time & 4.13 & 6.81 & 5.36 & - \\
ADA & SR & 19.76 & 15.78 & 12.13 & 1.64 \\
 & Return & 155.16 & 124.91 & 84.16 & 13.15 \\
 & Volatility & 7.85 & 7.92 & 6.94 & 8.03 \\
 & MaxDD & -5.69 & -5.69 & -5.69 & -6.58 \\
 & $\alpha$ & 176.12 & 144.73 & 91.75 & 9.17 \\
 & $t_{\alpha}$ & 20.12 & 16.38 & 20.67 & 3.92 \\
 & Active \% & 100.00 & 100.00 & 65.10 & 100.00 \\
 & OtC time & 4.32 & 6.21 & 5.63 & - \\
\bottomrule
\end{tabularx}
\end{center}
\caption{Comparison of Different Arbitrage Trading Strategies}
\label{port_strategy_comparison}
\bigskip
\footnotesize
This table compares the performance of 4 trading strategies. Correct is the strategy proposed in this paper. No Clamp is the strategy with the fundamental price given in our earlier draft and \citet{ackerer_perpetual_2023} which ignores the clamp feature. Long-only is our strategy with a restriction of no short sales in the spot market. Carry is the strategy studied in \citet{christin2022crypto}. Statistics reported are the annualized Sharpe ratio, return (\%), volatility (\%), max drawdown (\%), alpha (\%) relative to the three factors of \citet{liu2022common}, t-stat of the alpha, proportion of time the strategy is active (\%), and average open-to-close (OtC) position duration in hours. Sample periods are those of Table \ref{tab2}.
\end{table}

In Appendix \ref{sec:data_driven}, we also present the performance of a data-driven strategy that optimally determines the open and close thresholds using data from the past 6 months. The data-driven strategy has a performance similar to or worse than our theory-motivated strategy, suggesting that our proposed trading strategy is close to being optimal for arbitrage.

Comparing the results with \citet*{du_deviations_2018}, we find that deviations in crypto perpetual futures are considerably larger in magnitude. As a result, gains from the arbitrage strategies we study are also larger. Even though the volatility of the trading strategy also scales up, the Sharpe ratios in the crypto space are still larger than those in the traditional foreign exchange market, as reported in \citet*{du_deviations_2018}.

When a futures-spot gap opens up, gains from the trading strategy could potentially arise from two main sources: price convergence and funding rate payments. While industry publications usually emphasize the funding rate channel, we note that price convergence can generate quicker gains from arbitrage if dislocations are short-lived.
To examine these two sources empirically, in Table \ref{tab_decom_rma}, we provide a decomposition of the trading strategy's performance into price convergence versus funding rate payments. We find that price convergence plays a dominant role in total trading returns, while funding rate payments have a more minor role, which seems to diminish over time.

\begin{table}[!htb]
\begin{center}
\begin{tabularx}{\textwidth}{@{\hskip\tabcolsep\extracolsep\fill}ll*{6}{r}}
\toprule
 &  & 2020 & 2021 & 2022 & 2023 & 2024 & All \\
\midrule
BTC & Return & 49.62 & 64.81 & 60.02 & 41.68 & 21.44 & 52.55 \\
 & Price & 29.99 & 34.76 & 55.93 & 36.22 & 3.30 & 37.61 \\
 & Funding & 19.63 & 30.05 & 4.09 & 5.46 & 18.15 & 14.94 \\
ETH & Return & 67.53 & 75.56 & 70.73 & 44.81 & 25.11 & 62.83 \\
 & Price & 41.98 & 40.66 & 65.52 & 39.17 & 7.54 & 45.05 \\
 & Funding & 25.55 & 34.91 & 5.21 & 5.64 & 17.57 & 17.78 \\
BNB & Return & 159.83 & 166.86 & 134.73 & 87.08 & 102.83 & 134.91 \\
 & Price & 125.07 & 132.19 & 124.80 & 73.48 & 74.90 & 111.75 \\
 & Funding & 34.77 & 34.67 & 9.93 & 13.61 & 27.93 & 23.16 \\
DOGE & Return & 408.14 & 207.91 & 191.43 & 62.89 & 35.36 & 180.96 \\
 & Price & 393.74 & 165.86 & 188.89 & 57.44 & 12.59 & 164.26 \\
 & Funding & 14.39 & 42.05 & 2.54 & 5.45 & 22.78 & 16.70 \\
ADA & Return & 147.47 & 138.28 & 210.28 & 145.53 & 42.29 & 155.13 \\
 & Price & 117.41 & 98.57 & 206.19 & 140.72 & 19.16 & 135.50 \\
 & Funding & 30.06 & 39.71 & 4.09 & 4.80 & 23.13 & 19.62 \\
\bottomrule
\end{tabularx}
\end{center}
\caption{Return Decomposition: Price Convergence versus Funding Rate Payment}\label{tab_decom_rma}
\bigskip
\small
This table decomposes the portfolio return into the part due to price convergence and the part due to funding rate payment. The returns reported are excess returns from the random-maturity arbitrage strategies. 
\end{table}

The success of the trading strategy supports the theory of random-maturity arbitrage. Whenever there is a deviation larger than the gap implied by the theory, betting on convergence tends to generate a positive payoff at some uncertain future time. Therefore, the convergence arbitrage trade generates high Sharpe ratios. We next examine whether these gains survive the transaction costs faced by an arbitrageur in practice.

\subsection{Effective Bid-Ask Spreads and the Cost of Arbitrage}\label{sec:tradingcosts}

We next examine the execution costs of the random-maturity arbitrage strategy. This requires measuring bid-ask spreads over the full sample. We therefore construct daily effective bid-ask spreads from Binance transaction records and document their behavior over time. We then incorporate these spreads, together with exchange fees, into the arbitrage strategy.

For each of the five currencies, we obtain the complete daily record of
aggregated trades in Binance's spot market and in its USD-margined perpetual
futures market.\footnote{The archives are available at
\url{https://data.binance.vision/data/spot/daily/aggTrades/} and
\url{https://data.binance.vision/data/futures/um/daily/aggTrades/}. An
aggregated trade combines the fills of a single taker order at a single price.}
Each record reports the trade price $p_t$ and whether the taker, the side that
initiated the trade, was the buyer or the seller. We code the trade direction as
$q_t=+1$ for a taker buy and $q_t=-1$ for a taker sell.

To estimate the effective spread, let $m_t$ denote the log efficient price at
the time of trade $t$ and let $s$ denote the proportional effective
bid-ask spread. The transaction price satisfies
\begin{equation}
    \log p_t = m_t + \frac{s}{2}q_t.
    \label{eq:spread_price}
\end{equation}
If $m_t$ follows a random walk with innovations $\epsilon_t$ that are
uncorrelated with trade direction, taking first differences gives
\begin{equation}
    \Delta \log p_t
    =
    \frac{s}{2}\Delta q_t+\epsilon_t.
    \label{eq:spread_roll}
\end{equation}
This is the trade-indicator representation underlying \citet{roll1984simple}. In the original Roll estimator, trade direction is unobserved, so the effective spread is inferred from the negative serial covariance of transaction price changes. Binance, by contrast, reports whether the taker is the buyer or the
seller for every trade. We can therefore use the observed trade direction to
estimate Equation \eqref{eq:spread_roll} directly by ordinary least squares,
separately for each currency, market, and day:
\begin{equation}
    \widehat{s}
    = 2
    \frac{\widehat{\operatorname{Cov}}
    \left(\Delta\log p_t,\Delta q_t\right)}
    {\widehat{\operatorname{Var}}\left(\Delta q_t\right)}.
    \label{eq:spread_estimator}
\end{equation}
A taker order executes at or beyond the best quote, and quotes lie on the
exchange's price grid, so the spread paid by a trade is at least one tick. We
therefore floor each daily estimate at one tick, measured as the most common
nonzero price change during the day and expressed in basis points of the average
price. The floor rarely binds for BTC, on at most 5\% of days in either market,
or for the ETH perpetual.

The Kaiko order-book sample described in Section \ref{sec:data} also provides an external validation of our trade-based spread estimates. In these snapshots, the
quoted spread of the Bitcoin perpetual equals one tick 99\% of the time and
averages approximately 0.05 basis points. Over the same period, our effective
spread estimate averages approximately 0.1 basis points. The two measures are
therefore of the same order of magnitude. The effective spread is somewhat
larger, as expected, because taker orders can execute at several price levels
beyond the best quote and because the estimator can also absorb part of the
price impact of trades.

\begin{figure}[!htb]
\begin{center}
\noindent\includegraphics[width=\textwidth]{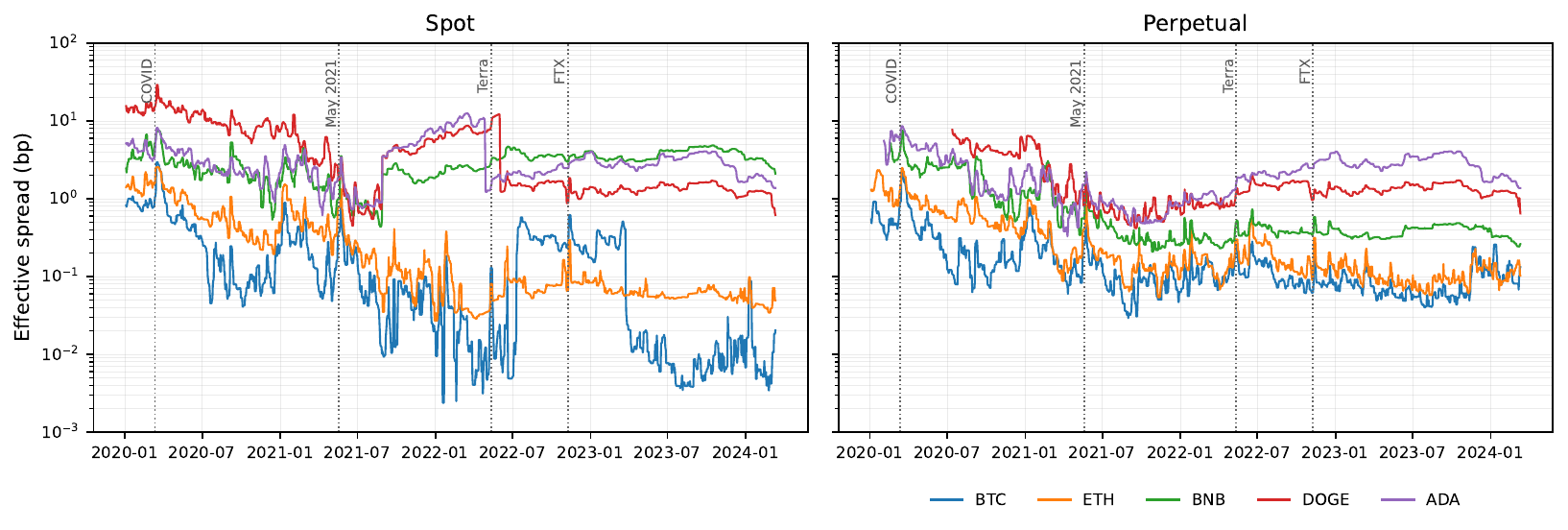}
\end{center}
\caption{Effective Bid-Ask Spreads of Spot and Perpetual Futures}
\label{fig:effective_spread}
\bigskip
\small
This figure plots the daily effective bid-ask spread on Binance of the spot
(left panel) and perpetual futures (right panel) markets, estimated from
Equation \eqref{eq:spread_estimator} and floored at one tick, in basis points.
Each series is smoothed with a seven-day moving median and plotted on a
logarithmic scale. Flat stretches are days on which the tick floor binds.
Dotted vertical lines mark the COVID crash of March 12, 2020, the crash of
May 19, 2021, the collapse of Terra in May 2022, and the collapse of FTX in
November 2022. Each series starts in January 2020 or on the first day with
trade records for the contract and ends on March 11, 2024.
\end{figure}

Figure \ref{fig:effective_spread} reveals three patterns. First, perpetual
futures are highly liquid. Over the full sample, the median daily effective
spread of the perpetual is 0.11 basis points for BTC, 0.18 for ETH, 0.44 for
BNB, 1.35 for DOGE, and 2.12 for ADA. The perpetual spread is below that of the
corresponding spot market for BNB, DOGE, and ADA and is comparable to it for
BTC and ETH. Second, effective spreads generally narrow over time. Comparing
the periods before and after 2022, the mean effective spread of the perpetual
declines by between 48\% and 73\% for BTC, ETH, BNB, and DOGE. Third, spreads
increase sharply during periods of market stress. On March 12, 2020, for
example, the effective spread of the Bitcoin perpetual reached 5.95 basis
points, more than ten times its median over the preceding weeks, and the
spreads of all other contracts trading at the time more than doubled.

We now incorporate these costs into the random-maturity arbitrage strategy.
In the presence of trading costs, as motivated by our theory, we use the
threshold trading rules implied by Table \ref{tab3}. Different fee tiers have
the same close-position threshold, equal to zero, while the open-position
threshold becomes wider as trading costs increase. We visualize these trading
rules in Figure \ref{trading_theory} in Appendix \ref{sec:figure}.

Beyond the baseline case with only exchange fees, we also report the performance of the strategy with both exchange fees and bid-ask spread. Let $c_S$ and $c_F$ denote the one-way spot and perpetual exchange fees in Table \ref{tab3}, and let $s^S_t$ and $s^F_t$ denote the corresponding daily effective spreads.

The proportional round-trip trading cost $c$ in Equation \eqref{tc_empirical} therefore becomes
\begin{equation}
    c_t
    =
    2\left(
        c_S
        +
        \frac{s^S_t}{2}
    \right)
    +
    2\left(
        c_F
        +
        \frac{s^F_t}{2}
    \right).
    \label{eq:total_trading_cost}
\end{equation}

Since the strategy under trading costs is a threshold trading rule, to
annualize Sharpe ratios, we follow \citet{lucca2015pre}, who consider a trading
strategy that is active only on FOMC announcement dates. We first calculate the
mean ($\mu$) and standard deviation ($\sigma$) of the strategy return during
the periods in which the strategy is active. We then scale $\mu/\sigma$ by the
number of active periods in a year:
\[
    SR = \frac{\mu}{\sigma}\sqrt{N_a},
\]
where $N_a$ is the average number of hours the strategy is active in a year.
We use the same convention to annualize returns and standard deviations:
$\mu_{\mathrm{ann}}=\mu N_a$ and
$\sigma_{\mathrm{ann}}=\sigma\sqrt{N_a}$.

\begin{table}[!htb]
\begin{spacing}{1}
\centering
\fontsize{9}{10.5}\selectfont
\setlength{\tabcolsep}{3pt}
\renewcommand{\arraystretch}{1.04}
\begin{tabular*}{\textwidth}{@{\extracolsep{\fill}}ll*{8}{r}@{}}
\toprule
 & & \multicolumn{2}{c}{No fee} & \multicolumn{2}{c}{Low} & \multicolumn{2}{c}{Medium} & \multicolumn{2}{c}{High} \\
\cmidrule(lr){3-4} \cmidrule(lr){5-6} \cmidrule(lr){7-8} \cmidrule(lr){9-10}
 & & Fee only & $+$ spread & Fee only & $+$ spread & Fee only & $+$ spread & Fee only & $+$ spread \\
\midrule
BTC & SR & 11.65 & 10.46 & 4.30 & 4.05 & 3.38 & 3.31 & 3.35 & 3.27 \\
 & Return & 52.55 & 46.63 & 18.04 & 16.94 & 13.88 & 13.53 & 12.96 & 12.55 \\
 & Volatility & 4.51 & 4.46 & 4.19 & 4.18 & 4.10 & 4.08 & 3.87 & 3.84 \\
 & MaxDD & -3.82 & -3.82 & -3.82 & -3.82 & -3.89 & -3.90 & -2.63 & -2.63 \\
 & $\alpha$ & 58.82 & 49.75 & 16.98 & 15.37 & 12.62 & 12.12 & 12.04 & 11.70 \\
 & $t_{\alpha}$ & 19.69 & 18.62 & 7.39 & 7.18 & 6.02 & 5.90 & 5.83 & 5.85 \\
 & Active \% & 100.00 & 95.01 & 45.44 & 43.83 & 33.69 & 32.38 & 24.19 & 23.22 \\
 & OtC time & 4.92 & 5.35 & 40.74 & 47.71 & 89.94 & 94.75 & 96.16 & 102.35 \\
\addlinespace[2pt]
ETH & SR & 12.77 & 11.42 & 5.83 & 5.49 & 4.69 & 4.66 & 4.56 & 4.45 \\
 & Return & 62.83 & 55.13 & 26.38 & 24.41 & 20.55 & 20.06 & 18.81 & 18.18 \\
 & Volatility & 4.92 & 4.83 & 4.52 & 4.45 & 4.38 & 4.31 & 4.12 & 4.08 \\
 & MaxDD & -2.62 & -2.63 & -2.67 & -2.65 & -2.66 & -2.62 & -2.64 & -2.65 \\
 & $\alpha$ & 72.39 & 60.91 & 26.96 & 23.91 & 20.13 & 19.19 & 18.20 & 17.02 \\
 & $t_{\alpha}$ & 17.45 & 16.70 & 7.41 & 7.40 & 6.35 & 6.33 & 5.62 & 5.77 \\
 & Active \% & 100.00 & 95.12 & 51.60 & 49.64 & 39.83 & 38.23 & 26.93 & 26.28 \\
 & OtC time & 5.05 & 5.52 & 33.89 & 41.84 & 73.21 & 82.26 & 80.74 & 85.05 \\
\addlinespace[2pt]
BNB & SR & 17.66 & 12.33 & 11.41 & 9.05 & 8.35 & 7.40 & 6.93 & 6.63 \\
 & Return & 134.74 & 88.36 & 81.83 & 60.40 & 56.43 & 46.34 & 43.68 & 38.39 \\
 & Volatility & 7.63 & 7.17 & 7.18 & 6.67 & 6.75 & 6.26 & 6.31 & 5.79 \\
 & MaxDD & -1.96 & -2.26 & -1.97 & -1.50 & -2.02 & -1.50 & -2.08 & -1.50 \\
 & $\alpha$ & 145.62 & 80.14 & 72.51 & 49.63 & 46.99 & 35.88 & 34.95 & 29.67 \\
 & $t_{\alpha}$ & 13.94 & 10.61 & 8.34 & 7.26 & 6.28 & 6.18 & 5.74 & 5.82 \\
 & Active \% & 100.00 & 80.49 & 73.41 & 58.34 & 52.28 & 43.42 & 38.55 & 28.60 \\
 & OtC time & 5.46 & 8.75 & 10.72 & 18.24 & 21.40 & 27.69 & 29.03 & 30.09 \\
\addlinespace[2pt]
DOGE & SR & 14.90 & 7.42 & 9.22 & 5.00 & 6.27 & 4.19 & 4.97 & 3.45 \\
 & Return & 181.12 & 84.53 & 108.46 & 55.45 & 71.35 & 45.18 & 55.22 & 35.45 \\
 & Volatility & 12.16 & 11.40 & 11.77 & 11.09 & 11.38 & 10.78 & 11.10 & 10.29 \\
 & MaxDD & -13.90 & -13.90 & -13.90 & -13.90 & -13.90 & -13.90 & -13.90 & -13.90 \\
 & $\alpha$ & 221.37 & 72.96 & 116.91 & 42.34 & 66.46 & 32.07 & 44.87 & 23.94 \\
 & $t_{\alpha}$ & 11.27 & 7.59 & 7.16 & 4.93 & 5.41 & 4.21 & 4.72 & 3.14 \\
 & Active \% & 100.00 & 65.04 & 60.85 & 40.51 & 40.48 & 30.91 & 30.13 & 20.25 \\
 & OtC time & 4.13 & 8.40 & 7.62 & 18.45 & 12.68 & 24.78 & 16.28 & 29.19 \\
\addlinespace[2pt]
ADA & SR & 19.76 & 8.29 & 10.72 & 4.97 & 6.03 & 3.94 & 4.14 & 3.80 \\
 & Return & 155.16 & 58.89 & 78.51 & 33.76 & 41.77 & 25.94 & 27.42 & 23.05 \\
 & Volatility & 7.85 & 7.10 & 7.32 & 6.79 & 6.92 & 6.58 & 6.62 & 6.06 \\
 & MaxDD & -5.69 & -5.69 & -5.69 & -5.69 & -5.69 & -5.69 & -5.69 & -5.69 \\
 & $\alpha$ & 176.12 & 52.23 & 77.94 & 27.68 & 36.32 & 21.23 & 22.88 & 19.90 \\
 & $t_{\alpha}$ & 20.12 & 13.59 & 13.57 & 7.53 & 8.79 & 6.01 & 6.23 & 5.59 \\
 & Active \% & 100.00 & 61.31 & 66.42 & 41.85 & 42.81 & 31.27 & 30.85 & 22.29 \\
 & OtC time & 4.32 & 10.68 & 8.47 & 27.51 & 19.39 & 49.40 & 42.24 & 63.71 \\
\bottomrule
\end{tabular*}
\caption{Random-Maturity Arbitrage with Exchange Fees and Effective Spreads}
\label{port_fee_theory}\label{tab:bidask}
\medskip
\footnotesize
\justifying
\noindent
Performance under different trading cost tiers, with and without bid-ask spread costs. The fees for spot (futures) are 0 (0) bps, 2.25 (0.18) bps, 4.5 (0.72) bps, and 6.75 (1.44) bps for the zero, low, medium, and high trading cost levels. For each fee tier, the spread-adjusted strategy additionally pays half the daily effective spread on each transaction in each leg. Spreads are estimated over the execution day using Equation \eqref{eq:spread_estimator}. Statistics reported are the annualized Sharpe ratio, return (\%), volatility (\%), max drawdown (\%), alpha (\%) relative to the three factors of \citet{liu2022common}, $t$-stat of the alpha, proportion of time the strategy is active (\%), and average open-to-close (OtC) position duration in hours. Sample periods are those of Table \ref{tab2}.
\end{spacing}
\end{table}

Table \ref{port_fee_theory} presents the strategy's performance under different
exchange fee tiers. As fees increase, the random-maturity arbitrage strategy
trades less frequently and the average open-to-close duration increases.
Nevertheless, even at the high fee tier, the strategy generates large Sharpe
ratios across all five currencies. It also delivers highly significant alphas
relative to the 3-factor model of \citet{liu2022common} and the 5-factor model
of \citet{cong2022staking}.

The effective spread has relatively little effect on BTC and ETH. At the high
fee tier, adding the spread reduces the Sharpe ratio from 3.35 to 3.27 for BTC
and from 4.56 to 4.45 for ETH. The effect is larger for BNB, DOGE, and ADA. As exchange fees increase, the spread represents a smaller share of total execution costs and the difference between the fee-only and fee-plus-spread strategies generally narrows.

Most importantly, incorporating effective spreads does not overturn the
economic conclusion from the arbitrage strategy. At the high fee tier, Sharpe
ratios net of both exchange fees and effective spreads range from 3.27 for BTC
to 6.63 for BNB. The corresponding alphas remain statistically significant for
all five currencies, with $t$-statistics of at least 3.14. Thus, although
effective bid-ask spreads are an economically important component of trading
costs, particularly for less liquid cryptocurrencies, they do not account
for the large deviations from the no-arbitrage benchmark.

Execution costs are only one potential impediment to arbitrage. Even a trade that is profitable upon convergence can generate interim losses if the deviation widens before the position is closed. Because perpetual futures positions are margined, sufficiently large interim losses could force an arbitrageur to liquidate the position before convergence.

The maximum drawdowns reported in Table \ref{port_fee_theory} provide a simple gauge of this risk. Across the specifications in the table, maximum drawdowns do not exceed 3.90\% for BTC, 2.67\% for ETH, and 2.26\% for BNB. They are larger for DOGE and ADA, reaching 13.90\% and 5.69\%, respectively.

Execution costs are only one potential impediment to arbitrage. Even a trade
that is profitable upon convergence can generate interim losses if the
deviation widens before it closes. We therefore next examine whether
liquidation risk can prevent arbitrageurs from exploiting these deviations. 

The maximum drawdowns reported in Table \ref{port_fee_theory} provide a simple gauge of this risk. Across the specifications in the table, maximum drawdowns do not exceed 3.90\% for BTC, 2.67\% for ETH, and 2.26\% for BNB. They are larger for DOGE and ADA, reaching 13.90\% and 5.69\%, respectively. Thus, interim losses are relatively modest for the more liquid cryptocurrencies, although liquidation risk can be more relevant for less liquid assets. Beyond drawdowns, liquidations also depend on leverage, maintenance-margin requirements, available collateral, and the applicable margining regime. We examine the broader role of margin constraints and arbitrage capital in Section \ref{sec:explaining}.

\section{Explaining Futures-spot Deviations}\label{sec:explaining}

Figure \ref{fig1} documents large and persistent futures-spot deviations that comove strongly across cryptocurrencies. What generates these deviations? We explore the hypothesis that they reflect the interaction between end-user demand for leveraged perpetual exposure and the limited capacity of arbitrageurs to intermediate that demand.

We proceed in four steps. Section \ref{sec:explain_reg} begins with a set of exploratory evidence on deviations $\rho$ in reduced-form panel regressions. Section \ref{sec:explain_model} then summarizes the economic mechanism and two empirical predictions of a simple model, which we develop formally in Appendix \ref{sec:model}. Sections \ref{sec:explain_pm} and \ref{sec:explain_impact} test these predictions using Binance's introduction of portfolio margining and the state dependence of the price impact of perpetual order flow and find support for the model in the data.

\subsection{Exploratory Evidence}\label{sec:explain_reg}

Two forces could plausibly generate the common variation in $\rho$. The first is variation in arbitrageurs' ability to intermediate between futures and spot markets. Since the same intermediaries operate across cryptocurrencies, common variation in their funding or balance-sheet conditions could generate common variation in deviations. The second force is variation in end-user demand for perpetual exposure relative to spot. Perpetuals allow for high leverage and therefore provide a natural venue for investors with strong directional beliefs. If such beliefs comove across cryptocurrencies, they can generate common demand pressure in perpetual markets.

We begin with a reduced-form regression designed to let the data speak to both channels. We proxy relative demand for perpetual exposure using trailing spot returns. If investors extrapolate recent returns, positive past performance should increase demand for perpetual exposure relative to spot. We proxy time-varying risk-based constraints on arbitrageurs using trailing spot-return volatility, following the Value-at-Risk logic of \citet{adrian2010liquidity}: higher volatility tightens risk constraints and should reduce intermediaries' willingness to absorb order flow.

Finally, Figure \ref{fig1} suggests that deviations decline substantially over the sample. Models of convergence trading such as \cite{kondor_risk_2009} predict that competition among arbitrageurs may reduce the gap gradually. Therefore, we add a time trend variable to the regression to test if the market becomes more efficient over time.

In sum, we estimate the following regression models:
\begin{align}
    & \rho_{i, t} = \beta_{0,i} + \beta_1 (t / 365) + \beta_2 R_{i, t - 120 : t-1} + \beta_3 \sigma_{i, t-120:t-1} + \epsilon_{i,t}, \\
    & |\rho_{i, t}| = \beta_{0,i} + \beta_1 (t / 365) + \beta_2 R_{i, t - 120 : t-1} + \beta_3 \sigma_{i, t-120:t-1} + \epsilon_{i,t},
\end{align}
where $\rho_{i, t}$ represents the futures-spot deviation, $\beta_{0, i}$ is the currency fixed effect, $t$ is the number of days since the beginning of the sample, $R_{i, t-120:t-1}$ is the annualized spot return to currency $i$ over the past 120 days, and $\sigma_{i, t-120:t-1}$ is the annualized standard deviation of spot return of currency $i$ over the past 120 days. Since we scale the time trend by the number of trading days in a year, the regression coefficient $\beta_1$ has an interpretation of the average annual change in the futures-spot deviation.

\begin{table}[!htb]
\begin{center}
\begin{tabularx}{\textwidth}{@{\hskip\tabcolsep\extracolsep\fill}ll*{6}{r}}
\toprule
Dependent Variable: & \multicolumn{3}{c}{$\rho$} & \multicolumn{3}{c}{$|\rho|$} \\
\cmidrule(lr){2-4} \cmidrule(lr){5-7}
Time        &       -0.20\sym{***}&       -0.13\sym{***}&       -0.23\sym{***}&       -0.22\sym{***}&       -0.16\sym{***}&       -0.23\sym{***}\\
            &     (-4.84)         &     (-3.59)         &     (-6.78)         &     (-6.09)         &     (-5.04)         &     (-8.22)         \\
Ret         &                     &        0.10\sym{***}&        0.17\sym{***}&                     &        0.08\sym{***}&        0.13\sym{***}\\
            &                     &      (4.45)         &     (10.79)         &                     &      (4.75)         &      (7.67)         \\
Vol         &                     &                     &       -0.03\sym{***}&                     &                     &       -0.02\sym{***}\\
            &                     &                     &     (-6.59)         &                     &                     &     (-4.52)         \\
Currency FE & Yes & Yes & Yes & Yes & Yes & Yes \\
\(R^{2}\)   &        0.07         &        0.17         &        0.22         &        0.12         &        0.21         &        0.25         \\
\(N\)       &        7236         &        7236         &        7236         &        7236         &        7236         &        7236         \\
\bottomrule
\end{tabularx}
\end{center}
\caption{Regression of the no-arbitrage deviations against explanatory variables}\label{reg_exp}
\bigskip
\small
We present regression results of $\rho$ and $|\rho|$ on a time trend (Time) measured in units of years, past 120 days' annualized returns (Ret), and volatility (Vol) in the spot market. We add a currency fixed effect to the regression and consider three specifications: (1) regressing $\rho$/$|\rho|$ on the time trend, (2) regressing $\rho$/$|\rho|$ on past returns, and (3) regressing $\rho$/$|\rho|$ on past returns and volatility. The observations are at a daily frequency. The sample period spans from January 8, 2020 to March 11, 2024, totaling 1,495 days. The table also reports the $R^2$ values for each regression model. We report Newey-West t-statistics in parentheses with a lag of 120 days. $*\text{ }p<.1$, $**\text{ }p<.05$, $***\text{ }p<.01$.
\end{table}

Table \ref{reg_exp} presents the regression results. Three facts stand out. First, deviations decline substantially over the sample. The coefficient on the time trend is negative and statistically significant at the one percent level in all six specifications. In the trend-only regression for $|\rho|$, the estimate of (-0.22) implies an average decline in the annualized absolute deviation of approximately 22 percentage points per year.

Second, deviations rise with past returns. The coefficients on trailing returns are positive and significant at the one percent level for both $\rho$ and $|\rho|$. This pattern is consistent with extrapolative demand: following high spot returns, investors increase their demand for leveraged perpetual exposure, and the portion of that demand not absorbed by arbitrageurs appears in the futures-spot deviation.

This finding is related to \citet{liu2021risks}, who document strong time-series momentum in cryptocurrency returns. Our evidence raises the possibility that some of this momentum is associated with leverage demand and price pressure originating in perpetual futures markets. We leave this as an open question for future research.

Third, the standard risk-based proxy for arbitrage constraints enters with the opposite sign. A simple Value-at-Risk mechanism predicts: higher volatility should tighten intermediary constraints and widen deviations. Therefore, the regressions provide no evidence that a time-varying, risk-based tightening of arbitrage constraints is the primary driver of the common variation in $\rho$.

\subsection{Economic Mechanism and Predictions}\label{sec:explain_model}

We use a simple model to organize the evidence and motivate the empirical tests that follow. The model is developed formally in Appendix \ref{sec:model}. 

The model has two types of market participants. End users demand perpetual futures because they provide leveraged exposure to the underlying cryptocurrency. Arbitrageurs take the other side of this demand and hedge their positions in the spot market. Their willingness to intermediate, however, is limited by trading costs, risk, and the amount of capital required to maintain the arbitrage position.

The key mechanism is that arbitrageurs absorb end-user demand as long as they have sufficient balance-sheet capacity. When capacity is abundant, an increase in demand can largely be accommodated by larger arbitrage positions, with relatively little effect on the futures-spot deviation. When arbitrageurs become capital constrained, they can no longer expand their positions sufficiently, and the market instead clears through larger price deviations. This mechanism delivers two predictions.

\noindent\textbf{Prediction 1 (arbitrage constraints).} \emph{Relaxing arbitrageurs' capital constraints should reduce futures-spot deviations.} Section \ref{sec:explain_pm} tests this prediction using Binance's introduction of portfolio margining, which allowed arbitrageurs to use their capital more efficiently by netting spot and futures positions within the same margin account.

\noindent\textbf{Prediction 2 (state-dependent price impact).} \emph{End-user demand should have a larger price impact when arbitrageurs are more constrained.} When arbitrageurs have spare capacity, they can absorb additional order flow with little movement in prices. When their capacity is limited, the same order flow should produce a larger change in the futures-spot deviation. Section \ref{sec:explain_impact} tests this prediction by examining whether perpetual-futures order flow has a larger price impact when the initial deviation is large.

Both tests provide support for these predictions and therefore for the suggested mechanism.

\subsection{Prediction 1: Arbitrage Constraints and Equilibrium Deviations}\label{sec:explain_pm}

Prediction 1 says that deviations should narrow when arbitrageurs can support a larger book with the same capital. Binance's introduction of portfolio margining on April 20, 2022 provides a direct test.\footnote{Details of the event are provided in Binance's announcement: \url{https://www.binance.com/en/support/announcement/detail/c013c3ceaa604a3eaa362647b06441a1}.} Before the change, arbitrageurs maintain separate margin accounts for spot and futures positions, so a gain on one leg could not offset a loss on the other, and each leg had to be capitalized against early liquidation on its own. Portfolio margining allowed the two legs to be netted inside a unified account. According to the model, this decline in margin requirements implies a rise in arbitrage capacity, therefore a compression of the deviation. 

\begin{table}[!htb]
\begin{center}
\begin{tabularx}{1\columnwidth}{@{\hskip\tabcolsep\extracolsep\fill}l*{6}{c}}
\toprule
Window & Pre Mean & Post Mean & Diff. & HAC $t$ & HAC $p$ & HAC Lag \\
\midrule
$20$ & $0.2437$ & $0.1657$ & $-0.0780$ & $-4.09$ & $0.0000$ & $6$ \\
$30$ & $0.2747$ & $0.2283$ & $-0.0464$ & $-1.71$ & $0.0873$ & $7$ \\
$40$ & $0.2668$ & $0.2271$ & $-0.0397$ & $-1.85$ & $0.0649$ & $7$ \\
$60$ & $0.2572$ & $0.2193$ & $-0.0378$ & $-2.36$ & $0.0182$ & $8$ \\
$90$ & $0.2333$ & $0.2008$ & $-0.0325$ & $-2.72$ & $0.0066$ & $9$ \\
\bottomrule
\end{tabularx}
\end{center}
\caption{Changes in $|\bar{\rho}|$ around the Portfolio Margin Event}\label{tab:portfolio_margin_event}
\bigskip
\small
This table presents changes in the absolute value of the average deviation from no-arbitrage benchmarks around the introduction of Binance portfolio margin on April 20, 2022. The windows are in units of days. The values in Pre Mean, Post Mean, and Diff are annualized. $|\bar{\rho}_t|$ is the absolute value of the cross-currency average of the hourly deviation $\rho$ across BTC, ETH, BNB, DOGE and ADA. Pre Mean and Post Mean are the means of $|\bar{\rho}_t|$ over the $\pm w$ days around the launch, and Diff (Post $-$ Pre) is the post-event change; a negative value is a compression. HAC $t$-stat and HAC $p$-value test Diff using \citet{neweywest1987} standard errors with a Bartlett kernel; the truncation lag (HAC Lag) follows $L = \lfloor 4(n/100)^{2/9}\rfloor$ for $n$ hourly observations.
\end{table}

Table \ref{tab:portfolio_margin_event} compares the cross-currency mean $|\bar{\rho}_t|$ over symmetric windows around the event. Over windows of 20 to 90 days, the annualized absolute deviation falls by $3.3$ to $7.8$ percentage points, with HAC $t$-statistics between $-1.71$ and $-4.09$. Against pre-event means of $23$ to $27$ annualized percentage points, this is a compression of roughly one seventh to one third. Appendix \ref{sec:portfolio_margin} plots the corresponding 90-day event window and shows the compression in deviations visually.

\subsection{Prediction 2: State Dependent Price Impact}\label{sec:explain_impact}

Prediction 2 says that the price impact of order flow depends on the state of arbitrage capacity. When arbitrage capital is slack, arbitrageurs absorb end-user demand with little movement in the futures-spot deviation. Once arbitrage capacity becomes constrained, however, additional order flow must be accommodated through a larger price response. We therefore test Prediction 2 by allowing the price impact of perpetual order flow to vary with the absolute deviation at the beginning of the interval.

We aggregate the hourly panel to eight-hour funding intervals and estimate the following specification on the pooled panel of five currencies:
\begin{equation}
\Delta P_{i,t+1}
= a_i + \sum_k \gamma_k B_{k,i,t}
+ \sum_k b_k \left(B_{k,i,t} \times OIB^{perp}{i,t+1}\right)
+ \varepsilon{i,t+1},
\label{eq}
\end{equation}
where $P_{i,t}=(F_{i,t}-S_{i,t})/F_{i,t}$ is the premium index, $\Delta P_{i,t+1}=P_{i,t+1}-P_{i,t}$ is measured in basis points, and $OIB^{perp}{i,t+1}$ is the perpetual taker order imbalance over the same interval, measured in percentage points. The specification includes currency fixed effects, $a_i$, and indicators $B_{k,i,t}$ for bins of the initial absolute premium, $|P_{i,t}|$. The coefficients $b_k$ therefore measure the price impact of order flow conditional on the initial size of the futures-spot deviation.

Table~\ref{tab:price-impact} first divides intervals within each currency into terciles of $|P_{i,t}|$. In the low-deviation tercile, where mean $|P|$ is 2.47 basis points, a one-percentage-point increase in taker order imbalance moves the premium index by 0.031 basis points. In the high-deviation tercile, where mean $|P|$ is 12.05 basis points, the same order flow moves the premium by 0.094 basis points, roughly three times as much. The high-minus-low difference is 0.062 basis points and is statistically significant.

\begin{table}[!htb]
\begin{center}
\begin{tabularx}{1\columnwidth}{@{\hskip\tabcolsep\extracolsep\fill}l*{4}{c}}
\toprule
& Low Deviation & Middle & High Deviation & High $-$ Low \\
\midrule
Price Impact
& $0.031^{**}$ & $0.022^{*}$ & $0.094^{***}$ & $0.062^{***}$ \\
& $(0.016)$ & $(0.012)$ & $(0.019)$ & $(0.023)$ \\
Mean Initial Premium (bp) & $2.47$ & $5.53$ & $12.05$ & \\
Observations & $22{,}129$ & $22{,}129$ & $22{,}129$ & $22{,}129$ \\
Currency FE & Yes & Yes & Yes & Yes \\
\bottomrule
\end{tabularx}
\end{center}
\caption{Price Impact of Perpetual Taker Order Flow by Initial Deviation}\label{tab:price-impact}
\bigskip
\small
This table presents the price impact of perpetual taker order imbalance across intervals with different initial futures-spot deviations. The panel is aggregated to eight-hour funding intervals. Within each cryptocurrency, intervals are divided into terciles based on the absolute premium index at the beginning of the interval. Price Impact is the change in the premium index, in basis points, associated with a one-percentage-point increase in perpetual taker order imbalance. The final column reports the difference between the high- and low-deviation terciles. Driscoll-Kraay standard errors with a Bartlett kernel and six lags are reported in parentheses. $^{*},p<.1$, $^{**},p<.05$, $^{***},p<.01$.
\end{table}

The results match the state dependence implied by Prediction 2. Over most of the deviation distribution, additional perpetual order flow produces little change in the futures-spot premium, consistent with arbitrage capital elastically absorbing end-user demand. Price impact rises sharply only in the upper tail, where large deviations indicate that arbitrage capacity is more likely to bind.

\section{Conclusion}\label{sec:conclusion}
Perpetual futures play an important role in today's crypto markets and could potentially be adopted in non-crypto markets in the future. Understanding the fundamental mechanism of this financial derivative is a crucial first step for understanding speculation and hedging dynamics in this fast-evolving area. We provide a comprehensive analysis of the arbitrage and funding rate payment mechanisms that underpin perpetual futures.

In an ideal, frictionless world, we show that arbitrageurs would trade perpetual futures in such a way that a constant proportional relationship would hold between the futures price and the spot price. In the presence of trading costs, the deviation of the futures price from the spot would lie within a bound.

Motivated by our theory, we empirically examine the comovement of the futures-spot spread across different cryptocurrencies and implement a theory-motivated arbitrage strategy. We find that this simple strategy yields substantial Sharpe ratios across various trading cost scenarios. The evidence supports our theoretical argument that perpetual futures-spot spreads exceeding trading costs represent a random-maturity arbitrage opportunity.

Finally, we develop a simple model in which end users demand leveraged perpetual exposure while arbitrageurs intermediate between futures and spot subject to limited balance-sheet capacity. The model predicts that looser arbitrage constraints compress deviations and that end-user demand has greater price impact when arbitrage capacity is constrained. Consistent with these predictions, deviations fall following Binance's introduction of portfolio margining, and perpetual order flow has greater price impact when initial deviations are large. We also find that past returns positively predict deviations, consistent with extrapolative demand for leveraged perpetual exposure. Together, these findings highlight the interaction between end-user demand and limited arbitrage capacity as an important force shaping perpetual futures prices.

\clearpage
    \newpage
     \onehalfspacing
    \bibliographystyle{jf}
    \bibliography{perp}

\begin{thebibliography}{39}
\expandafter\ifx\csname natexlab\endcsname\relax\def\natexlab#1{#1}\fi

\bibitem[Ackerer et~al.(2023)Ackerer, Hugonnier, and Jermann]{ackerer_perpetual_2023}
Ackerer, Damien, Julien Hugonnier, and Urban~J. Jermann, 2023, Perpetual {{Futures Pricing}}, Working Paper.

\bibitem[Adams et~al.(2021)Adams, Zinsmeister, Salem, Keefer, and Robinson]{adams2021uniswap}
Adams, Hayden, Noah Zinsmeister, Moody Salem, River Keefer, and Dan Robinson, 2021, Uniswap v3 core, Whitepaper.

\bibitem[Adrian and Shin(2010)]{adrian2010liquidity}
Adrian, Tobias, and Hyun~Song Shin, 2010, Liquidity and leverage, {\em Journal of financial intermediation\/} 19, 418--437.

\bibitem[Alexander et~al.(2020)Alexander, Choi, Park, and Sohn]{alexander2020bitmex}
Alexander, Carol, Jaehyuk Choi, Heungju Park, and Sungbin Sohn, 2020, Bitmex bitcoin derivatives: Price discovery, informational efficiency, and hedging effectiveness, {\em Journal of Futures Markets\/} 40, 23--43.

\bibitem[Amiram et~al.(2021)Amiram, Lyandres, and Rabetti]{amiram2020competition}
Amiram, Dan, Evgeny Lyandres, and Daniel Rabetti, 2021, Cooking the order books: Information manipulation and competition among crypto exchanges, Working Paper.

\bibitem[Angeris et~al.(2022)Angeris, Chitra, Evans, and Lorig]{angeris2022primer}
Angeris, Guillermo, Tarun Chitra, Alex Evans, and Matthew Lorig, 2022, A primer on perpetuals, Working Paper.

\bibitem[Brunnermeier and Pedersen(2009)]{brunnermeier2009market}
Brunnermeier, Markus~K, and Lasse~Heje Pedersen, 2009, Market liquidity and funding liquidity, {\em The review of financial studies\/} 22, 2201--2238.

\bibitem[Bryzgalova et~al.(2025)Bryzgalova, Pavlova, and Sikorskaya]{bryzgalova2025strategic}
Bryzgalova, Svetlana, Anna Pavlova, and Taisiya Sikorskaya, 2025, Strategic arbitrage in segmented markets, {\em Journal of Financial Economics\/} 166, 104008.

\bibitem[Christin et~al.(2022)Christin, Routledge, Soska, and Zetlin-Jones]{christin2022crypto}
Christin, Nicolas, Bryan~R Routledge, Kyle Soska, and Ariel Zetlin-Jones, 2022, The crypto carry trade, Working Paper.

\bibitem[Cochrane(2009)]{cochrane2009asset}
Cochrane, John, 2009, {\em Asset pricing: Revised edition\/} (Princeton university press).

\bibitem[Cong et~al.(2022{\natexlab{a}})Cong, He, and Tang]{cong2022staking}
Cong, Lin~William, Zhiheng He, and Ke~Tang, 2022{\natexlab{a}}, Staking, token pricing, and crypto carry, Working Paper.

\bibitem[Cong et~al.(2022{\natexlab{b}})Cong, Karolyi, Tang, and Zhao]{cong2022value}
Cong, Lin~William, George~Andrew Karolyi, Ke~Tang, and Weiyi Zhao, 2022{\natexlab{b}}, Value premium, network adoption, and factor pricing of crypto assets, Working Paper.

\bibitem[Cong et~al.(2022{\natexlab{c}})Cong, Li, Tang, and Yang]{cong2022crypto}
Cong, Lin~William, Xi~Li, Ke~Tang, and Yang Yang, 2022{\natexlab{c}}, Crypto wash trading, Working Paper.

\bibitem[Cox et~al.(1981)Cox, Ingersoll~Jr, and Ross]{cox1981relation}
Cox, John~C, Jonathan~E Ingersoll~Jr, and Stephen~A Ross, 1981, The relation between forward prices and futures prices, {\em Journal of Financial Economics\/} 9, 321--346.

\bibitem[De~Blasis and Webb(2022)]{de2022arbitrage}
De~Blasis, Riccardo, and Alexander Webb, 2022, Arbitrage, contract design, and market structure in bitcoin futures markets, {\em Journal of Futures Markets\/} 42, 492--524.

\bibitem[Du et~al.(2018)Du, Tepper, and Verdelhan]{du_deviations_2018}
Du, Wenxin, Alexander Tepper, and Adrien Verdelhan, 2018, Deviations from {{Covered Interest Rate Parity}}, {\em Journal of Finance\/} 73, 915--957.

\bibitem[Duffie(2010)]{duffie2010presidential}
Duffie, Darrell, 2010, Presidential address: Asset price dynamics with slow-moving capital, {\em The Journal of finance\/} 65, 1237--1267.

\bibitem[Engle and Granger(1987)]{engle1987co}
Engle, Robert~F, and Clive~WJ Granger, 1987, Co-integration and error correction: representation, estimation, and testing, {\em Econometrica: journal of the Econometric Society\/}  251--276.

\bibitem[Ferko et~al.(2022)Ferko, Moin, Onur, and Penick]{ferko2022trades}
Ferko, Alex, Amani Moin, Esen Onur, and Michael Penick, 2022, Who trades bitcoin futures and why?, {\em Global Finance Journal\/}  100778.

\bibitem[Garleanu and Pedersen(2011)]{garleanu2011margin}
Garleanu, Nicolae, and Lasse~Heje Pedersen, 2011, Margin-based asset pricing and deviations from the law of one price, {\em The Review of Financial Studies\/} 24, 1980--2022.

\bibitem[Grinblatt and Jegadeesh(1996)]{grinblatt1996relative}
Grinblatt, Mark, and Narasimhan Jegadeesh, 1996, Relative pricing of eurodollar futures and forward contracts, {\em The journal of finance\/} 51, 1499--1522.

\bibitem[Gromb and Vayanos(2018)]{gromb2018dynamics}
Gromb, Denis, and Dimitri Vayanos, 2018, The dynamics of financially constrained arbitrage, {\em The Journal of Finance\/} 73, 1713--1750.

\bibitem[Han(2024)]{han2024primer}
Han, David, 2024, A primer on perpetual futures, {\em Coinbase Institutional Trading Insights\/} .

\bibitem[Hertzog et~al.(2017)Hertzog, Benartzi, Benartzi, and Ross]{hertzog2017bancor}
Hertzog, Eyal, Guy Benartzi, Galia Benartzi, and Omri Ross, 2017, Bancor protocol, continuous liquidity for cryptographic tokens through their smart contracts, Whitepaper.

\bibitem[Koijen et~al.(2018)Koijen, Moskowitz, Pedersen, and Vrugt]{koijen2018carry}
Koijen, Ralph~SJ, Tobias~J Moskowitz, Lasse~Heje Pedersen, and Evert~B Vrugt, 2018, Carry, {\em Journal of Financial Economics\/} 127, 197--225.

\bibitem[Kondor(2009)]{kondor_risk_2009}
Kondor, P{\'e}ter, 2009, Risk in {{Dynamic Arbitrage}}: {{The Price Effects}} of {{Convergence Trading}}, {\em Journal of Finance\/} 64, 631--655.

\bibitem[Kozak et~al.(2018)Kozak, Nagel, and Santosh]{kozak2018interpreting}
Kozak, Serhiy, Stefan Nagel, and Shrihari Santosh, 2018, Interpreting factor models, {\em The Journal of Finance\/} 73, 1183--1223.

\bibitem[Liu and Tsyvinski(2021)]{liu2021risks}
Liu, Yukun, and Aleh Tsyvinski, 2021, Risks and returns of cryptocurrency, {\em The Review of Financial Studies\/} 34, 2689--2727.

\bibitem[Liu et~al.(2022)Liu, Tsyvinski, and Wu]{liu2022common}
Liu, Yukun, Aleh Tsyvinski, and Xi~Wu, 2022, Common risk factors in cryptocurrency, {\em The Journal of Finance\/} 77, 1133--1177.

\bibitem[Lucca and Moench(2015)]{lucca2015pre}
Lucca, David~O, and Emanuel Moench, 2015, The pre-fomc announcement drift, {\em The Journal of finance\/} 70, 329--371.

\bibitem[Lustig et~al.(2011)Lustig, Roussanov, and Verdelhan]{lustig2011common}
Lustig, Hanno, Nikolai Roussanov, and Adrien Verdelhan, 2011, Common risk factors in currency markets, {\em The Review of Financial Studies\/} 24, 3731--3777.

\bibitem[Makarov and Schoar(2020)]{makarov2020trading}
Makarov, Igor, and Antoinette Schoar, 2020, Trading and arbitrage in cryptocurrency markets, {\em Journal of Financial Economics\/} 135, 293--319.

\bibitem[Newey and West(1987)]{neweywest1987}
Newey, Whitney~K., and Kenneth~D. West, 1987, A simple, positive semi-definite, heteroskedasticity and autocorrelation consistent covariance matrix, {\em Econometrica\/} 55, 703--708.

\bibitem[Pedersen(2019)]{pedersen2019efficiently}
Pedersen, Lasse~Heje, 2019, {\em Efficiently inefficient: how smart money invests and market prices are determined\/} (Princeton University Press).

\bibitem[Protter(2012)]{protter2012stochastic}
Protter, Philip~E, 2012, Stochastic differential equations, in {\em Stochastic integration and differential equations\/},  249--361 (Springer).

\bibitem[Roll(1984)]{roll1984simple}
Roll, Richard, 1984, A simple implicit measure of the effective bid-ask spread in an efficient market, {\em The Journal of Finance\/} 39, 1127--1139.

\bibitem[Schmeling et~al.(2022)Schmeling, Schrimpf, and Todorov]{schmeling2022crypto}
Schmeling, Maik, Andreas Schrimpf, and Karamfil Todorov, 2022, Crypto carry, Working Paper.

\bibitem[Shiller(1993)]{shiller1993measuring}
Shiller, Robert~J, 1993, Measuring asset values for cash settlement in derivative markets: hedonic repeated measures indices and perpetual futures, {\em The Journal of Finance\/} 48, 911--931.

\bibitem[Streltsov and Ruan(2022)]{streltsov2022perpetual}
Streltsov, Artem, and Qihong Ruan, 2022, Perpetual price discovery and crypto market quality, Working Paper.

\end{thebibliography}

\newpage

\appendix
\counterwithin{figure}{section}
\renewcommand{\thefigure}{\thesection.\arabic{figure}}
\counterwithin{table}{section}
\renewcommand{\thetable}{\thesection.\arabic{table}}
\counterwithin{assumption}{section}
\renewcommand{\theassumption}{\thesection.\arabic{assumption}}

\begin{center}{\bf{\LARGE Appendix}}\end{center}
\section{Additional Figures}\label{sec:figure}

\begin{figure}[!htb]
\begin{center}
\begin{tikzpicture}
	\node[anchor=south west,inner sep=0] at (0,0) {\includegraphics[width=\textwidth]{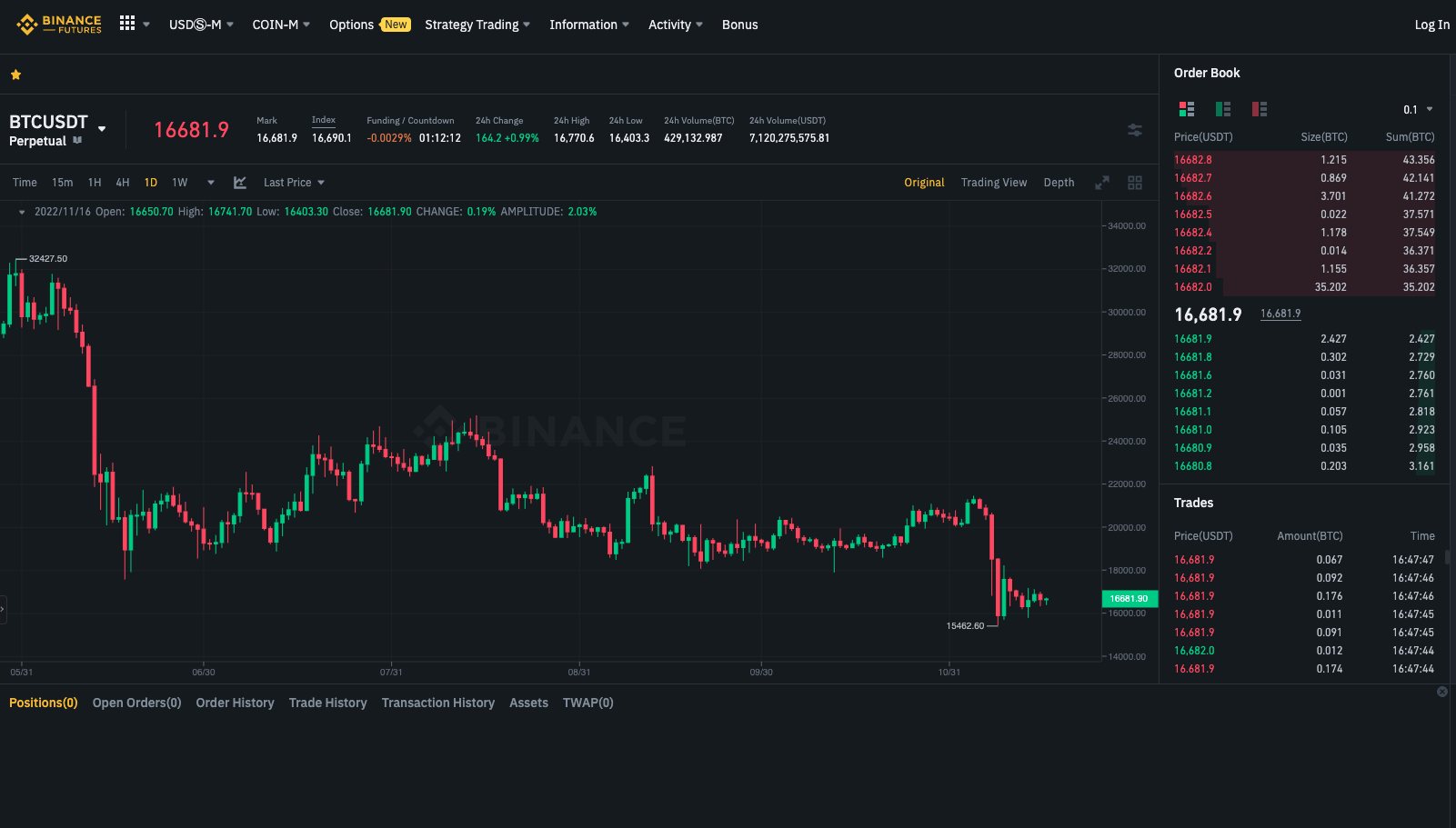}};
	\draw[red,ultra thick,rounded corners] (4.1,7.6) rectangle (5.3,8.2);
\end{tikzpicture}
\end{center}
\caption{Trading view of BTC perpetual futures on Binance}
\bigskip
\small
This figure presents the perpetual futures trading view on Binance. The key information includes the futures price (Mark), spot price (Index), real-time funding rate based on the rolling average of the past 8 hours' observations, and countdown toward funding rate payment (Funding / Countdown) (illustrated with a red rectangle). In this example, futures are trading at a lower price than the spot. The funding rate is negative and is to be paid in 1 hour and 12 minutes.
\end{figure}

\clearpage
\begin{figure}[!htb]
\begin{center}
\begin{tikzpicture}
    \node[anchor=south west,inner sep=0] at (0,0) {\includegraphics[width=\textwidth]{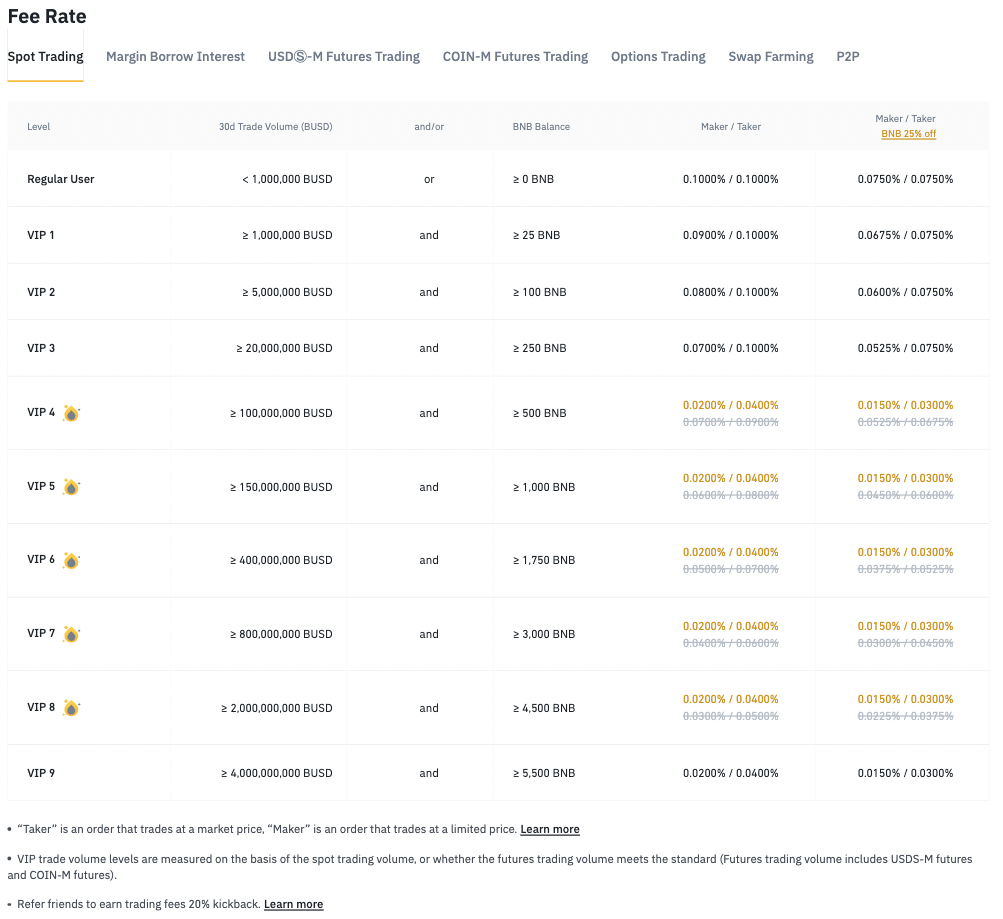}};
    \draw[red,ultra thick,rounded corners] (16,11.5) rectangle (13.8,11);
    \draw[red,ultra thick,rounded corners] (16,6.7) rectangle (13.8,7.5);
    \draw[red,ultra thick,rounded corners] (16,3.1) rectangle (13.8,3.9);
\end{tikzpicture}
\end{center}
\caption{Trading costs tiers for the spot market}\label{trading_cost_spot}
\bigskip
\small
This figure presents the trading cost tiers for the crypto spot market from Binance: \url{https://www.binance.com/en/fee/trading}. Our high, medium, and low trading cost specifications correspond to VIP 1, VIP 5, and VIP 8 tiers, as illustrated with red rectangles in the picture. The 30-day trading volume requirements for VIP 1, 5, and 8 are 1 million, 150 million, and 2 billion, respectively, in the spot market. We consider the trading costs for makers as institutions typically trade maker orders. Binance offers a temporary discount for VIP 4-8 to have the same trading cost as VIP 9. We consider the non-discounted trading costs to make the comparison more reliable and fair.
\end{figure}

\begin{figure}[!htb]
\begin{center}
\begin{tikzpicture}
    \node[anchor=south west,inner sep=0] at (0,0) {\includegraphics[width=\textwidth]{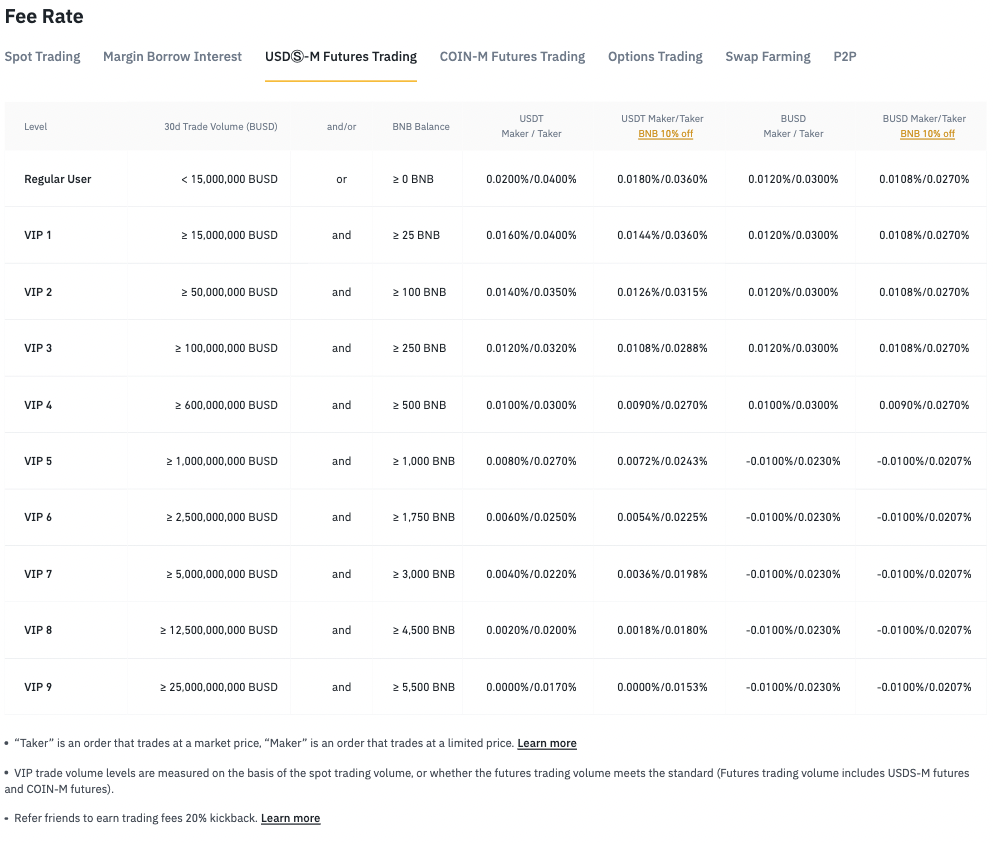}};
    \draw[red,ultra thick,rounded corners] (10,9.8) rectangle (12,10.5);
    \draw[red,ultra thick,rounded corners] (10,6) rectangle (12,6.7);
    \draw[red,ultra thick,rounded corners] (10,3.2) rectangle (12,3.9);
\end{tikzpicture}
\end{center}
\caption{Trading costs tiers for the perp market}\label{trading_cost_perp}
\bigskip
\small
This figure presents the trading cost tiers for the crypto perpetual market from Binance: \url{https://www.binance.com/en/fee/trading}. Our high, medium, and low trading cost specifications correspond to VIP 1, VIP 5, and VIP 8 tiers, as illustrated with red rectangles in the picture. The 30-day trading volume requirements for VIP 1, 5, and 8 are 15 million, 1 billion, and 12.5 billion, respectively, in the perpetual market. We consider the trading costs for makers as institutions typically trade maker orders.
\end{figure}

\clearpage
\begin{figure}[!htb]
    \begin{center}
    \includegraphics[width = \textwidth]{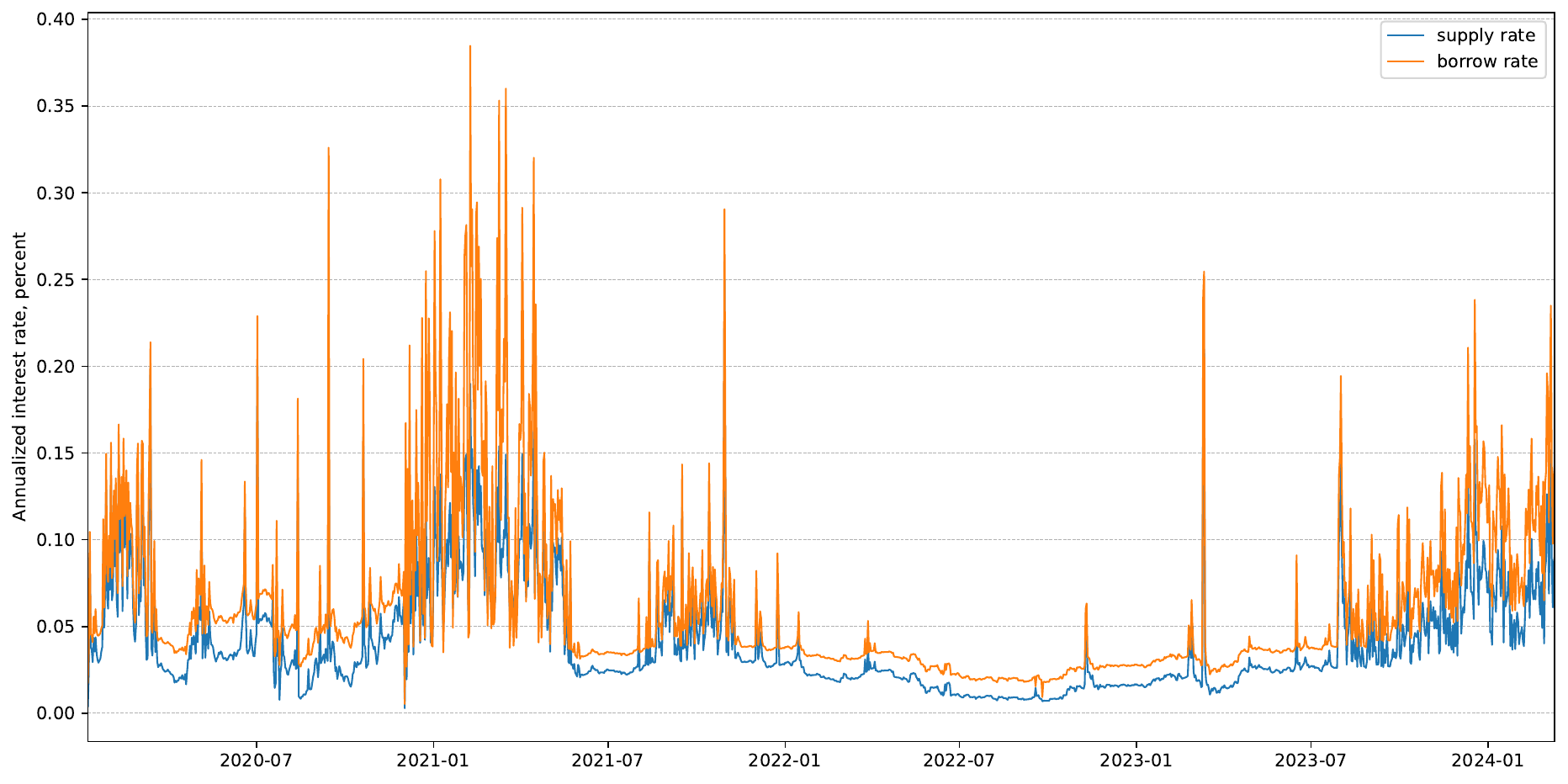}
    \end{center}
    \caption{Annualized interest rate from Aave}
    \bigskip
    \small
    This figure presents the daily supply and borrowing rate from Aave. We consider three stablecoins: USDT, USDC, and DAI. Each day, we take the average interest rate of the three crypto stablecoins to get a proxy for the risk-free funding rate of the arbitrageurs. The averaging removes the idiosyncratic noise in borrowing and supply in individual stablecoins. The sample period is from 2020-01-08 to 2024-03-11.
    \label{fig:interest_rate}
\end{figure}

\clearpage
\begin{figure}[!htb]
\begin{center}
\includegraphics[width = 0.95\textwidth]{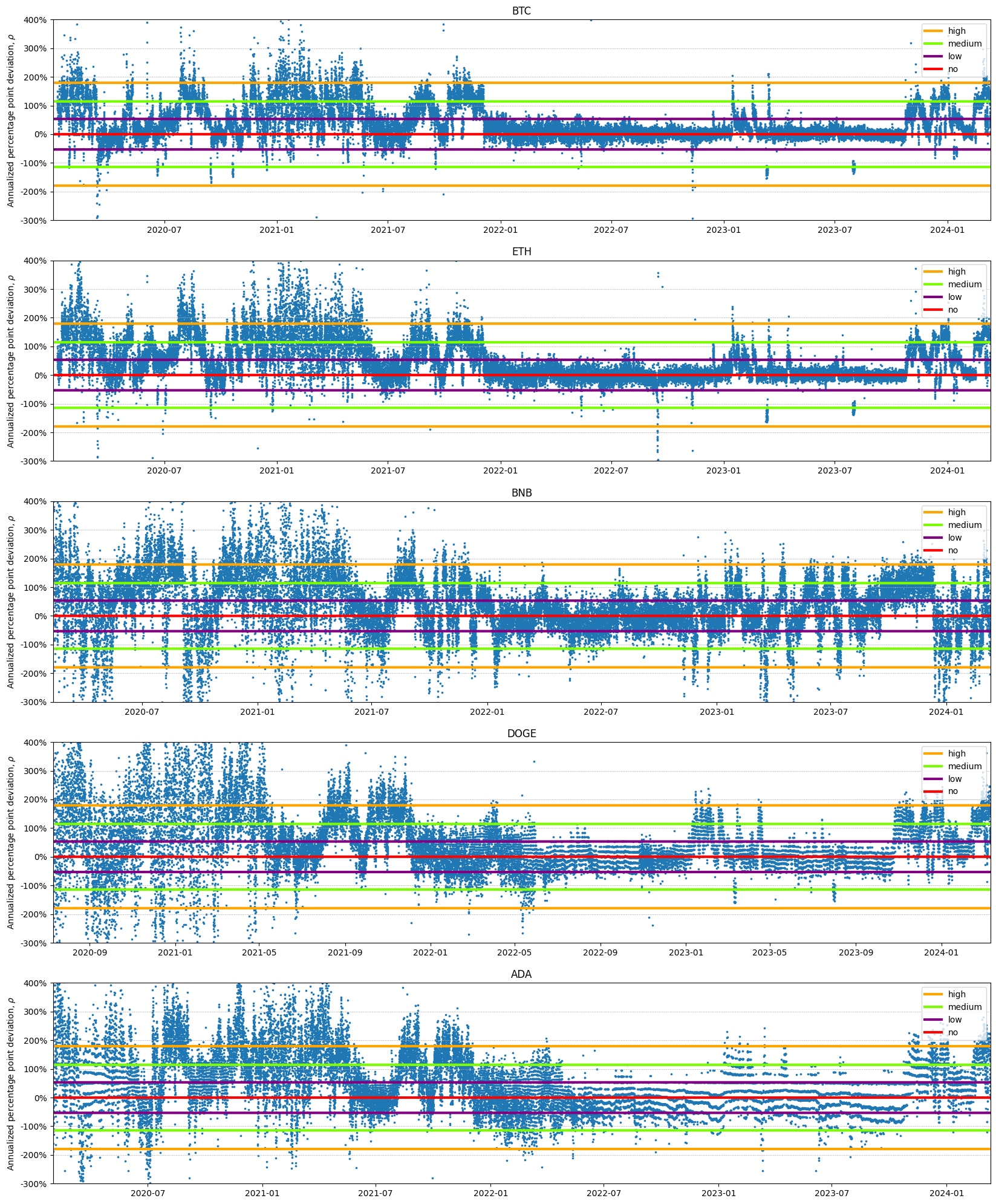}
\end{center}
\caption{Trading Strategy Visualization: Random Maturity Arbitrage}\label{trading_theory}
\bigskip
\small
This figure presents the random-maturity arbitrage strategy motivated by the theory. The orange, green, and purple lines correspond to the open-position threshold under high, medium, and low trading costs. The red line represents the close-position threshold.
\end{figure}

\newpage
\section{Funding Rate Calculation on Binance}\label{sec:fund_binance}

In this section, we describe the funding rate mechanism used by Binance in greater detail and discuss the economic role of its clamped funding rule. Our description follows the methodology provided by Binance.\footnote{\url{https://www.binance.com/en/support/faq/360033525031}}

There are five general steps in obtaining the funding rate. The first step is to find the Impact Bid/Ask Price Series in a given funding period by analyzing the order book and determining the average fill price to execute a certain notional value (Impact Margin Notional) on the bid and ask sides.

The impact margin notional is defined as the notional available to trade with a margin of 200 USD. So it can be calculated using the following equation:
$$
\text{IMN} = 200 / \text{Initial margin rate at maximum leverage level}.
$$

Given the definition of impact margin notional, the impact bid/ask price is defined as the average trade price after a market sell/buy order with the total quoted notional quantity given by the IMN.

Given impact bid and ask prices (denoted as IBP and IAP, respectively), the second step is to calculate the premium index (denoted as P) series every $5$ seconds using the following equation:
$$
    P = [\text{Max}(0, \text{IBP} - \text{Price Index}) - \text{Max}(0, \text{Price Index} - \text{IAP})] / \text{Price Index},
$$
where `Price Index' is the weighted average spot price of the underlying asset listed on major spot exchanges.

The rationale behind this formula is to capture the relative difference between the perpetual contract's price and the underlying asset's price, taking into account the actual market liquidity and depth represented by the IBP and IAP.

The third step is to calculate the time-weighted average premium index over the funding period (8 hours for Binance) using the premium index series:
$$
    \bar{P} = \sum_{t = 1}^T w_t P_t / \sum_{t = 1}^T w_t,
$$
where $P_t$ denotes the $t$-th 5-second in the 8-hour funding period, $w_t = t$ is the weight for each observation, and $T$ is the total number of 5-second intervals within the 8-hour period. For Binance,  $ T = 12\times 6 \times 8 = 5760$. With this weighting scheme, the observations closer to the funding rate payment receive larger weights. The weights increase linearly in time over the 8-hour interval.

Given the average premium index, the fourth step is to calculate the funding rate using the equation below:
$$
    fr = \bar{P} + \text{clamp}(\iota - \bar{P}, -0.05\%, 0.05\%),
$$
where fr is the funding rate, $\iota$ is the interest component set by Binance, and 
$$
    \text{clamp}(x, \text{min}, \text{max}) = \begin{cases}
        min, & \text{if } x < \text{min} \\
        x, & \text{if } \text{min} \leq x \leq \text{max} \\
        max, & \text{if } x > \text{max}
    \end{cases}
$$
Empirically, for Binance futures, the interest rate $r$ is fixed at 0.01\% per 8 hours for most perpetual futures. Figure \ref{figfr} plots the funding rate as a function of the premium index. As can be seen, when the premium index is close to the clamp region, the funding rate is fixed at the interest component.

This feature has several economic implications. First, it makes funding payments more stable and predictable when the perpetual futures price is close to the underlying spot price. Small fluctuations in the premium do not mechanically translate into changes in the funding payment. In this sense, the clamp introduces a ``no-adjustment region'' around the normal funding rate.

Second, the clamp preserves the convergence mechanism when deviations become economically meaningful. It is important that the clamp is not an outer cap on the funding rate. Once the premium index moves outside the no-adjustment region, the funding rate again increases one-for-one with the premium index. Large positive premiums therefore generate larger payments from long to short positions, while sufficiently large discounts generate payments in the opposite direction. The funding mechanism consequently becomes stronger precisely when the futures price moves sufficiently far from the spot benchmark.

Third, the design reduces the sensitivity of funding payments to small or transitory distortions in the measured premium. Under a purely linear rule, any change in the premium index immediately changes the funding transfer between long and short positions. Under the clamped rule, by contrast, the local response of the funding rate to the premium index is zero inside the clamp region. A trader attempting to alter the funding payment must therefore move the average premium sufficiently far to cross the boundary of the no-adjustment region before the funding payment begins to change.

This consideration is reinforced by the way the premium index itself is constructed. The index is based on impact prices rather than the best quoted prices, so changing it requires moving prices over some depth of the order book. It is also averaged over the funding period, so a sufficiently large or persistent distortion is required to have a material effect on the settlement premium. Together with the clamp, these features make funding payments less sensitive to small or short-lived price distortions. We view this as an additional economic rationale for the clamped design, although we do not attempt to quantify its importance for manipulation empirically.

Table \ref{tab:clamp} compares the realized Binance funding rate $\Phi(P)$ with the counterfactual linear funding rate $\kappa P$, where $P = (F - S)/F$ is the premium index, evaluated on the same hours. It reports how often the clamp binds, and the mean and standard deviation of the funding payment under each rule.

\begin{table}[!htb]
\begin{center}
\footnotesize
\begin{tabularx}{1\columnwidth}{@{\hskip\tabcolsep\extracolsep\fill}l*{6}{r}}
\toprule
 & \multicolumn{2}{c}{Clamp binds} & \multicolumn{2}{c}{Std (ann. \%)} & \multicolumn{2}{c}{Mean (ann. \%)} \\
\cmidrule(lr){2-3} \cmidrule(lr){4-5} \cmidrule(lr){6-7}
Asset & Hours (\%) & Settlements at $\iota$ (\%) & $\Phi$ & $\kappa P$ & $\Phi$ & $\kappa P$ \\
\midrule
BTC  & 45.2 & 41.7 & 27.8 & 87.1 & 15.3 & $-3.9$ \\
ETH  & 43.4 & 41.9 & 36.5 & 83.7 & 18.6 & 5.4 \\
BNB  & 38.0 & 43.8 & 55.8 & 104.8 & $-0.1$ & $-7.7$ \\
DOGE & 35.7 & 50.4 & 47.5 & 257.5 & 16.7 & 4.8 \\
ADA  & 36.5 & 49.3 & 43.8 & 107.6 & 18.4 & 3.3 \\
\bottomrule
\end{tabularx}
\end{center}
\caption{The clamp versus a linear funding rate}\label{tab:clamp}
\bigskip
\small
This table compares the realized Binance funding rate $\Phi(P)$ with the counterfactual linear rule $\kappa P$, where $P = (F-S)/F$ is the premium index. ``Hours (\%)'' is the share of hours in which the premium index lies inside the clamp region, $|\kappa P - \iota| \leq \gamma$, where the funding payment is fixed at $\iota$ and does not respond to the basis. ``Settlements at $\iota$ (\%)'' is the share of realized eight-hour funding settlements paid at exactly the interest component. BNBUSDT is the one contract in our sample with $\iota = 0$. Standard deviations and means are computed over settlement hours and annualized by a factor of $3 \times 365$. Sample periods are those of Table \ref{tab2}.
\end{table}

The table shows that the no-adjustment region is economically relevant rather than a minor institutional detail. Depending on the asset, the clamp region accounts for approximately 36--45\% of observations, and roughly 42--50\% of realized funding settlements occur exactly at the interest component. Moreover, funding payments under the clamped rule are substantially less volatile than under the counterfactual linear rule. These results are consistent with the interpretation that the clamp stabilizes funding payments during normal market conditions while preserving stronger funding incentives when the futures-spot deviation becomes sufficiently large.

Finally, Binance imposes an additional outer cap and floor on the funding rate based on the maximum leverage of the contract. For contracts with maximum leverage of 30x or above, the cap and floor depend on the Maintenance Margin Ratio. For contracts with maximum leverage of 25x or below, they are fixed at $\pm3\%$. This final restriction serves a different purpose from the clamp. Whereas the clamp makes the funding rate locally insensitive to small price deviations, the outer cap limits the size of funding transfers during extreme market conditions. Extremely large funding payments can generate substantial losses for highly leveraged positions and potentially trigger liquidations. The outer cap therefore limits this tail risk without eliminating the funding mechanism's response to economically meaningful price deviations.

\newpage
\section{Alternative Arbitrage Strategies}\label{sec:altstrat}
\subsection{Data-driven Arbitrage Strategy}\label{sec:data_driven}
In the main text of our paper, we demonstrate the profitability of a simple theory-motivated trading strategy. The trading threshold can also be potentially further improved using a data-driven approach. In this part, we implement a data-driven two-threshold arbitrage trading strategy in the perpetual futures market to verify that our proposed theory-motivated strategy is optimal. 

\begin{figure}[!htb]
\begin{center}
\includegraphics[width=\textwidth]{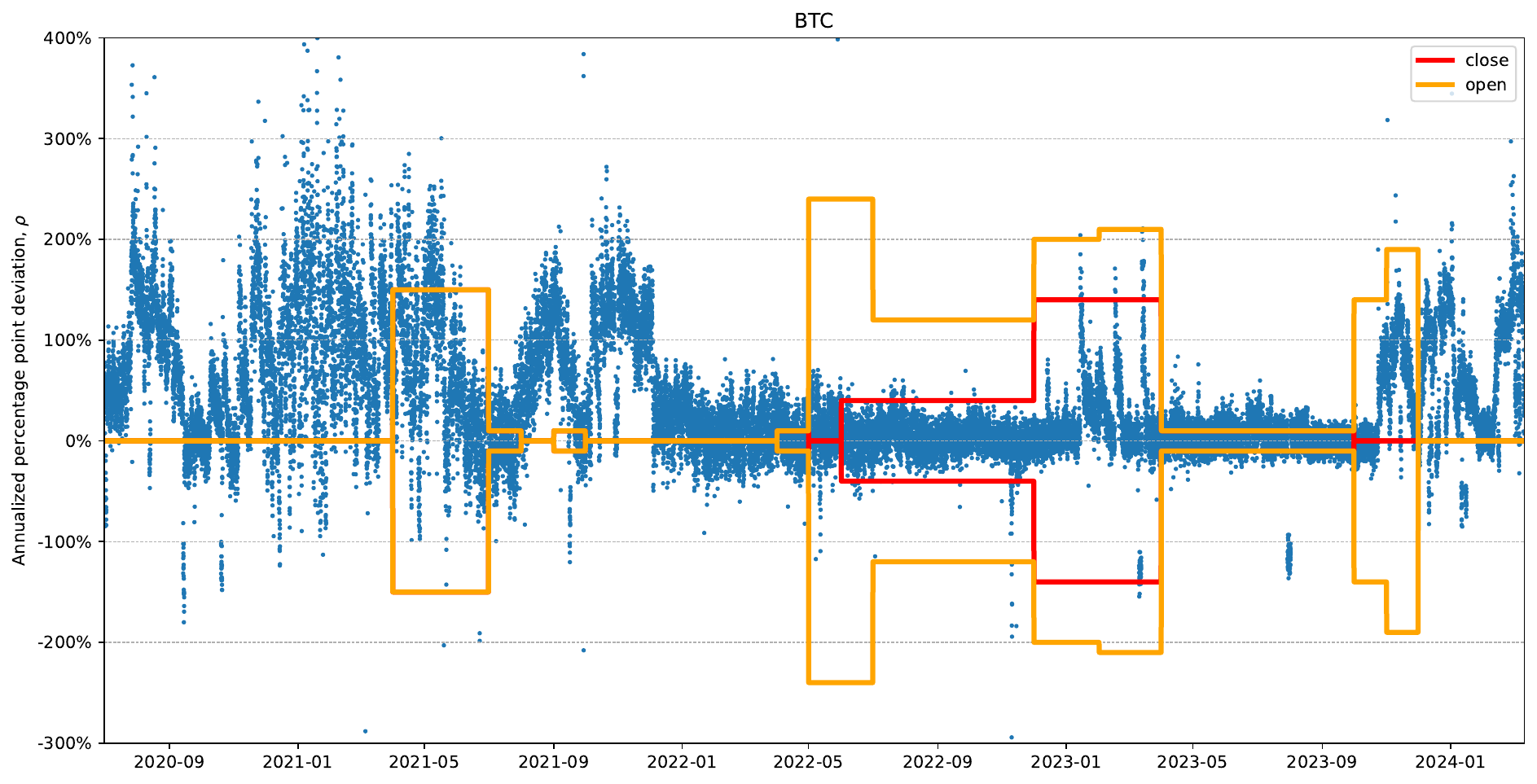}
\end{center}
\caption{Data-driven Random Maturity Arbitrage Strategy for Bitcoin}\label{fig_btc_data_driven}
\bigskip
\small
Deviation from the no-arbitrage price ($\rho$) and trading rules of the data-driven random-maturity arbitrage strategy we implement for BTC. Each blue dot in the figure represents the deviation of futures from the no-arbitrage price calculated using Equation \ref{eqdef} at hourly frequency. The number is annualized for ease of interpretation. The orange line is the open-position threshold, and the red line is the close-position threshold.
\end{figure}

The data-driven strategy can be characterized by a tuple of two thresholds: $(u,l)$, where $u$ denotes the upper bar and $l$ denotes the lower bar, $u > l$. When $\rho > u$, we long the spot and short the futures. When $-l < \rho < l$, we close the position. When $\rho < - u$, we long the futures and short the spot. Figure \ref{fig_btc_data_driven} presents an illustration of the strategy for Bitcoin with the trading thresholds estimated using real-time data.

To determine the optimal $(u, l)$, at the beginning of each month, we calculate the returns of the two-threshold trading strategy based on the past six months' data on a grid of parameters. From Table \ref{tab3}, we find that the theory-implied bounds for price deviation across all trading-cost specifications lie within $-200\%$ to $+200\%$. So we choose the grid as increasing from $0\%$ to $250\%$ with an incremental step of $10\%$ for $u$ and $l$ ($u \geq l$). In total, there are $351$ model specifications. We choose the model that delivers the highest Sharpe Ratio in the validation sample of the past six months.

We present the visualization of all trading strategies under zero to high trading costs in Appendix Figure \ref{trading_no} to Figure \ref{trading_high}. Under no trading costs, the strategy chooses a low open-position threshold, and the open- and close-position thresholds coincide with each other. Since there is no trading cost, the strategy no longer needs to wait for price convergence to avoid high turnover. On the contrary, with high trading costs, the strategy chooses a larger open-position threshold and a smaller close-position threshold. This reflects the automatic adjustment by the algorithm for trading costs and turnover.

Table \ref{port_mh_data_unrestricted} presents the performance statistics for this strategy under no trading costs over time. Table \ref{port_fee_data_driven} further presents the overall performance of the data-driven strategy across different trading costs. We find that the arbitrage trading strategy has a high Sharpe ratio, similar to and comparable to our theory-motivated ones. This suggests our theory-motivated strategy is close to optimal.

Similar to the theory-motivated case, we decompose the performance of the trading strategy into: (1) the price convergence and (2) the funding rate payment. The results are presented in Table \ref{port_decomp_data_driven}. Similar to the theory-motivated strategy, we find that the return from price convergence accounts for a larger proportion of the total profit.

Overall, our analysis of the data-driven trading strategy validates the optimality of our theory-driven approach, underscoring the value of our theoretical benchmark. Notably, introducing additional flexible adjustments does not significantly improve the performance of the arbitrage strategy.

\clearpage
\newpage
\begin{table}[!htb]
\begin{center}
\footnotesize
\begin{tabularx}{\textwidth}{@{\hskip\tabcolsep\extracolsep\fill}ll*{6}{r}}
\toprule
 &  & 2020 & 2021 & 2022 & 2023 & 2024 & All \\
\midrule
BTC & SR & 5.14 & 10.86 & 20.28 & 12.80 & 11.84 & 8.43 \\
 & Return & 35.93 & 53.76 & 21.00 & 21.93 & 21.44 & 32.18 \\
 & Volatility & 6.99 & 4.95 & 1.04 & 1.71 & 1.81 & 3.82 \\
 & MaxDD & -3.82 & -2.00 & -0.12 & -0.12 & -0.17 & -3.82 \\
 & Active \% & 100.00 & 76.42 & 30.00 & 32.71 & 100.00 & 56.47 \\
 & OtC time & 10.74 & 8.66 & 3.26 & 2.46 & 11.44 & 5.54 \\
 & N & 4,416 & 8,760 & 8,760 & 8,760 & 1,682 & 32,378 \\
ETH & SR & 8.60 & 12.01 & 20.01 & 11.90 & 13.90 & 11.78 \\
 & Return & 48.21 & 61.78 & 43.04 & 41.25 & 25.11 & 47.40 \\
 & Volatility & 5.61 & 5.14 & 2.15 & 3.47 & 1.81 & 4.02 \\
 & MaxDD & -1.09 & -1.43 & -0.67 & -0.29 & -0.16 & -1.43 \\
 & Active \% & 63.65 & 48.58 & 50.09 & 86.77 & 100.00 & 64.05 \\
 & OtC time & 11.52 & 5.89 & 2.55 & 3.59 & 8.54 & 4.15 \\
 & N & 4,416 & 8,760 & 8,760 & 8,760 & 1,682 & 32,378 \\
BNB & SR & 18.67 & 15.51 & 34.53 & 21.26 & 24.05 & 19.05 \\
 & Return & 140.41 & 124.74 & 108.01 & 93.76 & 53.50 & 109.56 \\
 & Volatility & 7.52 & 8.04 & 3.13 & 4.41 & 2.22 & 5.75 \\
 & MaxDD & -0.61 & -1.41 & -0.83 & -0.61 & -0.16 & -1.41 \\
 & Active \% & 45.26 & 56.69 & 57.45 & 82.92 & 19.56 & 60.86 \\
 & OtC time & 3.87 & 5.68 & 2.44 & 4.72 & 2.25 & 3.81 \\
 & N & 3,672 & 8,760 & 8,760 & 8,760 & 1,682 & 31,634 \\
DOGE & SR & - & 13.64 & 44.30 & 20.96 & 6.92 & 14.51 \\
 & Return & - & 208.60 & 159.41 & 41.62 & 13.23 & 129.13 \\
 & Volatility & - & 15.30 & 3.60 & 1.99 & 1.91 & 8.90 \\
 & MaxDD & - & -3.82 & -0.29 & -0.21 & -0.29 & -3.82 \\
 & Active \% & - & 80.17 & 55.59 & 40.55 & 16.41 & 56.22 \\
 & OtC time & - & 6.00 & 1.92 & 2.82 & 18.40 & 3.15 \\
 & N & - & 8,760 & 8,760 & 8,760 & 1,682 & 27,962 \\
ADA & SR & 13.53 & 12.34 & 35.09 & 26.92 & 13.03 & 16.15 \\
 & Return & 87.73 & 118.98 & 122.89 & 68.76 & 42.29 & 98.20 \\
 & Volatility & 6.48 & 9.64 & 3.50 & 2.55 & 3.25 & 6.08 \\
 & MaxDD & -0.89 & -5.69 & -0.22 & -0.50 & -0.18 & -5.69 \\
 & Active \% & 48.62 & 60.99 & 33.00 & 33.41 & 100.00 & 46.30 \\
 & OtC time & 5.73 & 5.24 & 2.12 & 2.45 & 13.90 & 3.68 \\
 & N & 4,416 & 8,760 & 8,760 & 8,760 & 1,682 & 32,378 \\
\bottomrule
\end{tabularx}
\end{center}
\caption{Data-driven Random Maturity Arbitrage Strategy over Time}\label{port_mh_data_unrestricted}
\bigskip
\small
This table presents the Sharpe ratios, annualized returns (\%), standard deviations (\%), maximum drawdowns (\%), active percentages (\%), and the average open-to-close time (hours) of the data-driven random maturity arbitrage strategies for five different cryptocurrencies. The returns reported are excess returns from the random maturity arbitrage strategies.
\end{table}

\clearpage
\newpage
\begin{table}
\begin{center}
\footnotesize
\begin{tabularx}{\textwidth}{@{\hskip\tabcolsep\extracolsep\fill}ll*{4}{r}}
\toprule
  &  & \multicolumn{4}{c}{Fee tiers} \\
\cmidrule{3-6}
    &          &      No &     Low & Medium &   High \\
\midrule
BTC & SR & 8.43 & 3.71 & 3.49 & 3.18 \\
 & Return & 32.18 & 13.18 & 12.01 & 10.88 \\
 & Volatility & 3.82 & 3.55 & 3.44 & 3.42 \\
 & MaxDD & -3.82 & -3.82 & -2.00 & -2.00 \\
 & $\alpha$ & 32.69 & 12.50 & 11.64 & 10.22 \\
 & $t_{\alpha}$ & 10.02 & 5.31 & 5.13 & 4.88 \\
 & Active \% & 56.47 & 26.38 & 24.79 & 23.55 \\
 & OtC time & 5.54 & 43.36 & 89.17 & 95.33 \\
ETH & SR & 11.78 & 5.88 & 4.50 & 4.22 \\
 & Return & 47.40 & 21.86 & 17.17 & 16.09 \\
 & Volatility & 4.02 & 3.72 & 3.81 & 3.81 \\
 & MaxDD & -1.43 & -1.47 & -2.62 & -2.66 \\
 & $\alpha$ & 49.84 & 20.56 & 16.39 & 14.91 \\
 & $t_{\alpha}$ & 10.79 & 5.72 & 5.28 & 5.07 \\
 & Active \% & 64.05 & 22.60 & 23.95 & 24.24 \\
 & OtC time & 4.15 & 16.01 & 69.85 & 89.17 \\
BNB & SR & 19.05 & 12.75 & 8.81 & 7.19 \\
 & Return & 109.56 & 66.86 & 44.35 & 36.76 \\
 & Volatility & 5.75 & 5.24 & 5.03 & 5.11 \\
 & MaxDD & -1.41 & -1.07 & -1.12 & -1.42 \\
 & $\alpha$ & 113.78 & 53.44 & 35.05 & 26.51 \\
 & $t_{\alpha}$ & 12.61 & 6.91 & 5.30 & 4.46 \\
 & Active \% & 60.86 & 41.44 & 24.69 & 21.51 \\
 & OtC time & 3.81 & 6.70 & 13.10 & 18.10 \\
DOGE & SR & 14.51 & 9.19 & 6.32 & 4.28 \\
 & Return & 129.13 & 80.53 & 52.43 & 37.25 \\
 & Volatility & 8.90 & 8.76 & 8.30 & 8.71 \\
 & MaxDD & -3.82 & -3.87 & -3.92 & -5.61 \\
 & $\alpha$ & 155.82 & 81.40 & 43.65 & 30.03 \\
 & $t_{\alpha}$ & 7.26 & 4.56 & 3.14 & 2.76 \\
 & Active \% & 56.22 & 47.00 & 26.97 & 22.88 \\
 & OtC time & 3.15 & 6.20 & 10.22 & 22.85 \\
ADA & SR & 16.15 & 12.08 & 6.44 & 3.86 \\
 & Return & 98.20 & 72.14 & 35.83 & 21.76 \\
 & Volatility & 6.08 & 5.97 & 5.56 & 5.64 \\
 & MaxDD & -5.69 & -5.67 & -5.69 & -5.58 \\
 & $\alpha$ & 109.86 & 65.07 & 27.92 & 18.26 \\
 & $t_{\alpha}$ & 9.55 & 9.78 & 7.09 & 5.22 \\
 & Active \% & 46.30 & 50.95 & 29.20 & 21.19 \\
 & OtC time & 3.68 & 5.40 & 12.81 & 40.13 \\
\bottomrule
\end{tabularx}
\end{center}
\caption{Data-driven Random Maturity Arbitrage across Different Trading Costs}
\label{port_fee_data_driven}
\bigskip
\small
Performance of data-driven random maturity arbitrage strategy under different trading cost tiers. The fees for spot (futures) are 2.25 (0.18) bps, 4.5 (0.72) bps, and 6.75 (1.44) bps for the low, medium, and high trading cost levels. Statistics reported are the annualized Sharpe ratio, return (\%), volatility (\%), max drawdown (\%), alpha (\%) relative to the three factors of \citet{liu2022common}, t-stat of the alpha, proportion of time the strategy is active (\%), and average open-to-close (OtC) position duration in hours. 
\end{table}

\clearpage
\newpage

\begin{table}[!htb]
\begin{center}
\begin{tabularx}{\textwidth}{@{\hskip\tabcolsep\extracolsep\fill}ll*{6}{r}}
\toprule
 &  & 2020 & 2021 & 2022 & 2023 & 2024 & All \\
\midrule
BTC & Return & 35.91 & 53.76 & 21.00 & 21.93 & 21.44 & 32.17 \\
 & Price & 17.53 & 28.50 & 19.25 & 19.28 & 3.30 & 20.70 \\
 & Funding & 18.38 & 25.27 & 1.75 & 2.65 & 18.15 & 11.48 \\
ETH & Return & 48.19 & 61.78 & 43.04 & 41.25 & 25.11 & 47.40 \\
 & Price & 25.20 & 42.92 & 39.62 & 36.37 & 7.54 & 36.00 \\
 & Funding & 22.99 & 18.87 & 3.43 & 4.88 & 17.57 & 11.40 \\
BNB & Return & 140.45 & 124.74 & 108.01 & 93.76 & 53.50 & 109.56 \\
 & Price & 114.53 & 100.82 & 99.69 & 81.75 & 45.62 & 93.88 \\
 & Funding & 25.92 & 23.92 & 8.32 & 12.01 & 7.88 & 15.68 \\
DOGE & Return & - & 208.40 & 159.41 & 41.62 & 13.23 & 129.06 \\
 & Price & - & 167.82 & 157.15 & 40.50 & 3.13 & 114.68 \\
 & Funding & - & 40.59 & 2.26 & 1.12 & 10.10 & 14.38 \\
ADA & Return & 87.49 & 118.98 & 122.89 & 68.76 & 42.29 & 98.17 \\
 & Price & 59.58 & 89.46 & 121.51 & 67.21 & 19.16 & 84.39 \\
 & Funding & 27.91 & 29.52 & 1.37 & 1.54 & 23.13 & 13.79 \\
\bottomrule
\end{tabularx}
\end{center}
\caption{Data-driven Strategy: Price Convergence vs Funding Rate Payment}\label{port_decomp_data_driven}
\bigskip
\small
This table decomposes the portfolio return into the part due to price convergence and the part due to funding rate payment.
\end{table}

\begin{figure}[!htb]
\begin{center}
\includegraphics[width=0.95\textwidth]{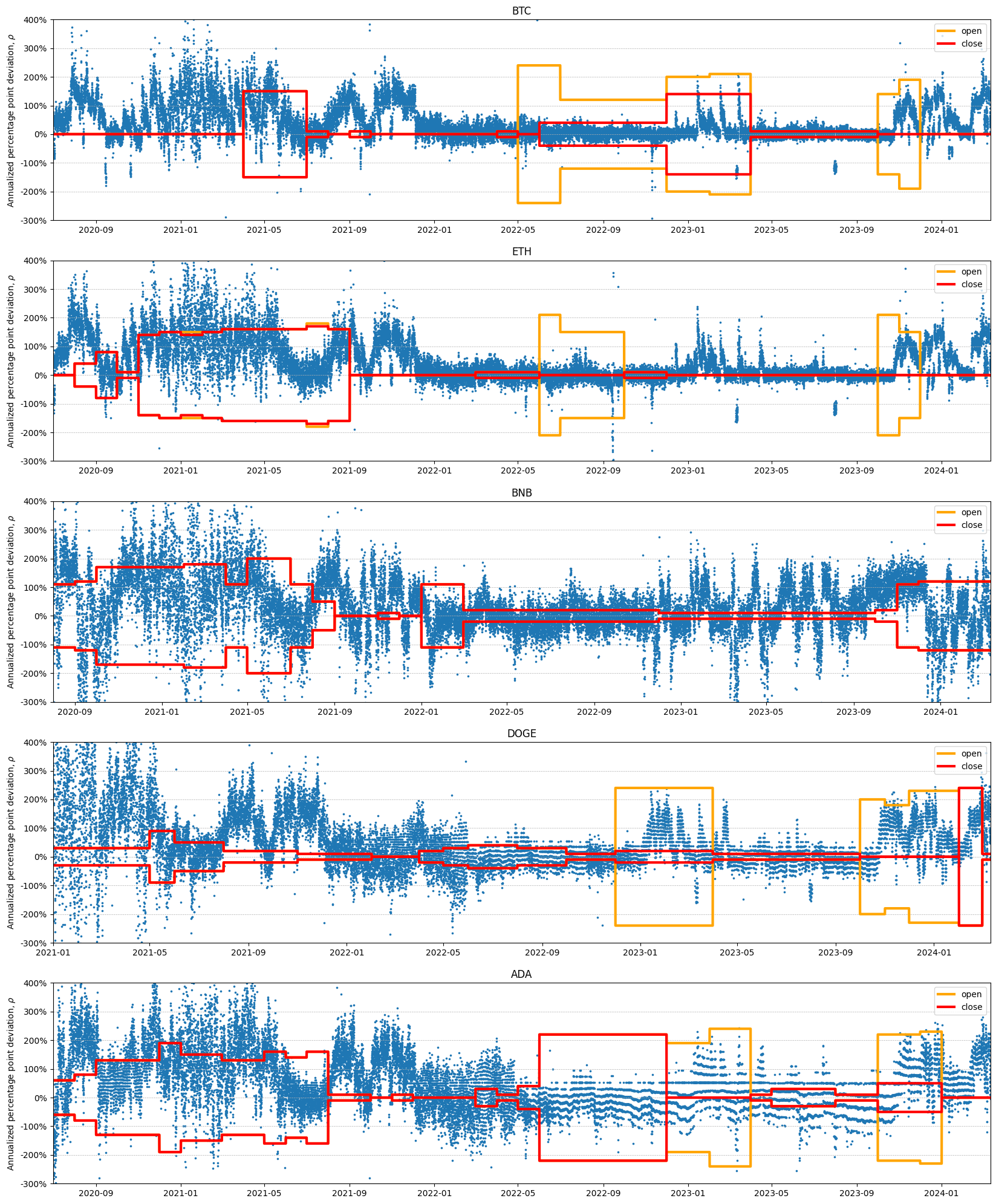}
\end{center}
\caption{Data-driven Trading Strategy Visualization: No Trading Costs} \label{trading_no}
\bigskip
\small
This figure presents the real-time trading thresholds under no trading costs. The orange line is the open-position threshold, and the red line is the close-position threshold. The trading thresholds are determined based on the adjusted SR from the past six months.
\end{figure}

\begin{figure}[!htb]
\begin{center}
\includegraphics[width=0.95\textwidth]{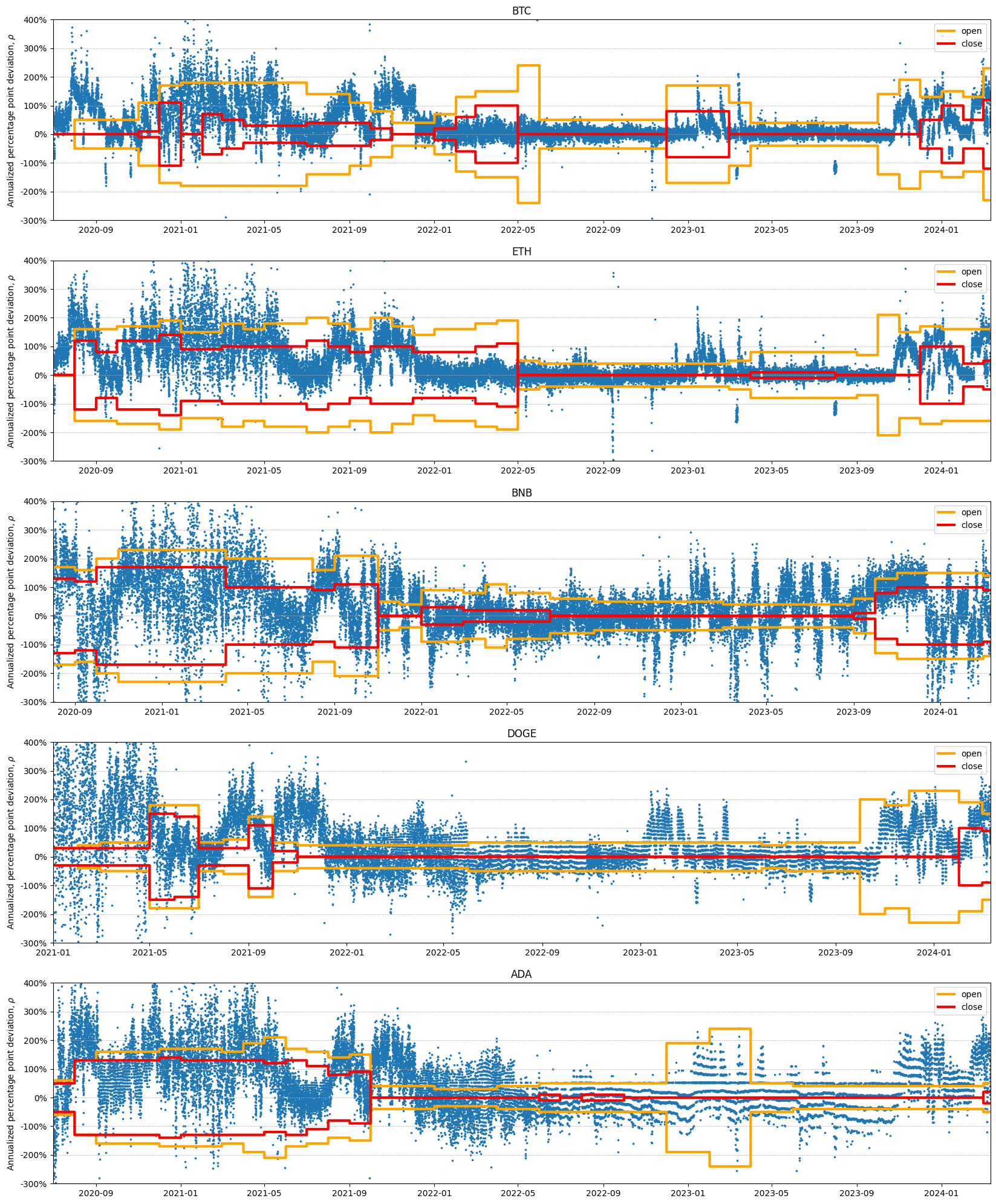}
\end{center}
\caption{Data-driven Trading Strategy Visualization: Low Trading Costs} \label{trading_low}
\bigskip
\small
This figure presents the real-time trading thresholds under low trading costs (2.25 bps for spot trading and 0.18 bp for futures trading). The orange line is the open-position threshold, and the red line is the close-position threshold. The trading thresholds are determined based on the adjusted SR from the past six months.
\end{figure}

\begin{figure}[!htb]
\begin{center}
\includegraphics[width=0.95\textwidth]{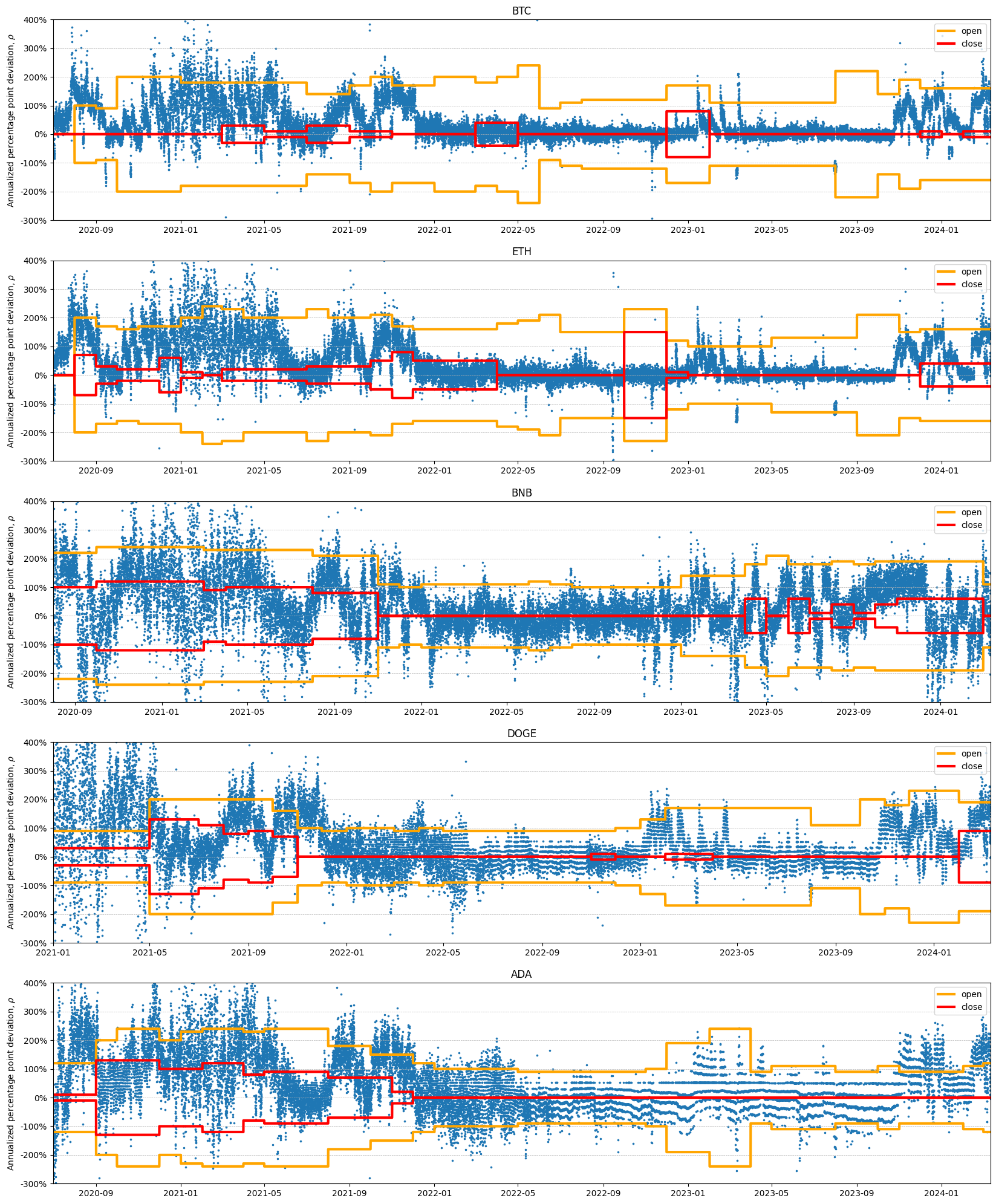}
\end{center}
\caption{Data-driven Trading Strategy Visualization: Medium Trading Costs} \label{trading_medium}
\bigskip
\small
This figure presents the real-time trading thresholds under medium trading costs (4.5 bps for spot trading and 0.72 bp for futures trading). The orange line is the open-position threshold, and the red line is the close-position threshold. The trading thresholds are determined based on the adjusted SR from the past six months.
\end{figure}

\begin{figure}[!htb]
\begin{center}
\includegraphics[width=0.95\textwidth]{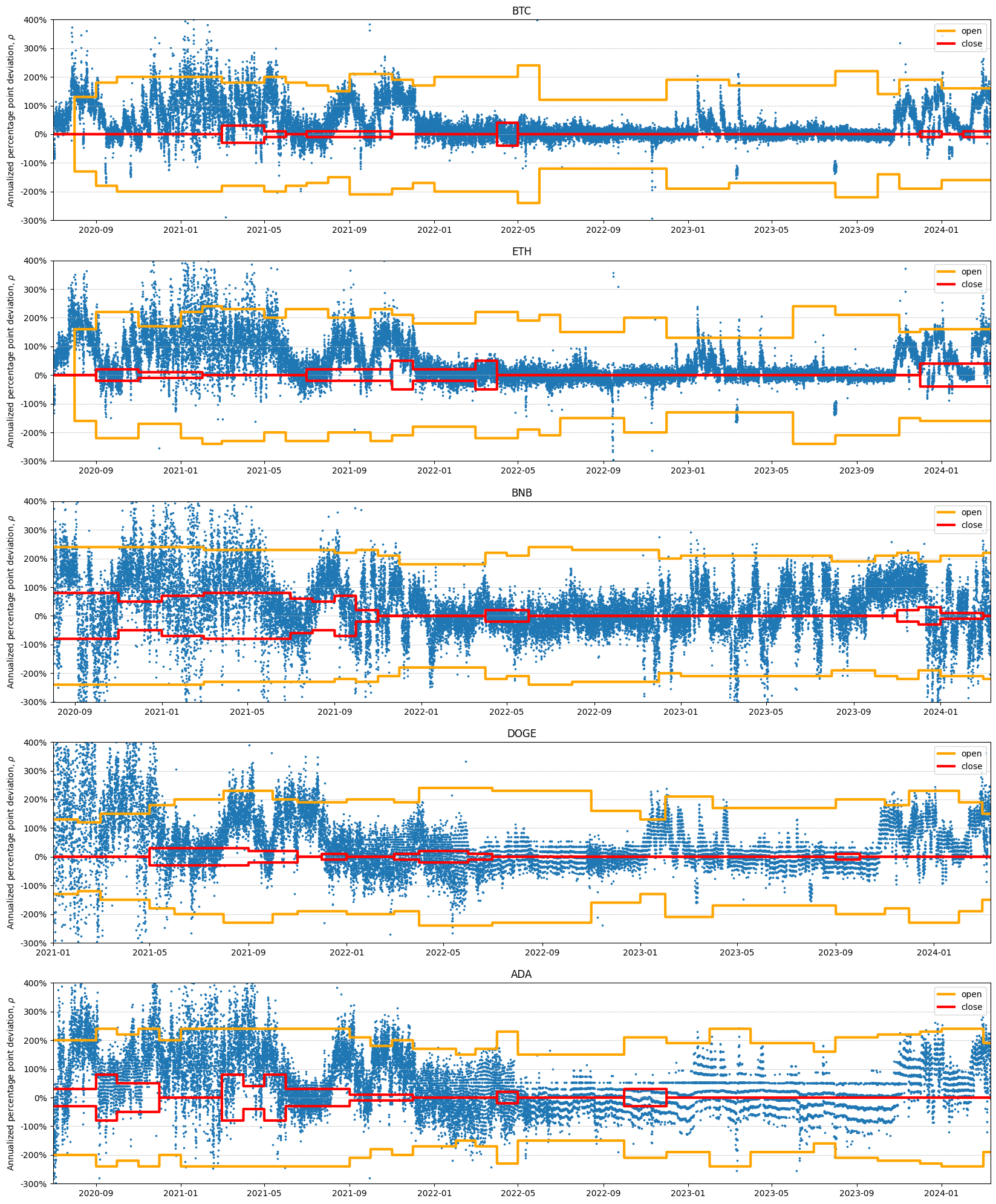}
\end{center}
\caption{Data-driven Trading Strategy Visualization: High Trading Costs} \label{trading_high}
\bigskip
\small
This figure presents the real-time trading thresholds under high trading costs (6.75 bps for spot trading and 1.44 bps for futures trading). The orange line is the open-position threshold, and the red line is the close-position threshold. The trading thresholds are determined based on the adjusted SR from the past six months.
\end{figure}

\clearpage
\newpage
\subsection{Statistical Arbitrage with Cointegration}\label{sec:statarb}

The random-maturity arbitrage (RMA) of Section \ref{sec:theory} and the statistical notion of cointegration with error correction describe the same object from two angles. We next consider a more parametric data-driven strategy inspired by statistical arbitrage with cointegration.

Proposition \ref{prop1} gives the frictionless price $F_t = \lambda S_t$, and Proposition \ref{prop3} confines the ratio to a bound around $\lambda$ once there are trading costs. Taking logs, the basis $b_t \equiv \log F_t - \log S_t$ is bounded and mean-reverting around $\log\lambda$, with the funding payment $\Phi$ supplying the restoring force. 

In the language of \citet{engle1987co}, if $\log F_t$ and $\log S_t$ are each $I(1)$, then they are cointegrated with cointegrating vector $(1,-1)$, and the funding rate plays the role of the error-correction term.

Table \ref{tab_coint} confirms this empirically. For every coin, the Augmented Dickey-Fuller (ADF) test fails to reject a unit root in the log price levels, while it rejects a unit root in the basis: the two series are individually $I(1)$ but cointegrated. 

\begin{table}[!htb]
\begin{center}
\begin{tabularx}{1\columnwidth}{@{\hskip\tabcolsep\extracolsep\fill}l*{5}{r}}
\toprule
 & \multicolumn{3}{c}{ADF \textit{p}-value} & & \\
\cmidrule{2-4}
Asset & $\log F$ & $\log S$ & basis & EG slope $\beta$ & Half-life (hrs) \\
\midrule
BTC  & 0.553 & 0.554 & $0.000$ & 1.0002 & 1.68 \\
ETH  & 0.209 & 0.209 & $0.000$ & 0.9999 & 2.34 \\
BNB  & 0.626 & 0.627 & $0.000$ & 0.9999 & 2.26 \\
DOGE & 0.275 & 0.274 & $0.000$ & 0.9999 & 9.53 \\
ADA  & 0.423 & 0.422 & $0.000$ & 1.0000 & 4.91 \\
\bottomrule
\end{tabularx}
\end{center}
\caption{Cointegration of Perpetual Futures and Spot Prices}\label{tab_coint}
\bigskip
\small
Tests for cointegration of $(\log F_t, \log S_t)$ over the full sample. The first three columns report ADF \textit{p}-values (null: unit root) for the log perpetual price, log spot price, and the log-basis $b_t=\log F_t-\log S_t$: levels are $I(1)$ while the basis is stationary. ``EG slope $\beta$'' is the slope from the Engle-Granger regression $\log F_t=\alpha+\beta\log S_t+u_t$; the residual ADF rejects at any conventional level for all coins. ``Half-life'' is the implied half-life of a basis shock from a VECM with one lagged difference.
\end{table}

The Engle-Granger residual test and the Johansen trace test both reject the no-cointegration null at any conventional level. The estimated cointegrating slope is essentially one, and a VECM implies a basis half-life of about two hours for the larger coins and up to roughly $9.5$ hours for DOGE, the natural empirical measure of the random horizon at which an RMA trade closes.

We also implement a purely statistical strategy that uses \emph{only} the two price series. Each hour we form the standardized basis $z_t = u_t/\sigma_t$ from a 30-day rolling cointegrating regression, open a position when $|z_t|$ exceeds an entry threshold of $1.96$ (shorting the perpetual and longing the spot when $z_t$ is high, and vice versa), and flatten when $z_t$ reverts through zero. Table \ref{tab_statarb} compares it to the theory-motivated and data-driven RMA strategies under the no-trading-cost tier.

\begin{table}[!htb]
\begin{center}
\begin{tabularx}{1\columnwidth}{@{\hskip\tabcolsep\extracolsep\fill}l*{6}{r}}
\toprule
 & \multicolumn{3}{c}{Sharpe ratio} & \multicolumn{3}{c}{Active \%} \\
\cmidrule{2-4} \cmidrule{5-7}
Asset & StatArb & RMA-theory & RMA-data & StatArb & RMA-theory & RMA-data \\
\midrule
BTC  & 4.16 & 11.62 &  8.43 & 25.7 & 100.0 & 60.8 \\
ETH  & 6.01 & 12.75 & 11.77 & 26.2 & 100.0 & 68.4 \\
BNB  & 9.29 & 17.66 & 18.95 & 27.4 & 100.0 & 71.5 \\
DOGE & 5.15 & 14.90 & 14.47 & 23.9 & 100.0 & 65.7 \\
ADA  & 7.82 & 19.76 & 16.09 & 21.8 & 100.0 & 52.6 \\
\bottomrule
\end{tabularx}
\end{center}
\caption{Statistical Arbitrage versus Random-maturity Arbitrage}\label{tab_statarb}
\bigskip
\small
Annualized Sharpe ratios and time-in-market for three strategies under the no-trading-cost tier and the DeFi risk-free rate. ``StatArb'' trades the standardized residual of a 30-day rolling cointegrating regression with no structural input; ``RMA-theory'' is the theory-motivated strategy of Section \ref{sec:empirical}, and ``RMA-data'' the data-driven strategy of Appendix \ref{sec:data_driven}. Because the strategies are active for different fractions of the sample (``Active \%''), Sharpe ratios are annualized following \citet{lucca2015pre}: active-subsample moments are scaled by the average number of active hours per year.
\end{table}

The statistical strategy is profitable on every coin using no structural input, confirming that the cointegrating relation alone is exploitable; the structural strategies nonetheless dominate, because they stay in the market far more of the time and so harvest the smaller, persistent funding-driven deviations that the residual rule discards as noise. 

\clearpage
\newpage
\section{Additional Empirical Results}\label{sec:addresults}
\subsection{Robustness to Alternative Risk-free Rate Measures}\label{sec:robustness}
We show robustness in the random-maturity arbitrage strategies' performance by considering alternative risk-free rates and present the results in Table \ref{port_robustness_rf}. We conduct our analysis using daily T-bill rates sourced from Kenneth French's data library (\url{https://mba.tuck.dartmouth.edu/pages/faculty/ken.french/Data_Library/f-f_factors.html}) and margin interest rates from Binance (\url{https://binance-docs.github.io/apidocs/spot/en/\#query-margin-interest-rate-history-user_data}). We find that the performance of the trading strategy is robust to these alternative risk-free rates and across different trading costs tiers.

\begin{table}[!htb]
\begin{center}
\footnotesize
\begin{tabularx}{\textwidth}{@{\hskip\tabcolsep\extracolsep\fill}ll*{4}{r}}
\toprule
  &  & \multicolumn{2}{c}{No Trading Cost} & \multicolumn{2}{c}{High Trading Cost} \\
\cmidrule{3-4} \cmidrule{5-6}
 &  & FF & Binance & FF & Binance \\
\midrule
BTC & SR & 11.65 & 11.73 & 3.79 & 2.88 \\
 & Return & 52.60 & 52.91 & 15.57 & 10.45 \\
 & Volatility & 4.52 & 4.51 & 4.11 & 3.63 \\
 & MaxDD & -3.82 & -3.82 & -3.82 & -2.63 \\
 & $\alpha$ & 59.60 & 54.40 & 14.28 & 12.11 \\
 & $t_{\alpha}$ & 18.59 & 15.65 & 5.91 & 7.03 \\
 & Active \% & 100.00 & 100.00 & 31.30 & 19.37 \\
 & OtC time & 5.20 & 5.12 & 200.86 & 97.05 \\
ETH & SR & 12.35 & 14.17 & 4.93 & 4.47 \\
 & Return & 60.83 & 69.50 & 21.23 & 15.98 \\
 & Volatility & 4.92 & 4.90 & 4.31 & 3.58 \\
 & MaxDD & -2.62 & -2.57 & -2.62 & -1.59 \\
 & $\alpha$ & 70.74 & 75.07 & 21.23 & 17.07 \\
 & $t_{\alpha}$ & 18.28 & 12.65 & 6.17 & 5.98 \\
 & Active \% & 100.00 & 100.00 & 35.27 & 22.25 \\
 & OtC time & 5.38 & 4.89 & 189.74 & 59.85 \\
BNB & SR & 16.94 & 19.40 & 6.63 & 7.44 \\
 & Return & 129.42 & 147.17 & 43.13 & 45.51 \\
 & Volatility & 7.64 & 7.59 & 6.50 & 6.11 \\
 & MaxDD & -1.97 & -1.96 & -2.09 & -2.08 \\
 & $\alpha$ & 142.69 & 154.89 & 36.18 & 38.04 \\
 & $t_{\alpha}$ & 15.19 & 13.78 & 6.50 & 6.25 \\
 & Active \% & 100.00 & 100.00 & 41.32 & 35.78 \\
 & OtC time & 5.79 & 5.03 & 35.45 & 30.05 \\
DOGE & SR & 14.45 & 14.87 & 4.91 & 5.33 \\
 & Return & 175.82 & 180.22 & 55.19 & 55.63 \\
 & Volatility & 12.17 & 12.12 & 11.23 & 10.43 \\
 & MaxDD & -13.90 & -13.87 & -13.90 & -10.08 \\
 & $\alpha$ & 218.01 & 205.74 & 45.21 & 48.22 \\
 & $t_{\alpha}$ & 11.30 & 10.24 & 4.80 & 5.38 \\
 & Active \% & 100.00 & 100.00 & 33.28 & 27.55 \\
 & OtC time & 4.29 & 4.13 & 19.67 & 13.88 \\
ADA & SR & 19.50 & 19.45 & 4.28 & 3.87 \\
 & Return & 153.28 & 152.70 & 28.75 & 24.91 \\
 & Volatility & 7.86 & 7.85 & 6.72 & 6.43 \\
 & MaxDD & -5.68 & -5.65 & -5.68 & -5.58 \\
 & $\alpha$ & 175.95 & 151.87 & 25.42 & 22.08 \\
 & $t_{\alpha}$ & 19.85 & 18.03 & 6.49 & 6.42 \\
 & Active \% & 100.00 & 100.00 & 33.70 & 23.79 \\
 & OtC time & 4.46 & 4.23 & 56.71 & 29.25 \\
\bottomrule
\end{tabularx}
\end{center}
\caption{Robustness Check: Alternative Risk-free Rates}
\label{port_robustness_rf}
\bigskip
\small
Performance under alternative risk-free rates. `FF' represents the risk-free rate from Ken French's data library. `Binance' represents the margin borrowing rate from Binance. We report performance under no trading costs and high trading costs. Statistics reported are the annualized Sharpe ratio, return (\%), volatility (\%), max drawdown (\%), alpha (\%) relative to the three factors of \citet{liu2022common}, t-stat of the alpha, proportion of time the strategy is active (\%), and average open-to-close (OtC) position duration in hours.
\end{table}

\clearpage
\newpage
\subsection{Deviations plus the Interest Rate}

\begin{figure}[!htb]
\begin{center}
\noindent\includegraphics[width=\textwidth]{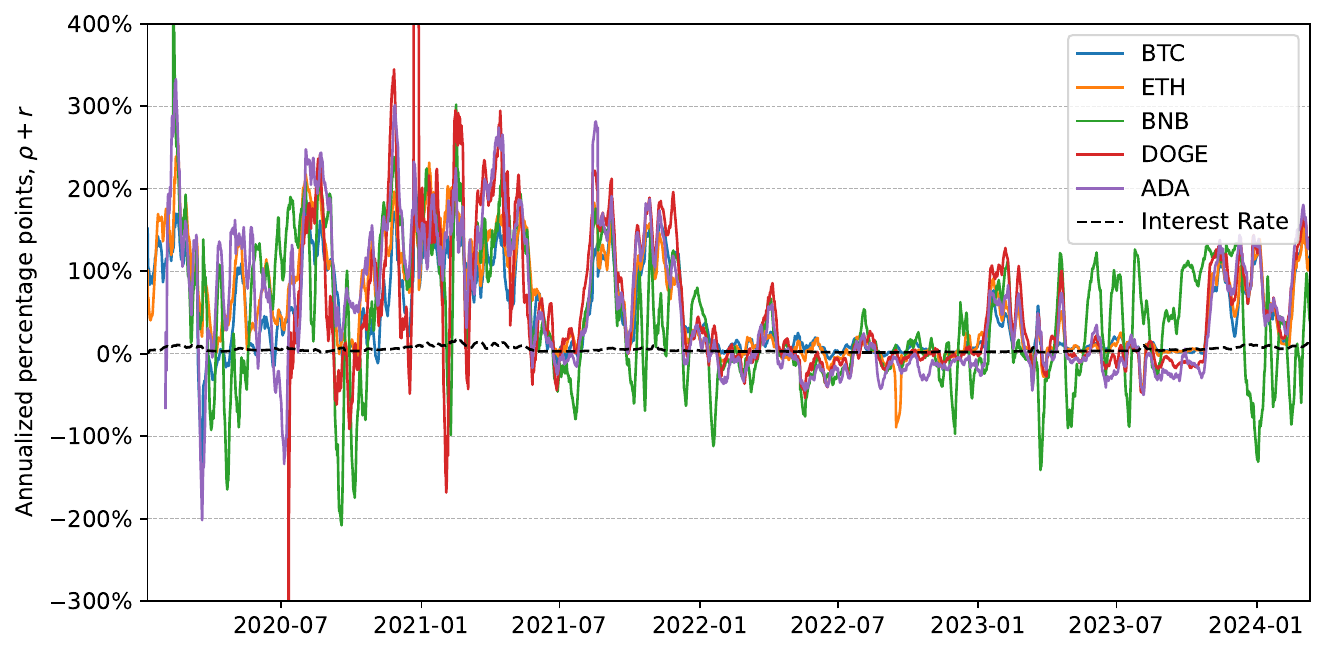}
\end{center}
\caption{Deviations of Perpetual Futures from No-arbitrage Benchmarks plus the Interest Rate}\label{fig:rhoplusr}
\bigskip
\small
This figure presents the 7-day moving averages of $\rho + r$, the annualized deviation of the perpetual futures-spot spread from the no-arbitrage benchmark plus the annualized interest rate, for the five cryptocurrencies. The vertical scale is the same as in Figure \ref{fig1}. The dashed line is the 7-day moving average of the annualized interest rate.
\end{figure}

The deviation $\rho$ in Equation \eqref{eqdef} incorporates the interest rate $r$. To verify that the variation in $\rho$ is driven by futures prices rather than by variation in the interest rate, we add $r$ back and examine
\[
\rho + r = \kappa \left( \frac{F-S}{F} \right) + \text{sgn}(\iota - r)\gamma.
\]
Figure \ref{fig:rhoplusr} plots $\rho+r$ using the same vertical scale as Figure \ref{fig1}. The two figures are nearly indistinguishable. Indeed, the correlation between $\rho$ and $\rho+r$ exceeds $0.999$ for every cryptocurrency. Thus, variation in the interest rate accounts for virtually none of the time-series variation in the measured deviations.

\clearpage
\newpage
\subsection{Portfolio Margin Event Study}\label{sec:portfolio_margin}

\begin{figure}[!htb]
    \begin{center}
    \includegraphics[width=\linewidth]{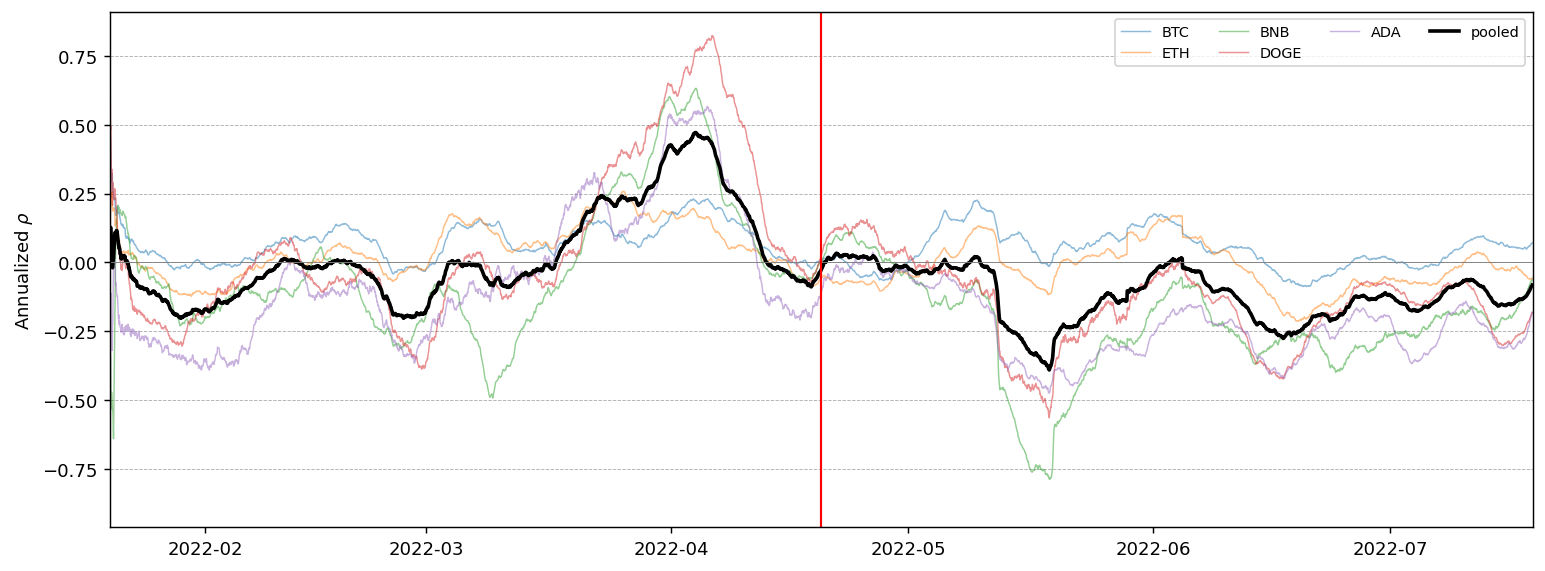}
    \end{center}
    \caption{Deviations of Perpetual Futures from No-arbitrage Benchmarks around the Introduction of Portfolio Margin}
    \label{fig:pm_event_90d}
    \bigskip
    \small
This figure presents the 7-day moving averages of the annualized deviation of the perpetual futures from the no-arbitrage benchmark for five cryptocurrencies around the introduction of the Binance portfolio margin program with an event window of 90 trading days. 
\end{figure}

To complement the portfolio-margin analysis in Section \ref{sec:explaining}, Figure \ref{fig:pm_event_90d} plots futures-spot deviations around Binance's introduction of portfolio margin on April 20, 2022. As discussed in the main text, portfolio margin allows gains and losses on spot and futures positions to be netted within a unified account, reducing the capital required to maintain the arbitrage trade.

Figure \ref{fig:pm_event_90d} shows that deviations are sizable and volatile before the introduction of portfolio margin and become visibly smaller afterward. This pattern is consistent across the five cryptocurrencies and provides a visual counterpart to the statistical evidence reported in Table \ref{tab:portfolio_margin_event} in the main text.

Together, the figure and the main-text estimates are consistent with Prediction 1: improving the capital efficiency of the arbitrage trade allows arbitrageurs to absorb larger positions and compresses deviations from the no-arbitrage benchmark.

\subsection{Distribution of the Time to Payoff Realization}\label{sec:horizondist}

Each round trip of the theory-motivated random-maturity arbitrage strategy is one realization of the random unwinding time $\tilde\tau$ of Definition \ref{def:random_maturity_arb}, the time at which the arbitrage payoff is realized. Table \ref{tab:horizondist} reports the distribution of these holding periods, in hours, at the zero-fee and the high-fee tier. We find that for a trader who pays no fees, the median holding period is two hours, and 98\% of trades are closed within one day. For a trader who pays the highest fees, the median holding period is 38 hours (longest for BTC), and 90\% of trades are closed within 10 days (224 hours).

\begin{table}[!htb]
\begin{center}
\footnotesize
\begin{tabularx}{1\columnwidth}{@{\hskip\tabcolsep\extracolsep\fill}ll*{9}{r}}
\toprule
Fee tier & Asset & Trades & p25 & Median & p75 & p90 & p99 & Max & $\le$ 1 day (\%) & $\le$ 1 week (\%) \\
\midrule
\multirow{5}{*}{No fee} & BTC  & 7,437 & 1 & 2 & 3 & 7 & 56 & 815   & 98.0 & 99.6 \\
 & ETH  & 7,243 & 1 & 2 & 3 & 7 & 61 & 1,538 & 97.9 & 99.6 \\
 & BNB  & 6,552 & 1 & 2 & 4 & 9 & 60 & 813   & 97.0 & 99.8 \\
 & DOGE & 7,786 & 1 & 2 & 3 & 7 & 37 & 878   & 98.3 & 99.8 \\
 & ADA  & 8,335 & 1 & 2 & 3 & 7 & 48 & 810   & 98.0 & 99.8 \\
\midrule
\multirow{5}{*}{High fee} & BTC  & 92  & 14 & 38 & 120 & 224 & 726 & 793   & 43.5 & 80.4 \\
 & ETH  & 122 & 5  & 31 & 91  & 238 & 561 & 1,059 & 45.1 & 85.2 \\
 & BNB  & 475 & 3  & 8  & 25  & 62  & 439 & 726   & 74.9 & 97.3 \\
 & DOGE & 594 & 2  & 4  & 10  & 29  & 228 & 794   & 89.1 & 98.0 \\
 & ADA  & 263 & 4  & 13 & 37  & 94  & 498 & 756   & 66.9 & 94.7 \\
\bottomrule
\end{tabularx}
\end{center}
\caption{The distribution of the time to payoff realization}\label{tab:horizondist}
\bigskip
\small
Each round trip of the theory-motivated random-maturity arbitrage strategy is one observation; holding periods are in hours. Every round trip in the sample is closed before the sample ends, with the single exception of one BTC position at the high-fee tier that is still open on the last day.
\end{table}

\subsection{Forward-looking Volatility}\label{sec:ivol}

Table \ref{reg_exp} measures arbitrageur risk using trailing 120-day realized volatility. Because this measure is backward-looking, we construct an ex ante HAR-RV forecast of realized volatility over the subsequent 120 days. Specifically, we regress future 120-day realized volatility on daily, weekly, and monthly lagged realized volatility and use the fitted value as the expected-volatility measure. The forecast uses only information available at the forecast date and is available for all five currencies over essentially the full sample.

Table \ref{tab:ivol} shows that the results are similar to those in Table \ref{reg_exp}. Expected volatility enters negatively and significantly for both $\rho$ and $|\rho|$. The time trend remains negative and significant, while past returns remain positively associated with deviations. Thus, replacing backward-looking realized volatility with an ex ante forecast does not provide support for the simple prediction that higher expected volatility tightens arbitrageurs' risk constraints and widens deviations.

\begin{table}[!htb]
\begin{center}
\footnotesize
\begin{tabularx}{1\columnwidth}{@{\hskip\tabcolsep\extracolsep\fill}ll*{4}{r}}
\toprule
Volatility measure & Dep.\ var. & Time & Ret & Volatility & $N$ \\
\midrule
\multirow{2}{*}{Realized Volatility} & $\rho$ & $-0.228$\sym{***} & $0.166$\sym{***} & $-0.031$\sym{***} & 7,236 \\
 &  & $(-6.83)$ & $(11.19)$ & $(-6.65)$ & \\
 & $|\rho|$ & $-0.234$\sym{***} & $0.132$\sym{***} & $-0.023$\sym{***} & 7,236 \\
 &  & $(-8.44)$ & $(8.20)$ & $(-4.66)$ & \\
\midrule
\multirow{2}{*}{HAR-RV forecast} & $\rho$ & $-0.197$\sym{***} & $0.138$\sym{***} & $-0.013$\sym{***} & 7,235 \\
 &  & $(-5.38)$ & $(6.95)$ & $(-5.89)$ & \\
 & $|\rho|$ & $-0.197$\sym{***} & $0.102$\sym{***} & $-0.007$\sym{***} & 7,235 \\
 &  & $(-6.64)$ & $(6.27)$ & $(-3.15)$ & \\
\bottomrule
\end{tabularx}
\end{center}
\caption{Deviations and forward-looking volatility}\label{tab:ivol}
\bigskip
\small
Daily panel regressions with currency fixed effects and Newey-West standard errors with $120$ lags, following the specification of Table \ref{reg_exp}. ``Time'' is a linear trend in years, ``Ret'' is the past $120$ days' annualized spot return, and ``Volatility'' is the stated volatility measure in annualized percentage points. Trailing realized volatility is the past $120$ days' annualized return standard deviation. The HAR-RV forecast is the fitted value from a daily regression of the next $120$ days' average realized volatility on the previous day's, week's and month's realized volatility. $*\text{ }p<.1$, $**\text{ }p<.05$, $***\text{ }p<.01$.
\end{table}

\subsection{Perpetual Order Flow and the Deviation}\label{sec:orderflow}

If deviations from the no-arbitrage benchmark induce convergence trading, liquidity-demanding order flow in the perpetual market should lean against the deviation. In particular, when the perpetual is expensive relative to the benchmark, selling pressure in the perpetual market should increase. We examine this prediction using hourly signed taker order imbalance on Binance.

Since every trade has a taker and a maker, we use the taker side to sign
each trade. Define hourly signed taker order imbalance in the perpetual
market as
\begin{equation}
    OIB^{perp}_{i,t}
    =
    \frac{
    V^{perp,\text{taker buy}}_{i,t} - V^{perp,\text{taker sell}}_{i,t}}{
    V^{perp,\text{taker buy}}_{i,t} + V^{perp,\text{taker sell}}_{i,t}
    } \in [-1,1],\label{eq:oib}
\end{equation}
where $V^{perp,\text{taker buy}}_{i,t}$ and
$V^{perp,\text{taker sell}}_{i,t}$ denote dollar volume initiated by
taker buys, and taker sells, respectively, for currency $i$ in hour $t$.
A higher value of $OIB^{perp}$ therefore indicates greater
liquidity-demanding buying pressure, while a lower value indicates
greater selling pressure.

Table \ref{tab:flow} strongly supports the prediction. A one-unit increase
in the annualized deviation, roughly one and a quarter standard deviations,
lowers perpetual taker order imbalance by $1.1$ percentage points, with a
$t$-statistic of $-13.3$. Thus, as perpetual futures become more expensive
relative to their no-arbitrage benchmark, liquidity-demanding order flow
shifts significantly toward selling the perpetual. This evidence is
consistent with market participants trading against deviations and exerting
price-convergence pressure.

\begin{table}[!htb]
\begin{center}
\footnotesize
\begin{tabularx}{0.52\columnwidth}{@{\hspace{6pt}}Xr@{\hspace{16pt}}}
\toprule
\multicolumn{2}{c}{Dependent variable: $OIB^{perp}_{i,t}$} \\
\midrule
$\rho_{i,t-1}$
    & $-0.0106$\sym{***} \\
    & $(-13.30)$ \\
\addlinespace[3pt]
\midrule
Currency fixed effects
    & Yes \\
Observations
    & 176,940 \\
\bottomrule
\end{tabularx}
\end{center}

\caption{Perpetual taker order flow and the no-arbitrage deviation}
\label{tab:flow}

\bigskip
\small
This table reports a pooled hourly panel regression of perpetual taker
order imbalance, defined in Equation \eqref{eq:oib}, on the lagged
annualized deviation. The regression includes currency fixed effects.
Newey-West standard errors using one week of hourly lags are used to compute
the $t$-statistic reported in parentheses. The underlying trade data are
obtained from the Binance public data archive.
$*\text{ }p<.1$, $**\text{ }p<.05$, $***\text{ }p<.01$.
\end{table}

\clearpage
\newpage
\section{Proofs}\label{sec:proof}
In this section, we provide proofs to Proposition \ref{prop1}, Proposition \ref{prop2}, Corollary \ref{corollary2}, and Proposition \ref{prop3}.

\textbf{Proof to Proposition \ref{prop1}:}
\vspace{-0.1in}
\begin{proof}
Define the futures-to-spot ratio and the premium index by
\begin{equation}
    R_t\equiv \frac{F_t}{S_t},
    \qquad
    p_t\equiv \frac{F_t-S_t}{F_t}=1-\frac{1}{R_t}.
\end{equation}
Because Proposition \ref{prop1} assumes that $\Phi^{-1}(r)$ is a singleton, let $x^*$ denote its unique element and set
\begin{equation}
    x^*\in\Phi^{-1}(r),
    \qquad
    \lambda\equiv (1-x^*)^{-1},
\end{equation}
and define the deviation from the no-arbitrage benchmark as
\begin{equation}
    u_t\equiv F_t-\lambda S_t=S_t(R_t-\lambda).
    \label{eq:alternative_u}
\end{equation}
Because $p(R)=1-R^{-1}$ is strictly increasing in $R$, and $\Phi$ is weakly increasing with $\Phi^{-1}(r)=\{x^*\}$, we have
\begin{equation}
    u_t>0
    \Longleftrightarrow R_t>\lambda
    \Longleftrightarrow p_t>x^*
    \Longleftrightarrow \Phi(p_t)>r,
    \label{eq:alternative_sign}
\end{equation}
and all inequalities reverse when $u_t<0$. Thus the funding mechanism rewards the appropriate convergence trade on either side of the benchmark: if the perpetual is expensive relative to $\lambda S$, long $\lambda$ units of spot and short one perpetual; if it is cheap, reverse the trade.

\medskip
\noindent\textit{Sufficiency.}
Suppose first that $F_t=\lambda S_t$ for all $t$. Then $p_t=x^*$ and $\Phi(p_t)=r$. The gains on one long perpetual are therefore
\begin{equation}
    dF_t-\Phi(p_t)F_t\,dt
    =\lambda\bigl(dS_t-rS_t\,dt\bigr),
\end{equation}
which are exactly the gains on $\lambda$ units of spot financed at the risk-free rate. The perpetual is therefore redundant given the spot and cash markets. Since the spot and cash markets admit no arbitrage, adding the perpetual cannot create a random-maturity arbitrage.

\medskip
\noindent\textit{Necessity.}
We now show that any deviation from $F=\lambda S$ generates a bounded-horizon random-maturity arbitrage. We begin with $R_0>\lambda$.

For one unit of the convergence trade, long $\lambda$ spot and short one perpetual, the discounted payoff from initiation to liquidation at time $t$ is
\begin{align}
    P_t
    &=-\int_0^t e^{-rs}\,du_s
      +\int_0^t e^{-rs}\bigl[\Phi(p_s)F_s-r\lambda S_s\bigr]ds \notag\\
    &=u_0-e^{-rt}u_t
      +\int_0^t e^{-rs}\bigl[\Phi(p_s)-r\bigr]F_s\,ds.
    \label{eq:alternative_oneunit}
\end{align}
The three terms have a direct interpretation. The first is the initial deviation captured by entering the convergence trade. The last is the cumulative funding received in excess of the financing cost of the spot leg. The middle term, $-e^{-rt}u_t$, is the discounted cost of unwinding while a deviation remains. Unlike a fixed-maturity futures contract, a perpetual has no expiration date at which this last term is mechanically zero.

When $u_t>0$, define the \emph{restoring rate}
\begin{equation}
    a_t
    \equiv
    \frac{[\Phi(p_t)-r]F_t}{u_t}
    =
    \frac{R_t}{R_t-\lambda}
    \left[\Phi\!\left(1-\frac{1}{R_t}\right)-r\right]
    >0.
    \label{eq:alternative_restore}
\end{equation}
The restoring rate is the net carry earned by the convergence trade per unit of the current deviation. It therefore has units of a rate: if the deviation and $a_t$ were held fixed, the excess funding flow would earn back an amount equal to the current deviation in approximately $1/a_t$ units of time. We call $a_t$ a restoring rate because it measures the strength of the contract's funding mechanism against a deviation. Importantly, it is not an assumption that $R_t$ itself has a mean-reverting drift.

This interpretation suggests scaling down the convergence position as the excess carry accumulates. Define
\begin{equation}
    H_t\equiv \exp\left(-\int_0^t a_s\,ds\right),
    \qquad dH_t=-a_tH_t\,dt,
    \label{eq:alternative_H}
\end{equation}
for as long as $u_t>0$, and dynamically hold $H_t$ units of the convergence trade. Because the perpetual position can be resized at zero entry cost and changes in the spot leg are offset in the cash account, this rebalancing is self-financing. Its discounted payoff is
\begin{align}
    V_t
    &=-\int_0^t H_s e^{-rs}\,du_s
      +\int_0^t H_s e^{-rs}
       \bigl[\Phi(p_s)F_s-r\lambda S_s\bigr]ds \notag\\
    &=u_0-H_te^{-rt}u_t.
    \label{eq:alternative_dynamic_payoff}
\end{align}
To see the second equality, let $G_t\equiv H_te^{-rt}$. By \eqref{eq:alternative_restore},
\begin{equation}
    \Phi(p_t)F_t-r\lambda S_t
    =[\Phi(p_t)-r]F_t+ru_t
    =(a_t+r)u_t,
\end{equation}
while $dG_t=-(a_t+r)G_t\,dt$. Integration by parts therefore gives
\begin{equation}
    -\int_0^t G_s\,du_s
    +\int_0^t (a_s+r)G_su_s\,ds
    =u_0-G_tu_t,
\end{equation}
which is Equation \eqref{eq:alternative_dynamic_payoff}. Economically, reducing the position at rate $a_t$ converts the running excess-funding gains into a smaller exposure to the deviation at unwinding. The entire accumulated funding term in \eqref{eq:alternative_oneunit} is absorbed by this dynamic scaling, leaving only the initial deviation and a scaled terminal deviation.

This calculation already allows jumps. For c\`adl\`ag semimartingales, integration by parts gives
\begin{equation}
    G_tu_t
    =G_0u_0+\int_0^t G_{s-}\,du_s
    +\int_0^t u_{s-}\,dG_s+[G,u]_t.
\end{equation}
Here $G$ is continuous and of finite variation, so $[G,u]_t=0$. Moreover, $dG_s$ is absolutely continuous with respect to $ds$ and therefore assigns zero weight to individual jump times; hence $u_{s-}$ can be replaced by $u_s$ in the $dG_s$ integral. Thus no additional jump term appears in \eqref{eq:alternative_dynamic_payoff}.\footnote{See, e.g., \citep{protter2012stochastic} for the integration-by-parts formula for c\`adl\`ag semimartingales. Funding is a time-flow, $\Phi(p_t)F_tdt$, so a jump in prices changes the contemporaneous premium and subsequent funding rate but does not itself generate a lump-sum funding payment.}

It remains to show that the scaled terminal deviation can be controlled within a deterministic finite horizon. Suppose
\begin{equation}
    R_0-\lambda\geq \delta>0,
\end{equation}
and set $c\equiv \delta/2$. Define the first \emph{crossing} time
\begin{equation}
    \tau_1\equiv
    \inf\left\{t>0:R_t-\lambda\leq c\right\}.
    \label{eq:alternative_tau1}
\end{equation}
We use a crossing time rather than a hitting time so that the argument also covers jumps. Before $\tau_1$, $R_t\geq \lambda+c$. Hence, by monotonicity of $\Phi$ and uniqueness of $x^*$,
\begin{equation}
    a_t
    \geq
    \Phi\!\left(1-\frac{1}{\lambda+c}\right)-r
    \equiv \alpha>0,
    \qquad t<\tau_1,
    \label{eq:alternative_alpha}
\end{equation}
where we have used $R_t/(R_t-\lambda)\geq1$. Consequently,
\begin{equation}
    H_t\leq e^{-\alpha t},
    \qquad t<\tau_1.
    \label{eq:alternative_Hbound}
\end{equation}

We now add $c=\delta/2$ units of spot to the strategy, financed through a matching position in the cash market. This extra spot leg is useful because it places a positive $c$ inside the terminal bracket that will dominate the residual scaled deviation. Using the same short-term financing convention as in Table~1, the spot position is financed through a cash-market position rolled at the current spot value. Its discounted payoff is
\begin{align}
    J_t
    &=c\int_0^t e^{-rs}\,dS_s
      -c\int_0^t r e^{-rs}S_s\,ds \notag\\
    &=c\bigl(e^{-rt}S_t-S_0\bigr).
    \label{eq:alternative_spot_ibp}
\end{align}
The second equality follows directly from integration by parts,
\begin{equation}
    d\bigl(e^{-rt}S_t\bigr)
    =e^{-rt}\,dS_t-r e^{-rt}S_t\,dt.
\end{equation}
Because $e^{-rt}$ is continuous and of finite variation, this identity is also unchanged when $S$ jumps. Equation \eqref{eq:alternative_spot_ibp} makes explicit why the extra spot position does not introduce an unaccounted financing term: the second integral is exactly the cumulative financing cost of the spot leg.

Combining \eqref{eq:alternative_dynamic_payoff} and \eqref{eq:alternative_spot_ibp}, and using $u_t=S_t(R_t-\lambda)$, gives
\begin{align}
    \Pi_t
    &=V_t+J_t \notag\\
    &=S_0\bigl[(R_0-\lambda)-c\bigr]
      +e^{-rt}S_t\bigl[c-H_t(R_t-\lambda)\bigr].
    \label{eq:alternative_total_payoff}
\end{align}
Choose the deterministic horizon
\begin{equation}
    T
    \equiv
    \frac{1}{\alpha}
    \log\left(\frac{\overline{R}-\lambda}{c}\right)
    =
    \frac{1}{\alpha}
    \log\left(\frac{2(\overline{R}-\lambda)}{\delta}\right),
    \label{eq:alternative_T}
\end{equation}
and liquidate at $\tau\equiv \tau_1\wedge T$.

There are two cases. If $\tau_1\leq T$, right continuity implies
$R_{\tau_1}-\lambda\leq c$. The inequality may be strict when the ratio jumps across the threshold; such an overshoot only improves the payoff. Since $H_{\tau_1}\leq1$,
\begin{equation}
    c-H_{\tau_1}(R_{\tau_1}-\lambda)\geq0.
\end{equation}
If instead $T<\tau_1$, then Assumption 2' and \eqref{eq:alternative_Hbound} imply
\begin{equation}
    H_T(R_T-\lambda)
    \leq e^{-\alpha T}(\overline{R}-\lambda)
    =c.
\end{equation}
Thus the second term in \eqref{eq:alternative_total_payoff} is nonnegative in either case, while the first satisfies
\begin{equation}
    S_0\bigl[(R_0-\lambda)-c\bigr]
    \geq \frac{\delta}{2}S_0>0.
\end{equation}
The strategy has zero initial cost, is self-financing, and delivers a strictly positive payoff on every path no later than the deterministic time $T$. It is therefore a random-maturity arbitrage.

The case $R_0<\lambda$ is symmetric, but we spell out the changes because Assumption \ref{assp2} now uses its lower bound. Let
\begin{equation}
    \widetilde{u}_t\equiv -u_t=S_t(\lambda-R_t)>0
\end{equation}
while the perpetual is below the benchmark, and use the reverse convergence trade: long one perpetual and short $\lambda$ spot. Its restoring rate is
\begin{equation}
    a_t^-
    \equiv
    \frac{[r-\Phi(p_t)]F_t}{\widetilde{u}_t}
    =
    \frac{R_t}{\lambda-R_t}
    \left[r-\Phi\!\left(1-\frac{1}{R_t}\right)\right]
    >0,
\end{equation}
with $H_t^-\equiv\exp(-\int_0^t a_s^-ds)$. The dynamically scaled reverse convergence trade then has discounted payoff
\begin{equation}
    V_t^-=\widetilde{u}_0-H_t^-e^{-rt}\widetilde{u}_t.
    \label{eq:alternative_reverse_dynamic}
\end{equation}
Suppose $\lambda-R_0\geq\delta>0$, again set $c=\delta/2$, and define
\begin{equation}
    \tau_1^-
    \equiv
    \inf\left\{t>0:\lambda-R_t\leq c\right\}.
\end{equation}
Before $\tau_1^-$, $\underline{R}\leq R_t\leq\lambda-c$. Therefore
\begin{equation}
    a_t^-
    \geq
    \frac{\underline{R}}{\lambda-\underline{R}}
    \left[
        r-\Phi\!\left(1-\frac{1}{\lambda-c}\right)
    \right]
    \equiv \alpha^->0.
    \label{eq:alternative_alpha_minus}
\end{equation}
Adding the same financed long position of $c$ units of spot gives
\begin{equation}
    \Pi_t^-
    =S_0\bigl[(\lambda-R_0)-c\bigr]
     +e^{-rt}S_t\bigl[c-H_t^-(\lambda-R_t)\bigr].
    \label{eq:alternative_reverse_total}
\end{equation}
Choose
\begin{equation}
    T^-
    \equiv
    \frac{1}{\alpha^-}
    \log\left(\frac{\lambda-\underline{R}}{c}\right),
\end{equation}
and liquidate at $\tau^-\equiv\tau_1^-\wedge T^-$. The same two-case argument gives
\begin{equation}
    \Pi_{\tau^-}^-
    \geq \frac{\delta}{2}S_0>0
\end{equation}
for every path. Hence $R_0<\lambda$ also creates a bounded-horizon random-maturity arbitrage.

We have shown that absence of random-maturity arbitrage requires $R_0=\lambda$. Because the same argument can be initiated after resetting the clock at any date, it follows that $R_t=\lambda$ for all $t$. Therefore
\begin{equation}
    F_t
    =\lambda S_t
    =\left(1-\Phi^{-1}(r)\right)^{-1}S_t,
\end{equation}
which proves necessity and completes the proof.
\end{proof}

\textbf{Proof to Proposition \ref{prop2}:}
\vspace{-0.1in}
\begin{proof}
By Assumption \ref{assp3}, the level set of the funding rate at the risk-free rate is a compact interval. Write
$$
    \Phi^{-1}(r) = [x_-, x_+],
$$
and define
$$
    \underline{\lambda}\equiv (1-x_-)^{-1},
    \qquad
    \overline{\lambda}\equiv (1-x_+)^{-1}.
$$
Since $x_- \leq x_+<1$, we have
$$
    0<\underline{\lambda}\leq \overline{\lambda}<\infty.
$$
Moreover, because $\Phi$ is weakly increasing,
$$
    p>x_+ \quad\Longrightarrow\quad \Phi(p)>r,
    \qquad
    p<x_- \quad\Longrightarrow\quad \Phi(p)<r.
$$
Thus, outside the interval $[\underline{\lambda},\overline{\lambda}]$, the funding mechanism rewards a convergence trade toward the nearest endpoint of the interval.

\textit{Necessity.}
We first consider a perpetual that is expensive relative to the upper
benchmark. Suppose that, at the date the trade is initiated,
$$
R_0-\overline{\lambda}\geq \delta>0,
$$
and set $c = \delta /2$. Define
$$
    u_t^+
    \equiv F_t-\overline{\lambda}S_t
    =S_t(R_t-\overline{\lambda}).
$$
As long as $u_t^+>0$, the appropriate convergence trade is long
$\overline{\lambda}$ units of spot and short one perpetual. Its restoring
rate is
$$
    a_t^+
    \equiv
    \frac{[\Phi(p_t)-r]F_t}{u_t^+}
    =
    \frac{R_t}{R_t-\overline{\lambda}}
    \left[
        \Phi\left(1-\frac{1}{R_t}\right)-r
    \right]
    >0.
$$
As in the proof of Proposition 1, define
$$
    H_t^+
    \equiv
    \exp\left(-\int_0^t a_s^+\,ds\right).
$$
Dynamically holding $H_t^+$ units of the convergence trade gives discounted
payoff
$$
    V_t^+
    =
    u_0^+-H_t^+e^{-rt}u_t^+.
$$
The economic role of the scaling is the same as in Proposition 1:
the excess funding received while the deviation remains open is used to
reduce the position, thereby shrinking the exposure to the remaining
deviation at liquidation.

Define the first crossing time
$$
    \tau_1^+
    \equiv
    \inf\left\{
        t>0:R_t-\overline{\lambda}\leq c
    \right\}.
$$
Before $\tau_1^+$, $R_t\geq \overline{\lambda}+c$, and hence
$$
    a_t^+
    \geq
    \Phi\left(
        1-\frac{1}{\overline{\lambda}+c}
    \right)-r
    \equiv \alpha^+>0.
$$
Therefore
\[
    H_t^+\leq e^{-\alpha^+t},
    \qquad t<\tau_1^+.
\]
As in Proposition 1, add $c$ units of spot financed through the cash
market. Combining this position with the dynamically scaled convergence
trade gives
\[
    \Pi_t^+
    =
    S_0\left[(R_0-\overline{\lambda})-c\right]
    +
    e^{-rt}S_t
    \left[
        c-H_t^+(R_t-\overline{\lambda})
    \right].
\]
Choose
\[
    T^+
    \equiv
    \frac{1}{\alpha^+}
    \log\left(
        \frac{\overline{R}-\overline{\lambda}}{c}
    \right),
    \qquad
    \tau^+\equiv \tau_1^+\wedge T^+.
\]
If $\tau_1^+\leq T^+$, then
$R_{\tau_1^+}-\overline{\lambda}\leq c$ and
$H_{\tau_1^+}^+\leq 1$, so
\[
    c-H_{\tau_1^+}^+
    (R_{\tau_1^+}-\overline{\lambda})\geq 0.
\]
If instead $T^+<\tau_1^+$, Assumption 2 implies
\[
    H_{T^+}^+(R_{T^+}-\overline{\lambda})
    \leq
    e^{-\alpha^+T^+}
    (\overline{R}-\overline{\lambda})
    =c.
\]
Thus in either case,
\[
    \Pi_{\tau^+}^+
    \geq
    S_0\left[(R_0-\overline{\lambda})-c\right]
    \geq
    \frac{\delta}{2}S_0>0.
\]
Hence a ratio strictly above $\overline{\lambda}$ generates a
bounded-horizon random-maturity arbitrage.

The argument below the lower benchmark is symmetric. Define
\[
    u_t^-
    \equiv
    \underline{\lambda}S_t-F_t
    =
    S_t(\underline{\lambda}-R_t)>0,
\]
and use the reverse convergence trade: long one perpetual and short
$\underline{\lambda}$ units of spot. Its restoring rate is
\[
    a_t^-
    \equiv
    \frac{[r-\Phi(p_t)]F_t}{u_t^-}
    =
    \frac{R_t}{\underline{\lambda}-R_t}
    \left[
        r-\Phi\left(1-\frac{1}{R_t}\right)
    \right]
    >0.
\]
If
\[
    \underline{\lambda}-R_0\geq\delta>0,
\]
again set $c=\delta/2$ and define
\[
    \tau_1^-
    \equiv
    \inf\left\{
        t>0:\underline{\lambda}-R_t\leq c
    \right\}.
\]
Before $\tau_1^-$,
$\underline{R}\leq R_t\leq \underline{\lambda}-c$, so
\[
    a_t^-
    \geq
    \frac{\underline{R}}
         {\underline{\lambda}-\underline{R}}
    \left[
        r-\Phi\left(
            1-\frac{1}{\underline{\lambda}-c}
        \right)
    \right]
    \equiv\alpha^->0.
\]
With
\[
    H_t^-=\exp\left(-\int_0^t a_s^-\,ds\right),
\]
the scaled reverse convergence trade plus $c$ units of financed spot has
payoff
\[
    \Pi_t^-
    =
    S_0\left[(\underline{\lambda}-R_0)-c\right]
    +
    e^{-rt}S_t
    \left[
        c-H_t^-(\underline{\lambda}-R_t)
    \right].
\]
Choosing
\[
    T^-
    \equiv
    \frac{1}{\alpha^-}
    \log\left(
        \frac{\underline{\lambda}-\underline{R}}{c}
    \right),
    \qquad
    \tau^-\equiv\tau_1^-\wedge T^-,
\]
the same two-case argument gives
\[
    \Pi_{\tau^-}^-
    \geq
    \frac{\delta}{2}S_0>0.
\]

The constructions above are admissible for the same reason as in
Proposition 1: Assumption 2 places deterministic upper and lower bounds on
$R_t$, which bound the possible loss from the residual deviation before
liquidation.

Finally, suppose that for some date $t_0$,
$\Pr(R_{t_0}>\overline{\lambda})>0$. Since
\[
    \{R_{t_0}>\overline{\lambda}\}
    =
    \bigcup_{n\geq1}
    \left\{
        R_{t_0}\geq\overline{\lambda}+\frac{1}{n}
    \right\},
\]
there exists a deterministic $\delta>0$ and an event
$A\in\mathcal{F}_{t_0}$ with positive probability on which
$R_{t_0}\geq\overline{\lambda}+\delta$. Initiating the strategy above at
$t_0$ on $A$ and doing nothing on $A^c$ yields a random-maturity
arbitrage. The same argument applies if
$\Pr(R_{t_0}<\underline{\lambda})>0$. Hence absence of
random-maturity arbitrage requires
\[
    R_t\in
    [\underline{\lambda},\overline{\lambda}]
    \qquad\text{for all }t.
\]

\textit{Sufficiency.}
Take any
$\lambda\in[\underline{\lambda},\overline{\lambda}]$. Then
\[
    1-\frac{1}{\lambda}
    \in[x_-,x_+]
    =\Phi^{-1}(r),
\]
so
\[
    \Phi\left(1-\frac{1}{\lambda}\right)=r.
\]
If $F_t=\lambda S_t$ for all $t$, the gain on one long perpetual is
therefore
\[
    dF_t-\Phi(p_t)F_t\,dt
    =
    \lambda(dS_t-rS_t\,dt),
\]
which is exactly the gain on $\lambda$ units of spot financed at the
risk-free rate. Thus the perpetual is redundant given the spot and cash
markets, and the sufficiency argument of Proposition 1 implies that no
random-maturity arbitrage exists.

Economically, this explains why the benchmark is an interval rather than
a point. Inside
$[\underline{\lambda},\overline{\lambda}]$, the funding rate equals the
financing rate exactly, so the excess carry and hence the restoring rate
are zero. Arbitrage therefore rules out prices outside the interval but
does not select a unique point within it.
\end{proof}

\textbf{Proof to Corollary \ref{corollary2}:} 
\vspace{-0.1in}
\begin{proof}
Write $p=(F-S)/F$ for the premium index. Substituting the definition of the
clamp into the Binance funding rule gives
\[
    \Phi(p)
    =
    \begin{cases}
        \kappa p+\gamma,
        & \kappa p<\iota-\gamma,\\[3pt]
        \iota,
        & \iota-\gamma\leq \kappa p\leq\iota+\gamma,\\[3pt]
        \kappa p-\gamma,
        & \kappa p>\iota+\gamma.
    \end{cases}
\]
The three branches agree at the two boundaries, so $\Phi$ is continuous.
It is also weakly increasing, with slope $\kappa>0$ outside the clamp and
slope zero inside it.

The level set at the risk-free rate is therefore
\[
    \Phi^{-1}(r)
    =
    \begin{cases}
        \left\{\dfrac{r-\gamma}{\kappa}\right\},
        & r<\iota,\\[8pt]
        \left[
            \dfrac{\iota-\gamma}{\kappa},
            \dfrac{\iota+\gamma}{\kappa}
        \right],
        & r=\iota,\\[10pt]
        \left\{\dfrac{r+\gamma}{\kappa}\right\},
        & r>\iota.
    \end{cases}
\]
These sets are nonempty and compact. Moreover, the condition
$|r|+\gamma<\kappa$ guarantees that their upper endpoints are strictly
smaller than one. Hence the Binance funding rate satisfies Assumption 3.

Suppose first that $r\neq\iota$. Then $\Phi^{-1}(r)$ is a singleton. Its
unique element can be written as
\[
    x^*
    =
    \frac{
        r-\operatorname{sgn}(\iota-r)\gamma
    }{\kappa}.
\]
Applying Proposition 1 gives
\[
    F_t=\lambda S_t,
    \qquad
    \lambda
    =
    (1-x^*)^{-1}
    =
    \frac{\kappa}
    {\kappa+\operatorname{sgn}(\iota-r)\gamma-r},
\]
which is Equation~\eqref{benchmark_binance}.

When $r=\iota$, the level set is the clamp interval. Proposition 2
therefore gives
\[
    F_t
    \in
    \left[
        \frac{\kappa}{\kappa+\gamma-r}S_t,
        \frac{\kappa}{\kappa-\gamma-r}S_t
    \right],
\]
and the interval cannot be narrowed.

It remains to calculate the restoring rate when $r\neq\iota$. Consider
first $r<\iota$, so that the benchmark lies on the lower linear branch,
where $\Phi(p)=\kappa p+\gamma$. Using $p_tF_t=F_t-S_t$,
\begin{align*}
    [\Phi(p_t)-r]F_t
    &=
    (\kappa p_t+\gamma-r)F_t\\
    &=
    \kappa(F_t-S_t)+(\gamma-r)F_t\\
    &=
    (\kappa+\gamma-r)
    \left(
        F_t-
        \frac{\kappa}{\kappa+\gamma-r}S_t
    \right).
\end{align*}
Hence, writing $u_t=F_t-\lambda S_t$,
\[
    [\Phi(p_t)-r]F_t
    =
    (\kappa+\gamma-r)u_t,
\]
and the restoring rate is
\[
    a_t
    =
    \kappa+\gamma-r
    =
    \frac{\kappa}{\lambda}.
\]
When $r>\iota$, the same calculation on the upper branch,
$\Phi(p)=\kappa p-\gamma$, gives
\[
    a_t
    =
    \kappa-\gamma-r
    =
    \frac{\kappa}{\lambda}.
\]
Combining the two cases,
\[
    a_t
    =
    \kappa+\operatorname{sgn}(\iota-r)\gamma-r
    =
    \frac{\kappa}{\lambda}.
\]

Thus, on the linear branch containing the benchmark, excess funding is
exactly proportional to the price deviation. Economically, this is why
the restoring rate is constant there: doubling the deviation doubles the
excess funding flow.

For completeness, when the premium lies inside the clamp,
$\Phi(p_t)=\iota$. Since
\[
    \frac{u_t}{F_t}
    =
    \frac{p_t-x^*}{1-x^*},
\]
the restoring rate becomes
\[
    a_t
    =
    \frac{(\iota-r)F_t}{u_t}
    =
    (\iota-r)
    \frac{1-x^*}{p_t-x^*}.
\]
Unlike on the linear branch, the funding flow inside the clamp is fixed
as a fraction of notional rather than proportional to the deviation, so
the restoring rate falls as the deviation becomes larger.
\end{proof}

\textbf{Proof to Proposition 3:}
\vspace{-0.1in}
\begin{proof}
By Assumption 3, write
\[
    \Phi^{-1}(r)=[x^-,x^+],
\]
and define
\[
    \lambda^L \equiv (1-x^-)^{-1},
    \qquad
    \lambda^U \equiv (1-x^+)^{-1}.
\]
Since $p(R)\equiv 1-R^{-1}$ is strictly increasing,
\[
    p(\lambda^L)=x^-,
    \qquad
    p(\lambda^U)=x^+.
\]
Because $\Phi$ is weakly increasing and $\Phi^{-1}(r)=[x^-,x^+]$, it follows that
\[
    R>\lambda^U
    \quad\Longrightarrow\quad
    \Phi(p(R))>r,
\]
and
\[
    R<\lambda^L
    \quad\Longrightarrow\quad
    \Phi(p(R))<r.
\]
Thus the funding mechanism rewards the corresponding convergence trade whenever the futures-to-spot ratio lies outside $[\lambda^L,\lambda^U]$.

We first establish the upper bound. Suppose, to the contrary, that for
some date $t_0$,
\[
    \Pr\left(
        F_{t_0}-\lambda^U S_{t_0}>C
    \right)>0.
\]
Because $S_{t_0}$ is finite almost surely, there exist constants
$\varepsilon>0$ and $M<\infty$ such that the event
\[
    A
    \equiv
    \left\{
        F_{t_0}-\lambda^U S_{t_0}\geq C+\varepsilon
    \right\}
    \cap
    \{S_{t_0}\leq M\}
\]
has positive probability.

On $A$,
\[
    R_{t_0}-\lambda^U
    =
    \frac{F_{t_0}-\lambda^U S_{t_0}}{S_{t_0}}
    \geq
    \frac{C+\varepsilon}{M}
    \equiv\delta>0.
\]
Choose
\[
    0<c<\delta
    \qquad\text{such that}\qquad
    cM<\varepsilon.
\]
The distinction from Proposition 1 is important: here we do not fix
$c=\delta/2$. The tolerance $c$ may be chosen arbitrarily small so that
almost all of the initial deviation remains available to pay the trading
cost.

Reset the clock to $t_0$ on the event $A$. Define
\[
    u_t^+
    \equiv
    F_t-\lambda^U S_t
    =
    S_t(R_t-\lambda^U).
\]
As long as $u_t^+>0$, dynamically scale the rich-side convergence trade
using the restoring rate
\[
    a_t^+
    =
    \frac{[\Phi(p_t)-r]F_t}{u_t^+},
    \qquad
    H_t^+
    =
    \exp\left(-\int_0^t a_s^+\,ds\right).
\]
The calculation in Proposition 1 gives
\[
    V_t^+
    =
    u_0^+-H_t^+e^{-rt}u_t^+.
\]
Add $c$ units of spot financed through the cash market. The gross
discounted payoff becomes
\[
    \Pi_t^+
    =
    S_0\left[(R_0-\lambda^U)-c\right]
    +
    e^{-rt}S_t
    \left[
        c-H_t^+(R_t-\lambda^U)
    \right].
\]

Define
\[
    \tau_1^+
    =
    \inf\left\{
        t>0:R_t-\lambda^U\leq c
    \right\}.
\]
Before $\tau_1^+$,
\[
    a_t^+
    \geq
    \Phi\left(
        1-\frac{1}{\lambda^U+c}
    \right)-r
    \equiv\alpha^+>0.
\]
The strict inequality follows because $\lambda^U$ corresponds to the
upper endpoint of $\Phi^{-1}(r)$, so any larger ratio implies a premium
index at which $\Phi>r$. Hence
\[
    H_t^+\leq e^{-\alpha^+t},
    \qquad t<\tau_1^+.
\]
Choose
\[
    T^+
    =
    \frac{1}{\alpha^+}
    \log\left(
        \frac{\overline{R}-\lambda^U}{c}
    \right),
    \qquad
    \tau^+=\tau_1^+\wedge T^+.
\]
Exactly as in Proposition 1, at $\tau^+$,
\[
    c-H_{\tau^+}^+
      (R_{\tau^+}-\lambda^U)
    \geq0.
\]
The gross payoff is therefore at least
\[
    \Pi_{\tau^+}^+
    \geq
    S_0\left[(R_0-\lambda^U)-c\right]
    =
    u_0^+-cS_0.
\]
On the event $A$,
\[
    u_0^+-cS_0
    \geq
    C+\varepsilon-cM
    >C.
\]
Thus the strategy earns strictly more than $C$ before trading costs.
Because the total discounted trading cost is bounded above by $C$, the
net payoff is strictly positive on $A$. The strategy does nothing on
$A^c$, has nonpositive initial cost, and is liquidated by a deterministic
finite horizon. Hence it is a random-maturity arbitrage, a contradiction.
Therefore
\[
    F_t\leq\lambda^U S_t+C
\]
for every $t$.

The lower bound is symmetric. Suppose instead that
\[
    \Pr\left(
        \lambda^L S_{t_0}-F_{t_0}>C
    \right)>0.
\]
Choose $\varepsilon>0$ and $M<\infty$ such that
\[
    A^-
    \equiv
    \left\{
        \lambda^L S_{t_0}-F_{t_0}\geq C+\varepsilon
    \right\}
    \cap
    \{S_{t_0}\leq M\}
\]
has positive probability, and set
\[
    \delta
    \equiv
    \frac{C+\varepsilon}{M}.
\]
Choose $c\in(0,\delta)$ so that $cM<\varepsilon$.

With
\[
    u_t^-
    \equiv
    \lambda^L S_t-F_t
    =
    S_t(\lambda^L-R_t),
\]
use the dynamically scaled reverse convergence trade,
\[
    a_t^-
    =
    \frac{[r-\Phi(p_t)]F_t}{u_t^-},
    \qquad
    H_t^-
    =
    \exp\left(-\int_0^t a_s^-\,ds\right).
\]
Adding $c$ units of financed spot gives
\[
    \Pi_t^-
    =
    S_0\left[(\lambda^L-R_0)-c\right]
    +
    e^{-rt}S_t
    \left[
        c-H_t^-(\lambda^L-R_t)
    \right].
\]
Define
\[
    \tau_1^-
    =
    \inf\left\{
        t>0:\lambda^L-R_t\leq c
    \right\}.
\]
Before $\tau_1^-$,
\[
    a_t^-
    \geq
    \frac{\underline{R}}
         {\lambda^L-\underline{R}}
    \left[
        r-\Phi\left(
            1-\frac{1}{\lambda^L-c}
        \right)
    \right]
    \equiv\alpha^->0.
\]
Choose
\[
    T^-
    =
    \frac{1}{\alpha^-}
    \log\left(
        \frac{\lambda^L-\underline{R}}{c}
    \right),
    \qquad
    \tau^-=\tau_1^-\wedge T^-.
\]
The same argument gives
\[
    \Pi_{\tau^-}^-
    \geq
    u_0^--cS_0
    \geq
    C+\varepsilon-cM
    >C
\]
on $A^-$. After trading costs, the payoff remains strictly positive,
again contradicting the absence of random-maturity arbitrage. Hence
\[
    F_t\geq\lambda^L S_t-C.
\]
Combining the upper and lower bounds gives
\[
    \lambda^L S_t-C
    \le F_t
    \le \lambda^U S_t+C,
\]
which proves the proposition.

For the Binance funding-rate specification, substituting the endpoints of $\Phi^{-1}(r)$ derived in Corollary \ref{corollary2} gives Equations \eqref{eq_no_arb_bound_binance_point} and \eqref{eq_no_arb_bound_binance_interval}.

\end{proof}

\clearpage
\newpage
\section{Stochastic Interest Rates}\label{sec:stochastic}

In this appendix, we relax Assumption \ref{assp1} and examine how stochastic interest rates change our analysis. We first show that the payoff to the arbitrage strategies changes by exactly one term: a rebalancing cash flow from changes in hedge ratios. Because the hedge ratio tracks interest rates, this term introduces additional interest rate risk.

We then ask how large instantaneous interest rate changes would have to be to rationalize the deviations we observe. We find empirically that the implied interest rate changes turn out to be one to two orders of magnitude larger than any observed movement in interest rates. This suggests stochastic interest rate risk is unlikely to account for the observed deviations of perpetual futures from the benchmark.

When the interest rate is stochastic, define the discount factor as
\begin{equation}
D_t = \exp\left( -\int_0^t r_s ds \right). \label{discounting}
\end{equation}
A stochastic rate affects the arbitrage strategies of Table \ref{tabcase1} through two channels. First, the hedging ratio of Corollary \ref{corollary2} depends on the interest rate, so the arbitrageur must adjust her spot holdings as the rate moves. Second, financing, reinvestment, and discounting run at the prevailing rate $r_s$.

Mirroring the proof of Proposition \ref{prop2}, for each strategy in Table \ref{tabrandomr}, we consider a hedge ratio tailored to its own side of the benchmark:
\begin{equation}
    \overline{\lambda}(r) \equiv \frac{\kappa}{\kappa - \gamma - r} \quad \text{for the first strategy}, \qquad \underline{\lambda}(r) \equiv \frac{\kappa}{\kappa + \gamma - r} \quad \text{for the second}. \label{eq:envelopes}
\end{equation}
The two expressions extend the two branches of the proportionality constant in Corollary \ref{corollary2} across all interest rate levels: $\overline{\lambda}(r) = \lambda(r)$ when $r > \iota$, and $\underline{\lambda}(r) = \lambda(r)$ when $r < \iota$. The advantage of this choice is smoothness: whereas $\lambda(r)$ switches branches discretely when the rate crosses $\iota$, each of $\overline{\lambda}$ and $\underline{\lambda}$ varies continuously with the rate, so tracking the hedge never requires a discrete jump in position.

Table \ref{tabrandomr} presents the cash flows of the two strategies with a rate-contingent hedge ratio $\lambda_s$, equal to $\overline{\lambda}(r_s)$ for the first strategy and to $\underline{\lambda}(r_s)$ for the second. Relative to Table \ref{tabcase1}, there is one new row: over each interval, the first strategy purchases $d\lambda_s$ units of spot at price $S_s$ to keep the hedge matched to the prevailing rate, a rebalancing cash flow of $-S_s d\lambda_s$ financed through the money market account.

\begin{table}[!htb]
\begin{center}
\footnotesize
\begin{tabularx}{\textwidth}{llYY}
\toprule
& & $F_0 > \lambda_0 S_0$ & $F_0 < \lambda_0 S_0$ \\
\midrule
Actions & & Long $\lambda_s$ spot, short 1 futures & Long 1 futures, short $\lambda_s$ spot \\
\midrule
& Futures & 0 & 0\\
Time $0$& Spot & $-\lambda_0 S_0$ & $+\lambda_0 S_0$ \\
& Cash & $+\lambda_0 S_0$ & $-\lambda_0 S_0$ \\
\midrule
\multirow{5}{*}{Time $t$}& Futures & $-\int_{0}^t D_s dF_s$ & $\int_{0}^t D_s dF_s$ \\
& Funding & $\int_0^t \Phi(p_s) D_s F_s ds$ & $-\int_0^t \Phi(p_s) D_s F_s ds$ \\
& Spot & $\int_0^t \lambda_s D_s dS_s$ & $-\int_0^t \lambda_s D_s dS_s$ \\
& Cash & $-\int_0^t r_s \lambda_s D_s S_s ds$ & $ \int_0^t r_s \lambda_s D_s S_s ds$ \\
\cmidrule{2-4}
& \multirow{2}{*}{Payoff} & $-\int_0^t D_s (d F_s - \lambda_s d S_s - S_s d\lambda_s)+$ & $\int_0^t D_s (d F_s - \lambda_s d S_s - S_s d\lambda_s)-$ \\
& & $\int_{0}^t D_s \left(\Phi(p_s) F_s - r_s\lambda_s S_s\right)ds - \int_0^t D_s S_s d\lambda_s$ & $\int_{0}^t D_s \left( \Phi(p_s) F_s - r_s\lambda_s S_s\right)ds + \int_0^t D_s S_s d\lambda_s$ \\
\bottomrule
\end{tabularx}
\end{center}
\caption{Discounted payoffs to arbitrage strategies under a stochastic short rate}\label{tabrandomr}
\bigskip
\small
This table presents the costs and benefits of two arbitrage trading strategies with the rate-contingent hedge ratio $\lambda_s$, equal to $\overline{\lambda}(r_s)$ of Equation \eqref{eq:envelopes} for the first strategy and to $\underline{\lambda}(r_s)$ for the second.
\end{table}

To derive the payoff of the first strategy, let $B_s$ denote its money market balance, which absorbs all interim cash flows:
\begin{equation}
    dB_s = r_s B_s\, ds - dF_s + \Phi(p_s) F_s\, ds - S_s\, d\lambda_s, \qquad B_0 = -\lambda_0 S_0,
\end{equation}
so that wealth is $W_s = B_s + \lambda_s S_s$, with $W_0 = 0$, and the discounted payoff is $\Pi_t = D_t W_t$. For interest rate paths of finite variation, the product rule contains no covariation term, $d\left(\lambda_s S_s\right) = \lambda_s dS_s + S_s d\lambda_s$. Discounting therefore gives
\begin{equation}
    d\left(D_s W_s\right) = D_s\left[-dF_s + \lambda_s dS_s + \left(\Phi(p_s) F_s - r_s \lambda_s S_s\right)ds\right].
\end{equation}
Defining the deviation relative to the strategy's hedge, $u_s \equiv F_s - \lambda_s S_s$, substituting $-dF_s + \lambda_s dS_s = -du_s - S_s d\lambda_s$, and integrating by parts, we obtain
\begin{equation}
    \Pi_t = u_0 - D_t u_t + \int_0^{t} D_s\left(\Phi(p_s) - r_s\right) F_s\, ds - \int_0^{t} D_s S_s\, d\lambda_s. \label{eq:payoff_stoch}
\end{equation}
The payoff of the second strategy is the negative of the same expression with $\lambda_s = \underline{\lambda}(r_s)$. When the short rate is constant, $d\lambda_s \equiv 0$, the hedge ratio of the profitable side coincides with the proportionality constant of Corollary \ref{corollary2}, and Equation \eqref{eq:payoff_stoch} reduces to the constant-rate payoff underlying Inequality \eqref{breakdown}.

The first three terms of Equation \eqref{eq:payoff_stoch} are the familiar components of the constant-rate payoff: the initial deviation, the discounted deviation at unwinding, and the present value of accumulated funding payments net of financing costs. Whenever the first strategy's position is open, $u_s > 0$, the premium index lies above the point at which the funding rate equals the prevailing interest rate, so $\Phi(p_s) > r_s$ and the funding flow is positive; the second strategy is symmetric.\footnote{For the first strategy, $u_s > 0$ is equivalent to $\frac{F_s - S_s}{F_s} > \frac{r_s + \gamma}{\kappa}$, the root of $\Phi(p_s) = r_s$ on the upper linear branch of the funding rate function. When $r_s \geq \iota$, the premium index then lies on the upper branch above this root, so $\Phi(p_s) = \kappa\frac{F_s - S_s}{F_s} - \gamma > r_s$. When $r_s < \iota$, the premium index lies above the lower edge of the clamp region (because $2\gamma > \iota$), so $\Phi(p_s) \geq \iota > r_s$. The verification for the second strategy is symmetric.} 

The genuinely new term is the rebalancing cash flow $\int_0^t D_s S_s d\lambda_s$: the arbitrageur adjusts a full spot notional as the rate moves her hedge ratio. The rebalancing term is closely related to changes in the interest rate. Because the hedge ratios in Equation \eqref{eq:envelopes} are smooth in the rate, $d\lambda_s = \overline{\lambda}{}'(r_s)\, dr_s$ with $\overline{\lambda}{}'(r) = \kappa/\left(\kappa - \gamma - r\right)^{2} \approx 1/\kappa$, and the term can be approximated by:
\begin{equation}
    \int_0^t D_s S_s\, d\lambda_s = \int_0^t D_s S_s \overline{\lambda}{}'(r_s) dr_s \approx \int_0^t (D_s S_s / \kappa) dr_s , \label{eq:leakage_bound}
\end{equation}
the analogous expression holding for $\underline{\lambda}$. We can understand this term as the discounted notional amount times the changes in interest rate. Since the interest rate, like the funding parameters, is measured per funding interval, a cumulative movement of one percentage point in the annualized interest rate over the life of a trade costs the arbitrageur at most about $0.1$ basis point of spot notional. 

Based on this framework, we then quantify the empirical relevance of interest rate risk for our results. We ask how large an instantaneous interest rate change would have to be to rationalize the deviations we observe.

The proof of Proposition \ref{prop1} shows that the fundamental price is pinned down by the condition that the funding rate equals the risk-free rate, $\Phi(p_t) = r$. Inverting this condition at observed prices defines the implied interest rate
\begin{equation}
    r^{*}_t \equiv \Phi\left(\frac{F_t - S_t}{F_t}\right) = \kappa \frac{F_t - S_t}{F_t} + clamp\left(\iota - \kappa \frac{F_t - S_t}{F_t}, -\gamma, \gamma\right), \label{eq:implied_rate}
\end{equation}
the level of the risk-free rate at which observed futures and spot prices would satisfy the no-arbitrage restrictions. 

We can also derive the measure using arbitrage payoff. Consider a one-time permanent change in the short rate from $r_0$ to $r'$ at time $0$. The last rebalancing term becomes:
\begin{equation}
    \int_0^t D_s S_s\, d\lambda_s = S_0\left[\overline{\lambda}(r') - \overline{\lambda}(r_0)\right],
\end{equation}
while the discount factor becomes $D_s = e^{-r' s}$. Substituting both into \eqref{eq:payoff_stoch}, the total payoff becomes:
\begin{equation}
    \Pi_t = F_0 - \overline{\lambda}(r')S_0 - e^{-r' t}\left[F_t - \overline{\lambda}(r')S_t\right] + \int_{0}^{t} e^{-r' s}\left(\Phi(p_s) - r'\right) F_s\, ds. \label{eq:payoff_jump}
\end{equation}
This is exactly the payoff to the arbitrage strategy with the initial deviation defined by the new rate $r'$. Based on the argument in Proposition \ref{prop1}, the smallest interest rate would equal the funding rate $\Phi(p_t)$ to erase future profits for the arbitrageur.

The difference $r^{*}_t - r_t = \Phi(p_t) - r_t$ is accordingly the instantaneous interest rate change that would erase the arbitrageur's gain: at the rate $r^{*}_t$, funding payments would exactly offset financing costs, and the hedged position would earn no reward for convergence. The interest rate cost would cancel the gains from the initial price gap. As a result, the total payoff becomes zero under the instantaneous interest rate change.

\begin{figure}[!htb]
    \begin{center}
    \includegraphics[width=\linewidth]{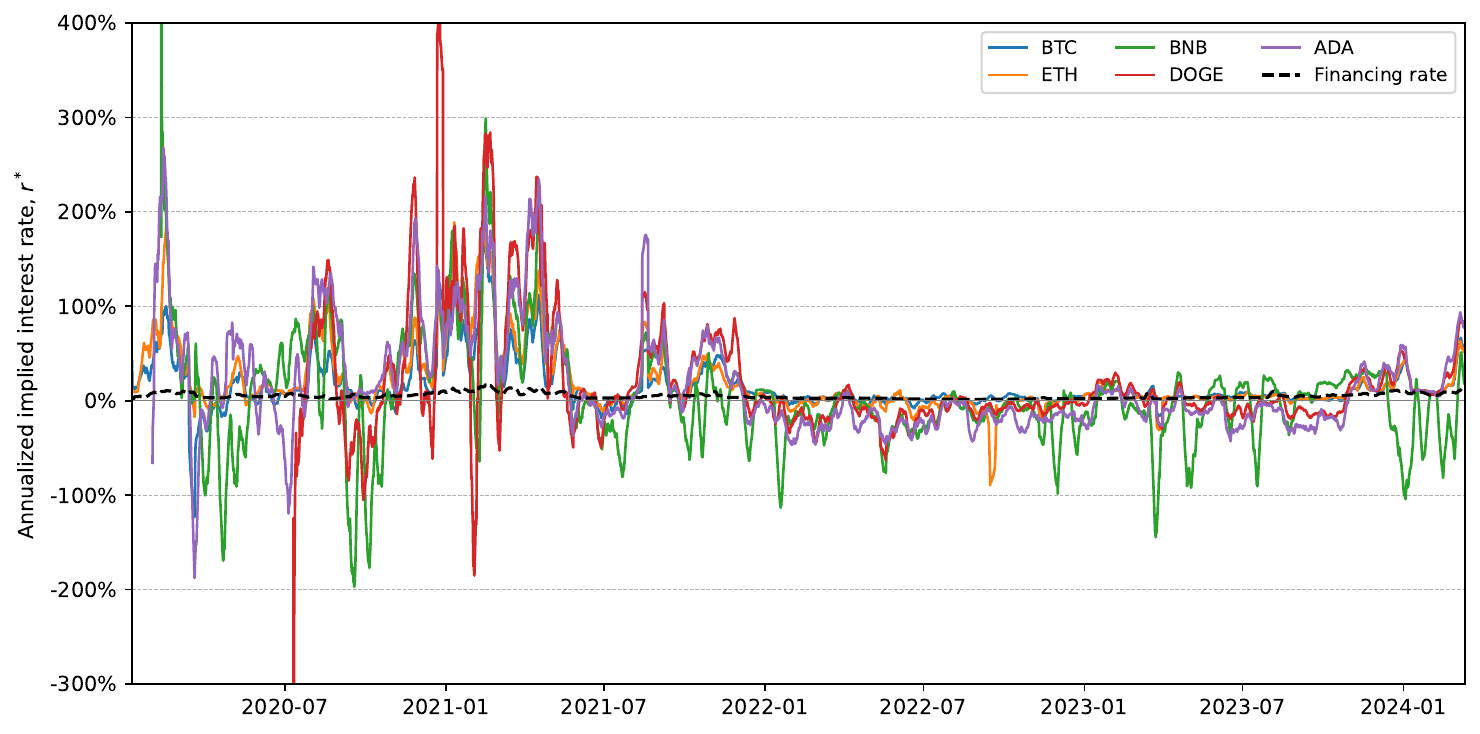}
    \end{center}
    \caption{Implied Interest Rates from Perpetual Futures Prices}
    \label{fig:implied_rate}
    \bigskip
    \small
    This figure presents the 7-day moving averages of the annualized implied interest rate $r^{*}$ of Equation \eqref{eq:implied_rate}, the interest rate level at which observed perpetual futures and spot prices would be arbitrage-free, for the five cryptocurrencies. The dashed black line is the annualized DeFi financing rate.
\end{figure}

Figure \ref{fig:implied_rate} plots the 7-day moving average of the implied interest rate for each of the five cryptocurrencies, together with the DeFi financing rate. The gap between the two is significant. The financing rate averages $4.9\%$ per year, with a standard deviation of $3.6\%$, and remains between $0.4\%$ and $25\%$ throughout the sample. The implied rate, by contrast, swings between roughly $-200\%$ and $+400\%$ per year in the early sample and remains far more volatile than the interest rate thereafter.

Table \ref{tab:implied_rate} summarizes the distribution of the implied interest rate $r^{*}$. Its standard deviation ranges from $62\%$ to $174\%$ per year across currencies, orders of magnitude larger than the level of the interest rate itself. The implied rate is also negative in $20\%$ to $34\%$ of hours, a region a nonnegative financing rate cannot reach.

These comparisons are unconditional. The next question is whether the required change is large relative to the interest rate risk prevailing at the moment. The distinction matters here because, as Figure \ref{fig:interest_rate} shows, the volatility of the financing rate is itself far from constant.

\begin{table}[!htb]
\begin{center}
\begin{tabularx}{1\columnwidth}{@{\hskip\tabcolsep\extracolsep\fill}l*{6}{c}}
\toprule
Asset & Mean & Std & p5 & p95 & $>2\sigma$ (\%) & $>6\sigma$ (\%) \\
\midrule
BTC & 0.15 & 0.62 & -0.18 & 0.78 & 84.4 & 50.6 \\
ETH & 0.20 & 0.64 & -0.20 & 1.07 & 86.2 & 53.0 \\
BNB & 0.04 & 0.95 & -1.21 & 1.25 & 92.1 & 68.5 \\
DOGE & 0.19 & 1.74 & -0.70 & 1.71 & 90.7 & 67.9 \\
ADA & 0.18 & 1.03 & -0.72 & 1.47 & 91.6 & 66.8 \\
\bottomrule
\end{tabularx}
\end{center}
\caption{Summary Statistics of Implied Interest Rates}\label{tab:implied_rate}
\bigskip
\small
This table summarizes the annualized implied interest rate $r^{*}$ of Equation \eqref{eq:implied_rate}, the level of the risk-free rate at which observed perpetual futures and spot prices would be arbitrage-free. Mean, Std, p5, and p95 describe the distribution of $r^{*}$ over the hourly samples laid out in Table \ref{tab2}. $>2\sigma$ (\%) and $>6\sigma$ (\%) are the shares of hours in which the instantaneous rate change required to erase the arbitrageur's gain exceeds two and six conditional standard deviations of the financing rate, $\left|r^{*}_t - r_t\right| > k\sigma_t$ for $k = 2$ and $k = 6$. The conditional volatility $\sigma_t$ is the residual standard deviation of the AR(1) of Equation \eqref{eq:vasicek}.
\end{table}

We therefore estimate the conditional volatility of the financing rate from a rolling Vasicek fit. For each date $t$, we estimate the AR(1) regression
\begin{equation}
    r_{s+1} = a_t + b_t r_s + \varepsilon_{s+1}, \qquad \varepsilon_{s+1} \sim \left(0, \sigma_t^2\right), \label{eq:vasicek}
\end{equation}
by OLS on the $90$ daily observations of the annualized financing rate strictly preceding $t$, and take the standard deviation of the fitted residuals, $\sigma_t$, as the conditional one-day volatility of the level of the rate. The procedure recovers the heteroskedasticity visible in Figure \ref{fig:interest_rate}: the median $\sigma_t$ is $1.9$ percentage points per year, but $\sigma_t$ ranges from $0.05$ to $4.7$ percentage points

The last two columns of Table \ref{tab:implied_rate} report the share of hours in which the required change exceeds two and six conditional standard deviations. The required change is beyond $2\sigma_t$ in $84\%$ to $92\%$ of hours and beyond $6\sigma_t$ in $51\%$ to $68\%$.

In sum, a stochastic short rate leaves the structure of the arbitrage payoff intact up to a small rebalancing term. The interest rate movements that would be required to rationalize observed deviations are implausibly large: they are frequently infeasible outright for defi rates, and where they are feasible they are routinely six or more conditional standard deviations of the prevailing financing rate. The evidence suggests that the deviation from our no-arbitrage benchmark is unlikely to come from stochastic interest rates.

\clearpage
\newpage
\section{A Model of Deviations in Perpetual Futures}\label{sec:model}

Section \ref{sec:explaining} attributes the futures-spot deviation to the interaction between end users' demand for leverage and the frictions faced by the arbitrageurs who accommodate it. This appendix uses the simple theoretical model to illustrate how these two forces influence the equilibrium deviation $\rho$. The model delivers two predictions that guide our empirical analysis. First, greater arbitrage supply sensitivity and looser capital constraints would reduce equilibrium deviations. Second, the relationship between end-user demand and the deviation can be kinked: once arbitrage capital becomes constrained, each additional dollar of demand generates a larger deviation.

Consider an environment with two groups of participants: end users, who demand leveraged exposure through perpetual futures, and arbitrageurs, who accommodate this demand by taking the opposite position and hedging in the spot market.

\paragraph{End users.} End users value the leverage provided by perpetual futures. For example, a trader who obtains one dollar of long exposure through the perpetual futures rather than purchasing the underlying token and financing the purchase in the cash market pays the deviation $\rho_t$ per unit of time. Thus, $\rho_t$ can be interpreted as the price of obtaining leveraged exposure through the perpetual market. We model aggregate net demand for perpetual exposure as
\begin{equation}
    Q^{d}_t = \theta_t - \eta \rho_t, \qquad \eta > 0, \label{eq:model_demand}
\end{equation}
where $\theta_t$ is a demand shifter and $\eta$ captures the sensitivity of end-user demand to the deviation. In our empirical analysis, we proxy for shifts in $\theta_t$ using past returns.

\paragraph{Arbitrageurs.} Arbitrageurs accommodate end-user demand by taking an opposite position in the perpetual with an offsetting position in the spot market, as in the strategies described in Table \ref{tabcase1}. Per dollar of intermediated notional, the trade earns the deviation $\rho_t$ but is subject to several costs and constraints. The first cost is the marginal cost $k$ of opening her position. There can be many forms of such cost, for example, the trading costs $\kappa c$ which we study extensively, as well as the expected cost of exchange default.

Arbitrageurs may also face a fixed cost of entry, as studied by \citet{bryzgalova2025strategic}. Our model focuses on the equilibrium after the entry decision has been made.

Secondly, arbitrageurs bear the risk that the deviation widens before the position can be unwound in the perpetual market. Let $\sigma^2$ denote the variance of the deviation per unit of time, $h_t$ the holding period, and $A$ the arbitrageur's risk aversion. Under mean-variance preferences, an arbitrage position with notional $X$ therefore entails a risk-bearing cost of $\frac{A}{2}X^2\sigma^2h_t$.

Third, arbitrageurs face margin constraints. Following \citet{brunnermeier2009market}, let $W_t$ denote the capital available to arbitrageurs and $m$ the amount of margin required per dollar of intermediated notional. The position must then satisfy $mX\leq W_t.$

Consider a positive deviation $\rho_t > k$, so arbitrageurs have an incentive to enter the market. An arbitrageur chooses the size of the arbitrage position to solve:
\begin{equation}
    \max_{X}\; X\left(\rho_t - k\right) - \frac{A}{2}X^{2}\sigma^{2}h_t \quad \text{subject to} \quad m X \leq W_t,
\end{equation}
so that her supply of the arbitrage trade is
\begin{equation}
    X_t = \min\left\{\Lambda_t\left(\rho_t - k\right),\; \frac{W_t}{m}\right\}, \qquad \Lambda_t \equiv \frac{1}{A\sigma^{2}h_t}. \label{eq:model_supply}
\end{equation}
We refer to $\Lambda_t$ as arbitrage supply sensitivity, defined as the amount of notional that arbitrageurs are willing to intermediate per unit increase in the deviation when they are unconstrained. We refer to $\frac{W_t}{m}$ as arbitrage capacity, the maximum amount of notional the arbitrageur can intermediate given the margin requirement.

The holding period $h_t$ is where the perpetual departs from a fixed-maturity contract. In a contract that expires at $T$, convergence is guaranteed at delivery, and the horizon is capped at $T - t$, so $h_t$ is bounded and known when the position is opened. In a perpetual, convergence occurs at a random time in the sense of Definition \ref{def:random_maturity_arb} and the compensation accrues as a flow of funding payments rather than at a terminal date, so $h_t$ is random.

\paragraph{Equilibrium.} Setting $Q^{d}_t = X_t$ gives
\begin{equation}
    \rho_t = \frac{\theta_t + \Lambda_t k}{\eta + \Lambda_t} \quad \text{if capital constraint is slack}, \qquad \rho_t = \frac{\theta_t - W_t/m}{\eta} \quad \text{if it binds}. \label{eq:model_eq}
\end{equation}

\noindent Which branch applies depends on the state of end-user demand relative to available arbitrage capital. The capital constraint is slack when $\Lambda_t (\rho_t - k) < W_t/m$. Holding $\Lambda_t$, $W_t$ and $m$ fixed, this condition is satisfied for sufficiently low realizations of $\theta_t$, whereas high realizations cause the capital constraint to bind. The equilibrium deviation is therefore an increasing, kinked function of end-user demand: its sensitivity to $\theta_t$ rises once arbitrage capital becomes constrained. This equilibrium characterization delivers two empirical predictions.

\subsection*{Prediction 1: Greater arbitrage supply sensitivity and looser capital constraints reduce equilibrium deviations}

In the slack branch of Equation \eqref{eq:model_eq},
\begin{equation}
    \frac{\partial \rho_t}{\partial \Lambda_t} = \frac{k\eta - \theta_t}{\left(\eta + \Lambda_t\right)^{2}} < 0,
\end{equation}
since $\rho_t > k$ is equivalent to $\theta_t > k\eta$. As $\Lambda_t \rightarrow \infty$ the equilibrium deviation approaches the cost floor $k$, which is the no-arbitrage bound of Proposition \ref{prop3}: with perfectly elastic arbitrage capital, the deviation collapses to the compensation for trading costs and venue risk. In the binding branch,
\begin{equation}
    \frac{\partial \rho_t}{\partial \left(W_t/m\right)} = -\frac{1}{\eta} < 0,
\end{equation}
so relaxing the margin requirement $m$, holding arbitrage capital $W_t$ fixed, compresses the deviation directly.

Section \ref{sec:explain_pm} tests this prediction on Binance's introduction of portfolio margining in April 2022, which allowed the spot and futures legs of the arbitrage to be collateralized inside a single account and therefore reduces $m$. The test provides evidence consistent with the model's prediction.

\subsection*{Prediction 2: Demand has a larger price impact when arbitrage capacity binds}

Differentiating Equation \eqref{eq:model_eq} with respect to the demand shifter gives
\begin{equation}
    \frac{\partial \rho_t}{\partial \theta_t} = \frac{1}{\eta + \Lambda_t} \quad \text{if capital constraint is slack}, \qquad \frac{\partial \rho_t}{\partial \theta_t} = \frac{1}{\eta} \quad \text{if it binds},
\end{equation}
and $1/\eta > 1/(\eta + \Lambda_t)$. When the constraint is slack, an increase in demand is partly absorbed by an expansion in arbitrage positions, dampening its effect on the deviation. Once the constraint binds, arbitrageurs can no longer expand their positions, so market clearing requires a larger adjustment in the deviation. The arbitrage supply curve is therefore kinked, with demand shocks producing larger price impacts in the constrained region.

The constrained region is also associated with larger equilibrium deviations, motivating the use of the initial deviation as an empirical proxy for proximity to the constrained regime. 

Section \ref{sec:explain_impact} takes this prediction to the data by estimating the price impact of perpetual order flow across different sizes of the initial premium index. Consistent with the model, order flow has a larger price impact when the initial deviation is larger, when arbitrage capital is more likely to be constrained. The evidence therefore supports the model's prediction that capital constraints amplify the price impact of end-user demand.

\end{document}